\documentclass[12pt]{article}
\usepackage{amsmath,amssymb}
\usepackage{graphicx,psfrag,epsf}
\usepackage{enumerate}
\usepackage{natbib}
\usepackage{soul}
\usepackage[normalem]{ulem}
\usepackage{subfigure,lscape}
\usepackage{float,multirow,booktabs}
\usepackage{placeins}
\usepackage[table,xcdraw]{xcolor}
\usepackage{url} % not crucial - just used below for the URL
\usepackage{rotating,titlesec}
\usepackage{algorithm,amsmath,tabularx}
\usepackage[noend]{algpseudocode}
\newcounter{algsubstate}
\renewcommand{\thealgsubstate}{\alph{algsubstate}}
\newenvironment{algsubstates}
{\setcounter{algsubstate}{0}%
	\renewcommand{\State}{%
		\stepcounter{algsubstate}%
		\Statex {\footnotesize\thealgsubstate:}\space}}

\makeatletter
\newcommand{\multiline}[1]{%
	\begin{tabularx}{\dimexpr\linewidth-\ALG@thistlm}[t]{@{}X@{}}
		#1
	\end{tabularx}
}
\makeatother

\usepackage[utf8]{inputenc}
\usepackage[english]{babel}

\newcommand{\blind}{1}

\def\0{\mbox{\boldmath{$\mathbf{0}$}}}
\def\1{\mbox{\boldmath{$\mathbf{1}$}}}
\def\bzeta{\mbox{\boldmath$\zeta$}}

\def\bmu{\mbox{\boldmath$\mu$}}
\def\bbeta{\mbox{\boldmath$\beta$}}
\def\bchi{\mbox{\boldmath$\chi$}}

\def\bzeta{\mbox{\boldmath$\zeta$}}

\def\bGamma{\mbox{\boldmath$\Gamma$}}

\def\bSigma{\mbox{\boldmath$\Sigma$}}

\def\blambda{\mbox{\boldmath$\lambda$}}
\def\bLambda{\mbox{\boldmath$\Lambda$}}
\def\calF{\mbox{$\mathcal{F}$}}
\def\calH{\mbox{$\mathcal{H}$}}

\def\c{\mbox{\boldmath{$\mathbf{c}$}}}

\def\x{\mbox{\boldmath{$\mathbf{x}$}}}
\def\y{\mbox{\boldmath{$\mathbf{y}$}}}

\begin{document}
	\setstcolor{red}

	\def\spacingset#1{\renewcommand{\baselinestretch}%
		{#1}\small\normalsize} \spacingset{1}
	%\input{definitions}
	
	%%%%%%%%%%%%%%%%%%%%%%%%%%%%%%%%%%%%%%%%%%%%%%%%%%%%%%%%%%%%%%%%%%%%%%%%%%%%%%
	
	\if1\blind
	{
		\title{\bf Tumor Radiogenomics with Bayesian Layered Variable Selection}
		\author{Shariq Mohammed$^{a,b,*}$, Sebastian Kurtek$^{c}$, Karthik Bharath$^{d}$,\\
			 Arvind Rao$^{a,b}$ and Veerabhadran Baladandayuthapani$^{a}$\\~\\
			$^{a}$Department of Biostatistics, University of Michigan\\
			$^{b}$Department of Computational Medicine \& Bioinformatics, University of Michigan\\
			$^{c}$Department of Statistics, Ohio State University\\
			$^{d}$School of Mathematical Sciences, University of Nottingham\\
			$^{*}${\small corresponding author. E-mail: shariqm@umich.edu}
			}
		\maketitle
	} \fi
	
	\if0\blind
	{
		\bigskip
		\bigskip
		\begin{center}
			{\Large\bf Tumor Radiogenomics with Bayesian Layered Variable Selection}
		\end{center}
		\medskip
	} \fi
	
	\bigskip
	\begin{abstract}
		We propose a statistical framework to integrate radiological magnetic resonance imaging (MRI) and genomic data to identify the underlying radiogenomic associations in lower grade gliomas (LGG). We devise a novel imaging phenotype by dividing the tumor region into concentric spherical layers that mimics the tumor evolution process. MRI data within each layer is represented by voxel--intensity-based probability density functions which capture the complete information about tumor heterogeneity. Under a Riemannian-geometric framework these densities are mapped to a vector of principal component scores which act as imaging phenotypes. Subsequently, we build Bayesian variable selection models for each layer with the imaging phenotypes as the response and the genomic markers as predictors. Our novel hierarchical prior formulation incorporates the interior-to-exterior structure of the layers, and the correlation between the genomic markers. We employ a computationally-efficient Expectation--Maximization-based strategy for estimation. Simulation studies demonstrate the superior performance of our approach compared to other approaches. With a focus on the cancer driver genes in LGG, we discuss some biologically relevant findings. Genes implicated with survival and oncogenesis are identified as being associated with the spherical layers, which could potentially serve as early-stage diagnostic markers for disease monitoring, prior to routine invasive approaches. %We provide a \texttt{R} package that can be used to deploy our framework to identify radiogenomic associations.
	\end{abstract}
	
	\noindent%
	{\it Keywords:} cancer driver genes, lower grade gliomas, radiogenomic associations, spike-and-slab prior.
	\vfill
	
	\newpage
	\spacingset{1.45} % DON'T change the spacing!
	
	\section{Introduction}\label{sec: intro}
	Low-grade gliomas (LGG) are infiltrative brain neoplasms characterized as World Health Organization (WHO) grade II and III neoplasms. LGG is a uniformly fatal disease of young adults (mean age: 41 years), with survival times averaging approximately seven years \citep{claus2015survival}. LGG patients usually have better survival than patients with high-grade (WHO grade IV) gliomas. However, some of the LGG tumors recur after treatment and progress into high-grade tumors, making it essential to understand the underlying etiology, and to improve the treatment management and monitoring for LGG patients.
	
	Due to recent technological advances, large multi-modal datasets are being produced; these include imaging as well as multi-platform genomics data. In complex disease systems such as cancer,  integrative analyses of such data can reveal important biological insights into specific disease mechanisms and its subsequent clinical translation \citep{wang2013ibag,morris2017statistical}. Genomic profiling technologies, such as microarrays, next-generation sequencing, methylation arrays and proteomic analyses, have facilitated thorough investigations at the molecular level. Such multi-platform genomic data resources have been used to develop models to better understand the molecular characterization across cancers and especially in gliomas \citep{ohgaki2004genetic, ohgaki2007genetic}. While genomic data provides information on the molecular characterization of the disease, radiological imaging, such as magnetic resonance imaging (MRI), computed tomography (CT) and positron emission tomography (PET), provide complementary information about the structural aspects of the disease. In this context, \emph{radiomic} analysis involves mining and extraction of various types of quantitative imaging features from different modalities obtained through high-throughput radiology images \citep{bakas2017advancing}. These image-derived features describe various characteristics such as morphology and texture, among others. Integrative analyses of genomic and radiomic features, commonly referred to as {\it radiogenomic} analysis, capture complementary characteristics of the underlying tumor \citep{gevaert2014glioblastoma,mazurowski2017radiogenomics}. However, such integrative modeling approaches present multiple analytical and computational challenges including incorporating complex biological structure, and high-dimensionality of quantitative imaging/genomic markers, necessitating principled and biologically-informed dimension reduction and information extraction techniques.
	
	Existing radiomic and radiogenomic studies typically consider  voxel-level summary features (based on intensity histogram, geometry, and texture analyses) as representations of a tumor image, to assess the progression (or regression) of tumors using statistical or machine learning models \citep{lu2018machine,elshafeey2019multicenter}. The main drawbacks of using voxel-intensity summary statistics/features are: (i) they do not capture the entire information in the voxel intensity histogram; (ii) there is subjectivity in the choice of the number of features and the location where they are computed. Consequently, the radiomic/radiogenomic analyses using summary features are unable to detect potential small-scale and sensitive changes in the tumor \citep{just2014improving}. To address these challenges, we propose a \textit{de novo} strategy for quantifying the tumor heterogeneity using voxel-intensity based imaging phenotypes that mimic the tumor evolution process.
	
	\subsection{De Novo Quantification of Imaging Phenotypes}\label{subsec: imaging}
	
\begin{figure}[t]
		\centering
		\begin{tabular}{|c|}
			\hline
			\includegraphics[trim = 3.25cm 2cm 8cm 1.5cm, clip, scale = 0.45, page = 2]{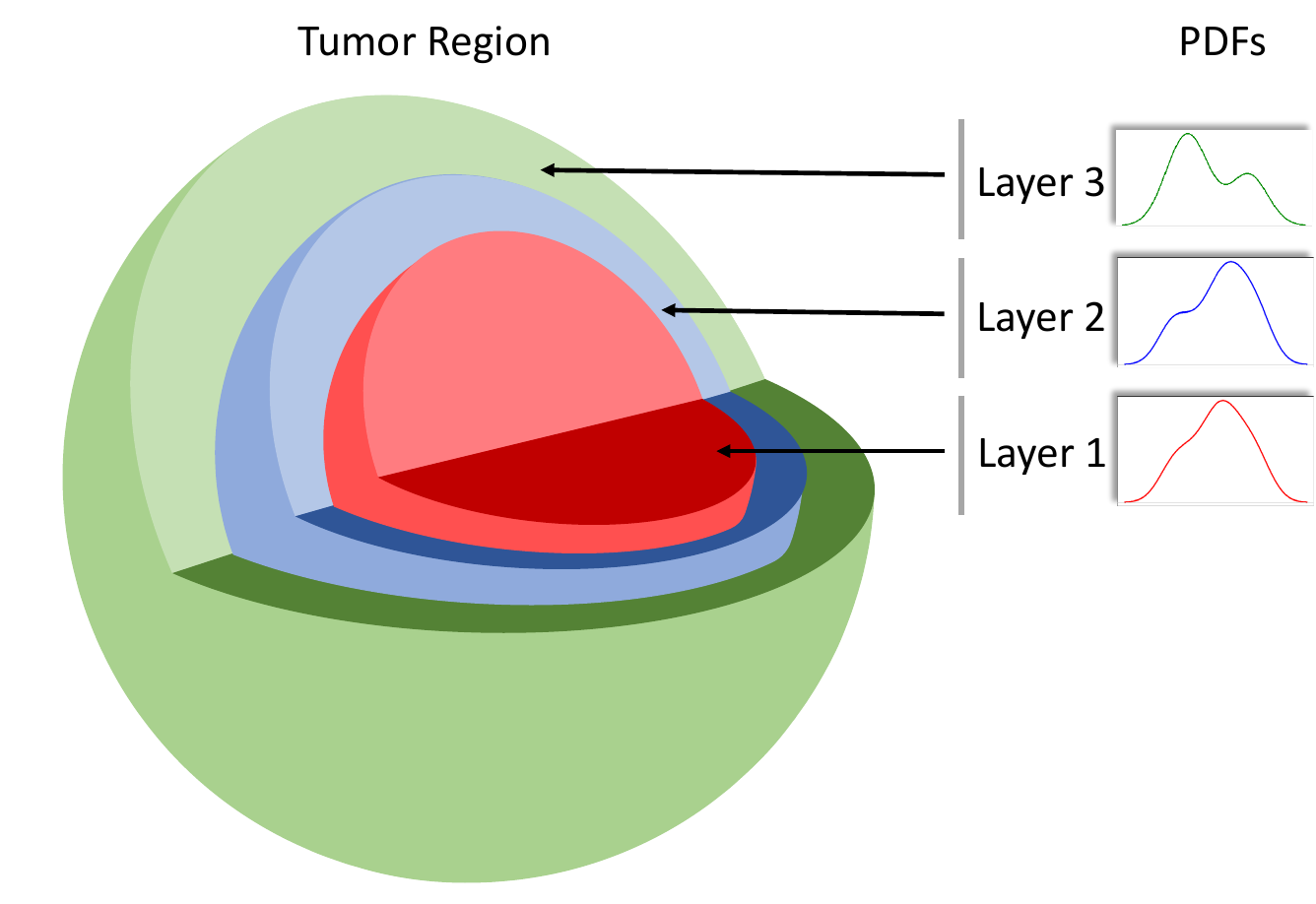}\\
			\hline
		\end{tabular}
		\caption{A graphic to visualize an example tumor region divided into three spherical layers. The object on the left shows three spherical layers colored in shades of red, blue and green. For each of these layers, the curves on the right depict corresponding PDFs.} 
		\label{fig: 3dsphere}
	\end{figure}

	Tumors typically evolve from a single (or group of) cancerous cell(s) and grow multiplicatively via cross-talk with other nearby cells, thereby producing a central necrosis region (dead cells) \citep{swanson2003virtual}. Hence, as we move from the inner core of a tumor toward the exterior, different layers of the tumor possess distinct tissue characteristics. Therefore, it is important to construct imaging phenotypes which (a) mimic the underlying tumor evolution process, and (b) efficiently harness the tumor heterogeneity from the imaging scans, for downstream modeling and interpretations. To address (a), we divide the tumor region into one inner sphere followed by concentric spherical shells which potentially capture the tumor growth process. In Figure \ref{fig: 3dsphere}, we show a visualization of an example tumor region divided into three spherical layers. Tumor spheroids (cell cultures with necrotic core and peripheral layer of proliferative cells) have been used as reliable models of in vitro solid tumors \citep{breslin2013three,costa20163d}. To address (b), we propose to work with the probability density functions (smoothed histograms) constructed using layer-wise tumor voxel intensities from three-dimensional (3D) MRI scans. These probability density functions (PDFs) are the data objects of interest and capture the entire information about the distribution of voxel intensities, no longer requiring any summaries of the distribution. Recent studies have successfully utilized density-based approaches in unsupervised clustering \citep{saha2016demarcate} and regression models for radiogenomic analyses \citep{mohammed2021radiohead,yang2020quantile}.
	
\begin{figure}[!t]
		\centering
		\resizebox{\textwidth}{!}{
			\includegraphics[trim=0cm 0cm 0cm 0cm, clip, scale=1, page=1]{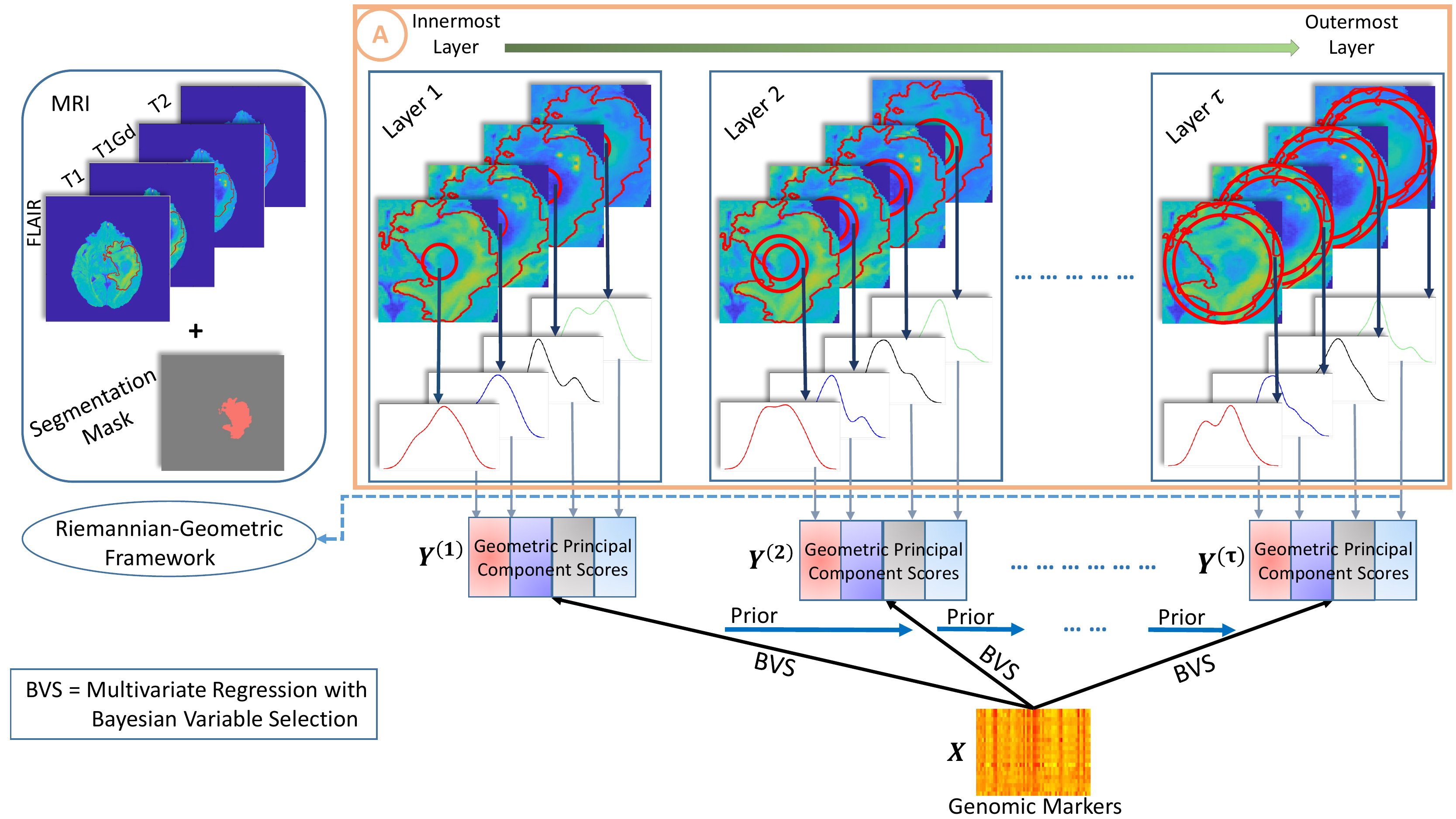}
		}
		\caption{Schematic representation of the proposed modeling approach. The tumor voxels are divided into spherical layers across all of the MRI sequences. Under a Riemannian-geometric framework, we construct principal component (PC) scores for each spherical layer across all four MRI sequences. With these PC scores as the multivariate response, and the genomic markers as predictors, we build a layer-wise Bayesian variable selection model. The estimation occurs sequentially, allowing for borrowing of information between layers through prior specification.}
		\label{fig: workflow}
	\end{figure}

	Briefly, we propose a novel approach for integrating multimodal data to identify the genomic markers that have significant associations with the PDF-based radiomic phenotypes. The imaging phenotype we consider arises from MRI, as it provides high-resolution images with a wide range of contrasts through multiple imaging sequences (see Section \ref{sec: radiomic}). The radiomic phenotypes are constructed by considering the aforementioned topological structure of the tumor and mimicking its evolution process. Our statistical framework identifies significant genomic associations on the different concentric spherical shells. These associations better inform the complex interplay between the molecular signatures and imaging characteristics in LGGs, and can provide effective non-invasive diagnostic options by monitoring the associated genomic markers prior to invasive alternatives such as biopsy. A schematic representation of our approach is presented in Figure \ref{fig: workflow} (see Section \ref{sec: radiomic} for details). This figure depicts an end-to-end workflow, which includes (i) constructing spherical layers from the tumor voxels in MRI scans and computing the corresponding imaging phenotypes, and (ii) building a statistical framework which includes the structural information from the data to identify radiogenomic associations, elements of which are summarized below.	
	
	Constructing one global PDF for each tumor disregards structural information about tissue characteristics of the tumor. This would prohibit us from analyzing any potential associations of genomic markers with local (regional) aspects of the tumor. We propose to evaluate associations between genomic signatures of the tumor through large (or subtle) changes in the overall shape of the voxel-intensity PDFs from biologically motivated tumor compartments/layers. We address this by constructing $\tau$ PDFs corresponding to $\tau$ tumor layers. Tumors for different subjects will have distinct characteristics of these layers, i.e., not all tumors might have structural information about all of the layers. However, tumor spheroids have been used as in vitro models of solid tumors \citep{breslin2013three,costa20163d}. Hence, we divide the tumor region into $\tau$ spherical shells to mimic the tumor growth process. The number of layers the tumor is divided into remains the same across subjects. We effectively have $\tau$ voxel-intensity PDFs as $x \mapsto \big(f_1(x),\ldots,f_\tau(x)\big)$, a multivariate functional data object with each component as a PDF. The space of PDFs is a non-linear, infinite-dimensional manifold, which poses significant challenges in their analysis. We use the Riemannian-geometric framework based on the Fisher-Rao metric on the space of PDFs (Section \ref{sec: radiomic}) that (i) leads to a tractable representation of each PDF, and (ii) engenders dimension reduction of the functional objects, in a manner compatible with the requirements of downstream statistical analyses.  This process yields a set of imaging phenotypes based on PDFs corresponding to each spherical layer in each MRI sequence. The natural dependence between the component functions is incorporated through a novel Bayesian prior formulation (Section \ref{sec: model}).
	
	\subsection{Statistical Framework}\label{subsec: statistical}
	We build sequential regression models for each spherical layer starting from the inner-most sphere and moving outwards towards the edge of the tumor. The sets of imaging phenotypes from a spherical layer across multiple MRI sequences are then used as the multivariate response in the regression corresponding to that spherical layer, with the genomic markers as the corresponding predictors. This specific structured form of the imaging phenotypes as well as assessing their interaction with genomic covariates leads to the following challenges:
	\begin{enumerate}
		\item While handling high-resolution imaging data, the number of parameters to estimate is usually much higher than the number of subjects, leading to the curse of dimensionality ($p \gg n$).

		\item Splitting the tumor voxels into concentric spherical layers induces a natural sequential ordering imitating tumor growth, which needs to be incorporated into the model formulation. Additionally, correlation between genomic markers needs to be accounted for as well. 
		
		\item High-dimensionality and sequential structure in imaging data, and correlation structure in genomic markers, create severe computational hurdles warranting scalable approaches. 
	\end{enumerate}
	
	In this paper, we build a novel statistical framework in a multiple-multivariate regression set-up, which addresses the above mentioned challenges incorporating the aforementioned dependence structures within and between the responses and covariates. Specifically, we leverage the hierarchical model specification to incorporate the sequential nature of the spherical layer-based imaging phenotypes as well as the correlation between the genomic markers. We propose a Bayesian variable selection approach that allows us to incorporate prior structural information in a natural and principled manner within a decision-theoretic framework, and deals with the issue of high-dimensionality. Specifically, we use continuous structured spike-and-slab priors \citep{george1997approaches,ishwaran2005spike,andersen2014bayesian} within the regression model for a specific layer to (i) blend prior information about the correlation between the genomic markers, and (ii) borrow information about the selected genomic markers from the estimation at one spherical layer and propagate the information  through the prior of the model to the subsequent spherical layer(s). As opposed to existing approaches that incorporate apriori structural information  \citep{li2010bayesian, vannucci2010bayesian}, our prior formulation incorporates structural information not only for the dependence between the predictors, but also for the dependence between the multivariate response and across the spherical layers.
	
	Finally, the complex structure of the model warrants scalable estimation approaches. Toward this end, as an alternative to computationally expensive  Markov chain Monte Carlo (MCMC) sampling techniques, we employ an Expectation--Maximization (EM)-based estimation strategy \citep{rovckova2014emvs}, where we only search for the posterior mode to iteratively estimate the posterior selection probabilities. Our EM-based estimation is an generalization of \cite{rovckova2014emvs} to a multiple-multivariate regression setting, and provides a scalable alternative to sampling strategies; it also facilitates model selection. While developed in the context of this specific imaging-genomics case study, our sequential modeling framework is general, has independent methodological value and could also be used in other applications, which exhibit a natural ordering, e.g. multivariate data collected over time, space, or other axes.
	
	The rest of this paper is organized as follows. In Section \ref{sec: radiomic}, we describe the radiological imaging data and the construction of the imaging phenotypes. We present our statistical modeling framework in Section \ref{sec: model}, including model specification, details about the EM-based estimation (Section \ref{sec: estimation}), and model selection (Section \ref{subsec: modelselection}). In Section \ref{sec: simulation}, we include results from simulation studies under different scenarios to assess the performance of the proposed modeling approach. In Section \ref{sec: radiogenomic}, we provide details of the genomic data, present results of our radiogenomic analysis of LGGs, and shed light upon some important biological findings (Section \ref{subsec: biological}). Finally, in Section \ref{sec: discussion}, we conclude with a discussion and some directions for future work. The indices to sections, figures, tables, and equations in the supplementary material are preceded by `S' (e.g. Section S1).
	
	\section{Data Characteristics and Imaging Phenotypes}\label{sec: radiomic}
	
	In this section, we describe the MRI data characteristics and present the explicit quantification of density-based imaging phenotypes. MRI scans are a rich source of imaging data as they provide a wide range of imaging contrasts. Consequently, different tissue characteristics in the brain are highlighted differently by the various MRI sequences. Primary MRI sequences include (i) native (T1), (ii) post-contrast T1-weighted (T1Gd), (iii) T2-weighted (T2), and (iv) T2 fluid attenuated inversion recovery (FLAIR). The utility of these imaging sequences in providing complementary information can be seen in Figure S1, where the tumor sub-regions appear with varying contrasts. MRI scans typically include data on all four imaging sequences and segmentation labels that generate a mask indicating the tumor and non-tumor regions in the MRI scan (top-left panel in Figure \ref{fig: workflow}). We specifically consider MRI scans of 65 LGG subjects obtained from The Cancer Imaging Archive (TCIA) \citep{clark2013cancer}. Further details about the data are provided in Section \ref{sec: radiogenomic}. Each MRI scan is structured as a 3D array where the third axis corresponds to the different axial slices of the image. Hence, we have four 3D arrays, each corresponding to one of the four imaging sequences, and an additional 3D array for the accompanying segmentation mask. All four MRI scans, and the segmentation mask, have a voxel-to-voxel correspondence which allows for clear specification of the tumor region. In Figure S1, we show an example of the axial slice from a brain MRI for a subject with LGG for each of the four imaging sequences. The region inside the red boundary overlaid on these images indicates the segmented tumor region. To integrate MRI data into our modeling, we first construct imaging phenotypes (or radiomic phenotypes) based on the voxel intensities of the MRI scans.	
	
	\textbf{Construction of PDF-based imaging phenotypes.} Consider MRI scans for $n$ LGG subjects across all four imaging sequences and the corresponding tumor segmentation masks. Let $N_i$ denote the number of voxels in the tumor region for each subject $i$. Let $\c_i$ denote array indices of the mid-point for the minimal bounding box of the tumor region, and $\tau \in \mathbb{N}$ denote the number of spherical shells we divide the tumor region into. In Algorithm S1 of Section S1, we describe construction of kernel density estimates $f^{(t,m)}_i$ corresponding to spherical shell $t \in \{1,\ldots,\tau\}$ from MRI sequence $m \in M = \{T1,T1Gd,T2,FLAIR\}$ for subject $i$. We divide the tumor region into $\tau$ spherical shells based on the total number of voxels in the tumor region, so that we have the same number of spherical shells for all subjects. While the innermost layer is a sphere, we refer to all of the $\tau$ layers as \textit{spherical shells or layers} hereafter. Hence, for each subject, we have the tumor region divided into $\tau$ spherical shells across all four imaging sequences. This is represented in panel A of Figure \ref{fig: workflow}, with spherical layers from four MRI sequences and their corresponding PDFs. Via Algorithm S1, we obtain kernel density estimates $f^{(t,m)}_i$ based on the voxels from MRI sequence $m$ in spherical shell $D^{(t)}_i$ for subject $i$. Using a Riemannian-geometric framework for analyzing PDFs \citep{srivastava2007riemannian}, we implement a principal component analysis (PCA) at each layer $t=1,\ldots,\tau$, which generates layer-specific PC scores (as shown in Figure \ref{fig: workflow}). The geometric framework for PDFs is necessary in order to ensure that the PCA is compatible with the nonlinear structure of the space of the voxel-intensity PDFs, and has been successfully used in several works \citep{saha2016demarcate,mohammed2021radiohead}. For ease of exposition, we omit the mathematical details here and describe them in Section S1.2. 
	
	Layer-specific PCA of PDFs thus maps each PDF to a Euclidean vector of PC scores, which efficiently captures the variability in the voxel-intensity PDFs. The vector of PC scores is representative of tumor heterogeneity, and constitutes the imaging phenotype, which is the response in our Bayesian model. For each imaging sequence $m$ and tumor layer $t$, we perform PCA using the sample of PDFs $f^{(t,m)}_1,\ldots,f^{(t,m)}_n$ to obtain the PC scores $Y^{(t,m)}$. The number of PCs included is decided such that the included PCs cumulatively explain 99\% of the total variance. Note that unlike the case for traditional functional data on linear Hilbert spaces, the corresponding principal directions of variability for PDFs are not orthogonal on the space of PDFs---they are orthogonal in the linear tangent space at the mean. Thus, one needs to be careful when interpreting the PC scores as capturing magnitude associated with PDF eigenfunctions; details in Section S1 elucidate on this. Although we consider the PDF-based PC scores as response in our modeling framework, the methods developed in Section \ref{sec: model} are broadly applicable to multivariate Euclidean responses. 	
	
	\section{Models and Methods}\label{sec: model}
	
\begin{figure}[!h]
		\centering
		\includegraphics[trim= 1.6cm 6.6cm 3cm 6.5cm, clip,scale=0.5]{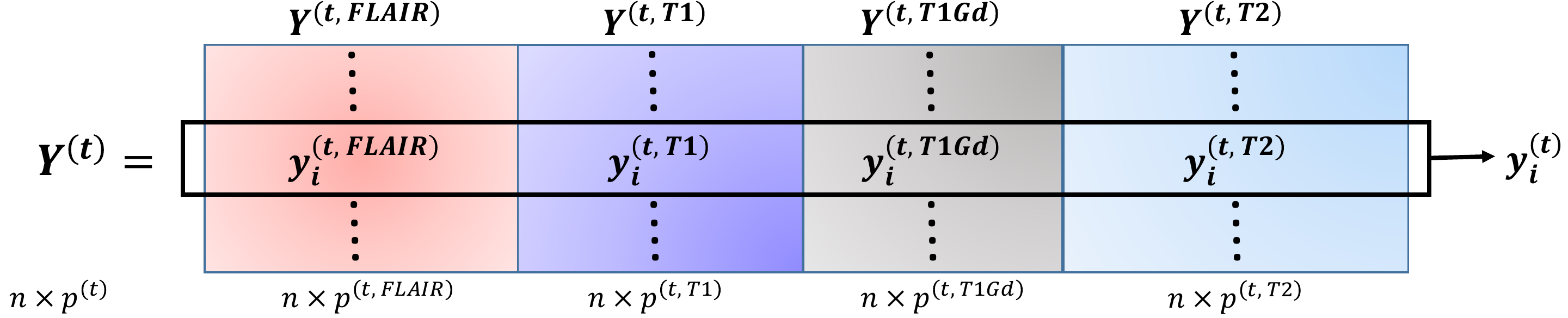}
		\caption{Pictorial representation of $Y^{(t)}$ by augmenting PC scores across imaging sequences.}
		\label{fig: reponse}
	\end{figure}

	Following the above quantification, our data structure is of the following format. Let $\y_i^{(t,m)} = (y_{i1}^{(t,m)},\ldots,y_{ip^{(t,m)}}^{(t,m)})$ represent a row vector of PC scores from imaging sequence $m \in M$ and layer $t \in T = \{1,\ldots,\tau\}$ for subject $i$. Here, $p^{(t,m)}$ represents the number of PCs included for layer $t$ and imaging sequence $m$. Let $\y_i^{(t)} = (\y_i^{(t,T1)},\ldots,\y_i^{(t,FLAIR)})$ represent the PC scores concatenated across all four imaging sequences and define $Y^{(t)} = ({\y_1^{(t)}}^\top,\ldots,{\y_n^{(t)}}^\top)^\top \in \mathbb{R}^{n \times p^{(t)}}$ where $p^{(t)} = \sum_{m \in M} p^{(t,m)}$ is the total number of PCs included across all four imaging sequences. A pictorial representation of this augmentation is provided in Figure \ref{fig: reponse}. 
	
	Let $X \in \mathbb{R}^{n \times g}$ correspond to the predictors from $g$ genomic markers for $n$ subjects. For each $t \in T$, we model the PC scores $Y^{(t)}$ using the genomic markers $X$ as 
	\begin{eqnarray}
		Y^{(t)} & \sim & N_{n,p^{(t)}}(XB^{(t)}, I_n \otimes \Delta^{(t)}), \text{ or equivalently}, \nonumber \\
		vec(Y^{(t)}) & \sim & N_{np^{(t)}}(vec(XB^{(t)}), I_n \otimes \Delta^{(t)}),	\label{eq: model_y}
	\end{eqnarray}  
	where $B^{(t)} = [B^{(t,T1)},\ldots,B^{(t,FLAIR)}] \in \mathbb{R}^{g \times p^{(t)}}$ and $B^{(t,m)} = [\bbeta_1^{(t,m)};\ldots;\bbeta_{p^{(t,m)}}^{(t,m)}] \in \mathbb{R}^{g \times p^{(t,m)}}$ for all $m \in M$. Here, $\bbeta_j^{(t,m)} = (\beta_{1j}^{(t,m)},\ldots,\beta_{gj}^{(t,m)})^\top$ for all $j=1,\ldots,p^{(t,m)}$ and $m \in M$, and $vec(\cdot)$ is the row-wise vectorization of a matrix. The coefficient $\beta_{kj}^{(t,m)}$ corresponds to the association of genomic marker $k$ and principal component $j$ from the imaging sequence $m$ in the layer $t$. Also, $I_n$ is the identity matrix of dimension $n$, and $\otimes$ denotes the Kronecker product. Note that $\Delta^{(t)} \in \mathbb{R}^{p^{(t)} \times p^{(t)}}$ is a covariance matrix that captures the covariance between the PCs across imaging sequences, which naturally exists as these sequences are imaging the same region/tumor. In other words, although the PC scores corresponding to a specific imaging sequence are uncorrelated (by construction), cross-sequence correlation could still exist (e.g. between PC scores from T1 and PC scores from T2); this correlation is captured through $\Delta^{(t)}$. We can succinctly rewrite $vec(XB^{(t)}) = Z\bbeta^{(t)}$, where $Z = X \otimes I_{p^{(t)}} \in \mathbb{R}^{np^{(t)} \times gp^{(t)}}$ \citep{gupta2018matrix} and $\bbeta^{(t)} = vec(B^{(t)}) \in \mathbb{R}^{gp^{(t)}}$. Let $\y^{(t)} = vec(Y^{(t)})$ and $\Theta^{(t)} = I_n \otimes \Delta^{(t)}$; then, Equation (\ref{eq: model_y}) is equivalent to $\y^{(t)} \sim N_{np^{(t)}}(Z\bbeta^{(t)}, \Theta^{(t)})$. 
	
	\textbf{Structured variable selection prior.} Our aim is to identify genomic markers associated with the PC scores, which translates to identifying the nonzero coefficients $B^{(t)}$ in the model in Equation (\ref{eq: model_y}). The genomic markers could themselves be correlated or have an inherent dependence structure (e.g. similar biological function or signaling pathway). We incorporate this information within the variable selection framework by proposing a generalization of structured continuous spike-and-slab priors as follows \citep{george1997approaches,andersen2014bayesian}: 
	\begin{eqnarray}
		\beta_{kj}^{(t,m)} \Big| \zeta_k^{(t,m)}, {\nu^2_{kj}}^{(t,m)} & \stackrel{ind}{\sim} & N\big(0, \big[(1-\zeta_k^{(t,m)})v_0 + \zeta_k^{(t,m)}v_1\big]{\nu^2_{kj}}^{(t,m)}\big) ~~\forall ~ k,j,t,m, \nonumber \\
		\zeta_k^{(t,m)} \Big| \lambda_k^{(t,m)} & \stackrel{ind}{\sim} & Ber(\Phi(\lambda_k^{(t,m)})) ~~\forall ~ k,t,m, \label{eq: prior} \\
		\blambda^{(t,m)} & \stackrel{ind}{\sim} & N(\bmu^{(t,m)}, \Lambda) ~~\forall ~ t,m, \nonumber 
	\end{eqnarray}
	where ${\eta^2_{kj}}^{(t,m)} := \big[(1-\zeta_k^{(t,m)})v_0 + \zeta_k^{(t,m)}v_1\big]{\nu^2_{kj}}^{(t,m)}$ is the hypervariance of $\beta_{kj}^{(t,m)}$ with $0 < v_0 \ll v_1 $. The indicator $\zeta_k^{(t,m)}$ takes values $1$ and $0$ with probability $\Phi(\lambda_k^{(t,m)})$ and $(1-\Phi(\lambda_k^{(t,m)}))$, respectively. Here $\Phi(\cdot)$ is the cumulative distribution function of a standard normal distribution. Note that any suitable injective function could be chosen instead of $\Phi$. We use a Gamma prior on ${\nu^{-2}_{kj}}^{(t,m)}$ with $a_1$ and $a_2$ as the shape and rate parameters, respectively. If $\zeta_k^{(t,m)} = 1$, the hypervariance for $\beta_{kj}^{(t,m)}$ is dictated by the prior on $v_1{\nu^{-2}_{kj}}^{(t,m)}$; when $\zeta_k^{(t,m)} = 0$, the hypervariance $v_0{\nu^2_{kj}}^{(t,m)}$ is small in magnitude allowing for shrinkage of the coefficient $\beta_{kj}^{(t,m)}$ toward zero. For this purpose, $v_0$ is chosen to be a much smaller positive number than $v_1$, encouraging the exclusion of much smaller nonzero effects. We use an inverse-Wishart (IW) prior on the covariance matrix $\Delta^{(t)}$ as $\Delta^{(t)} \stackrel{iid}{\sim} IW(\delta, \Psi)$, where $\delta$ is the degrees of freedom and $\Psi$ is the scale matrix.
	
	The indicator $\zeta_k^{(t,m)}$ corresponds to the genomic marker $k$ across all PC scores $j$. That is, at each level $t$ and imaging modality $m$, genomic marker $k$ is assumed to be correlated with the entire voxel intensity PDF, and not just some of its features. This structure is built-in with the assumption that we are interested in associations of genomic markers with imaging sequences as a whole, and not just associations with specific PCs within an imaging sequence. 
	
\begin{figure}[!t]
		\centering
		\resizebox{\textwidth}{!}{
			\includegraphics{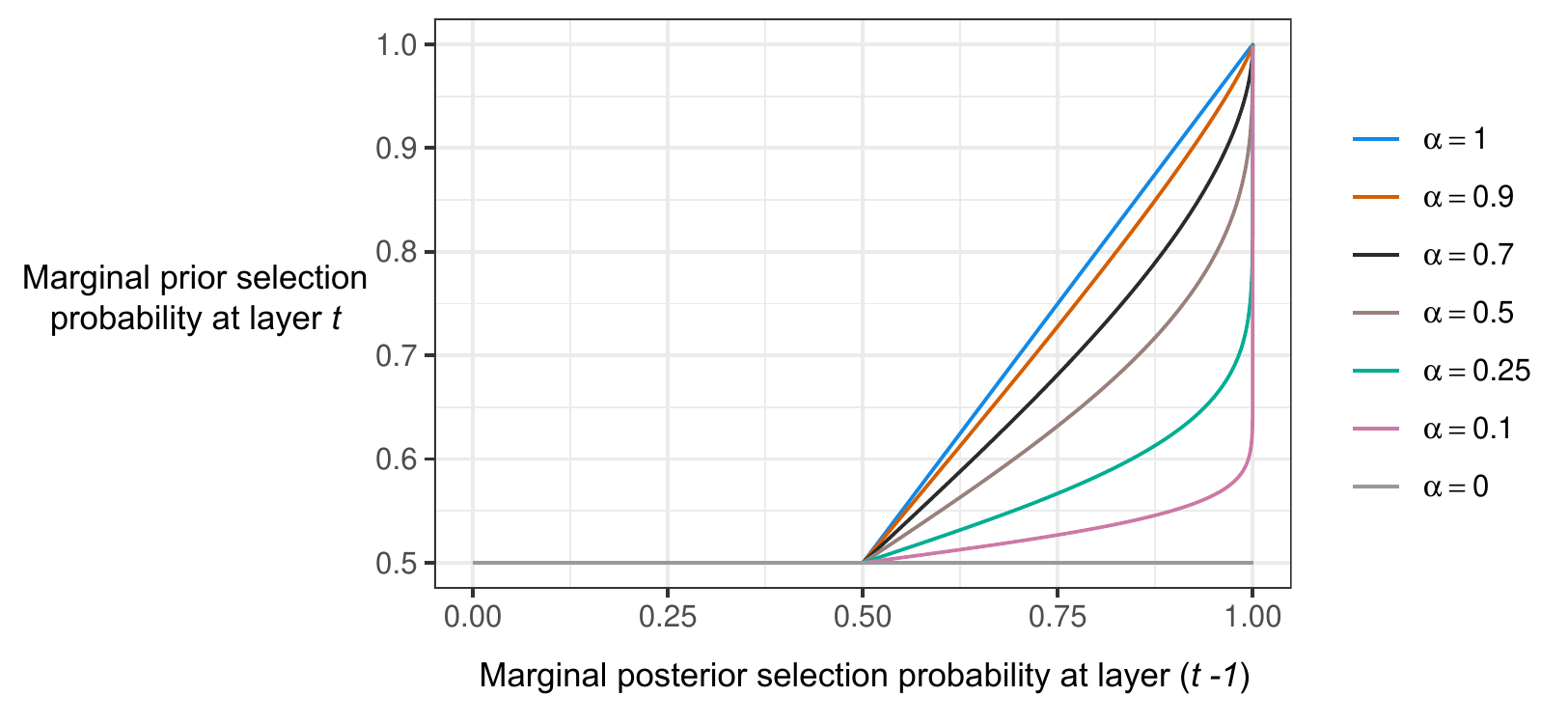}
		}
		\caption{Marginal prior selection probability $\Phi(\lambda^{(t,m)}_k)$ (evaluated at $\lambda^{(t,m)}_k = \mu^{(t,m)}_k = \alpha\hat{\lambda}^{(t-1,m)}_{k,+}$) \textit{versus} the marginal posterior selection probability $\Phi(\hat{\lambda}^{(t-1,m)}_k)$ when $\Lambda_{kk} = 1$.}
		\label{fig: marg_prob}
	\end{figure}

	\textbf{Borrowing strength using $\bmu^{(t,m)}$ and $\Lambda$.} In essence, our model breaks down the components of a usual multiple-multivariate regression, into a sequence of such separate models for each layer $t \in T$. The dependence structure between the responses and hence, the corresponding regression coefficients is captured  through the hyperparameters in our model specification. For $t = 1$, we choose a fairly vague prior by considering $\bmu^{(1,m)} = \0$. However, for the model at layer $t>1$, we incorporate information from the estimation for models at the previous layers $1,\ldots,(t-1)$. We do so through a sequential posterior updating of the layer-specific parameters. Let $\hat{\blambda}^{(s,m)}$ be the posterior estimate of $\blambda^{(s,m)}$ for $s<t$; then, some potential choices for $\bmu^{(t,m)}$ include 
	\begin{enumerate}
		\item $\bmu^{(t,m)} = \alpha\hat{\blambda}^{(t-1,m)} + (1-\alpha)\0 $, where $\alpha \in [0,1]$;
		\item $\bmu^{(t,m)} = \sum\limits_{s=1}^{t-1} \alpha_s \hat{\blambda}^{(t-s,m)} + \alpha_t\0$, such that $\sum\limits_{s=1}^{t} \alpha_s = 1$;
		\item $\bmu^{(t,m)} = \sum\limits_{s=1}^{t-1} \alpha_s \hat{\blambda}^{(t-s,m)}$ for $\alpha_s \in \mathbb{R}$; 
		\item $\bmu^{(t,m)} = \alpha\hat{\blambda}^{(t-1,m)}_+ $, where $\hat{\lambda}^{(t-1,m)}_{k,+} = \max(\hat{\lambda}^{(t-1,m)}_k, 0)$ for $k=1,\ldots,g$ and $\alpha \in [0,1]$.
	\end{enumerate}
	The first choice centers the prior on $\blambda^{(t,m)}$ based only on the estimates from the previous layer $(t-1)$. The second choice centers the prior in the convex hull created by estimates from all previous layers $1,\ldots,(t-1)$. The third choice induces an auto-regressive structure on the estimates from all previous layers $1,\ldots,(t-1)$. The fourth choice places higher prior probability of selection on those columns of $X$ which were selected during the estimation for the previous layer $(t-1)$. That is, predictor $k$ will have a higher chance of selection ($\zeta_k^{(t,m)} = 1$) in layer $t$ for imaging sequence $m$ when $\lambda_k^{(t,m)} > 0$. However, $\lambda_k^{(t,m)} > 0$ has higher probability of occurrence if $\mu_k^{(t,m)} > 0$. The construction of $\bmu^{(t,m)}$ in the fourth choice above ensures $\mu_k^{(t,m)} \geq 0$. Note that this choice does not penalize the selection of the predictor $k$ in case it was not selected for the previous layer $(t-1)$. In Figure \ref{fig: marg_prob}, for different choices of $\alpha$, we plot the marginal prior selection probability $\Phi(\lambda^{(t,m)}_k)$ (evaluated at $\lambda^{(t,m)}_k = \mu^{(t,m)}_k = \alpha\hat{\lambda}^{(t-1,m)}_{k,+}$)  of gene $k$ for imaging sequence $m$ at layer $t$, versus the marginal posterior selection probability $\Phi(\hat{\lambda}^{(t-1,m)}_k)$ at layer $(t-1)$ when $\Lambda_{kk} = 1$. For example, if $\hat{\lambda}^{(t-1,m)}_k = 1$, then predictor $k$ has marginal posterior selection probability $\Phi(\hat{\lambda}^{(t-1,m)}_k) = 0.84$ for imaging sequence $m$ at layer $(t-1)$. In this case, the marginal prior selection probability $\Phi(\lambda^{(t,m)}_k)$ at layer $t$ is $0.60, 0.69$ and $0.84$ when $\alpha = 0.25, 0.5$ and $1$, respectively. The marginal prior selection probability $\Phi(\lambda^{(t,m)}_k) = 0.5$ at layer $t$ when $\hat{\lambda}^{(t-1,m)}_k \leq  0$.
	
	We focus on the fourth case for the simulations and case study. Nevertheless, our estimation process proceeds in a sequential manner in $t$, where the estimation for model at layer $t$ borrows information from the model for layer $(t-1)$, and possibly other previous layers. In the prior specification in Equation (\ref{eq: prior}), the parameter $\lambda_k^{(t,m)}$ serves as a structural parameter as its prior can be used to incorporate any dependence in the columns of $X$. The correlation structure between the genomic markers is incorporated into the prior for $\blambda^{(t,m)} = (\lambda_1^{(t,m)}, \ldots, \lambda_g^{(t,m)})^\top$ through its covariance specification $\Lambda$. If a certain genomic marker $k$ is associated with the PC scores from imaging sequence $m$ and layer $t$, then we assume that other positively correlated genomic markers might also have a similar association. This assumption is driven by the fact that voxels from the boundaries of the spherical layers are contiguous objects with potentially similar characteristics. This can be incorporated through a graph Laplacian or other choice of network structure-based covariance matrix $\Lambda$. If $\bmu^{(t,m)} = \0$ and $\Lambda = I_g$, then the joint prior on $\zeta^{(t,m)}_k$ and $\blambda_{(t,m)}$ reduces to $\zeta_k^{(t,m)} \stackrel{ind}{\sim} Ber(\upsilon^{(t,m)}_k)$ with $\upsilon^{(t,m)}_k \stackrel{iid}{\sim} U(0,1)$.
	
	\textbf{Role of $\bmu^{(t,m)}$ and $\Lambda$.} While  $\bmu^{(t,m)}$  incorporates the layered dependence structure between responses for the selection of regression coefficients, $\Lambda$ incorporates the correlation between the columns of $X$, and hence places higher prior probability of joint selection for any two correlated columns $k$ and $k'$. On the other hand, $\bmu^{(t,m)}$ induces prior dependence for the same column across various layers $t$. In other words, $\bmu^{(t,m)}_k$ centers the prior probability of selection for column $k$ in layer $t$, based on its selection in previous layers. As a result, $\bmu^{(t,m)}$ and $\Lambda$ contribute complementary information to the model and jointly contribute towards the selection of the regression coefficients.

	\section{Estimation and Model Selection}\label{sec: estimation}
	The full posterior distribution corresponding to the parameters of the model in Equation (\ref{eq: model_y}) with the prior structure considered in Equation (\ref{eq: prior}) is given in Section S2.1. Due to the multiple-multivariate regression setup, we have $g \times p^{(t)}$ coefficients ($\beta_{kj}$) and/or $|M| \times g$ posterior inclusion probabilities to sequentially estimate for each model at layer $t$. A full MCMC-based estimation can be employed, but the computation time would be additive with each additional spherical layer included, as the estimation is required to be sequential based on the model formulation. As a faster alternative to MCMC, we consider the EM-based strategy to variable selection as proposed by \cite{rovckova2014emvs}. This EM-based estimation approach is a deterministic alternative to commonly used stochastic search algorithms, which exploits the prior formulation to rapidly find posterior modes. We emphasize that the EM-based estimation presented in this work is an adaptation of \cite{rovckova2014emvs} to the multiple-multivariate regression setting. We discuss sequential estimation using EM and describe the model selection procedure below. 
	
	\subsection{Sequential EM Updating}
	Our implementation of the EM-based estimation approach indirectly maximizes the parameter posterior given by $\pi(\beta_{kj}^{(t,m)}, \blambda^{(t,m)}, {\nu^{-2}_{kj}}^{(t,m)}, \Delta^{(t)}| \y^{(t)})$, by proceeding iteratively over the complete-data log posterior $\log \pi(\beta_{kj}^{(t,m)}, \zeta_k^{(t,m)}, \blambda^{(t,m)}, {\nu^{-2}_{kj}}^{(t,m)}, \Delta^{(t)} | \y^{(t)})$, where $\zeta_k^{(t,m)}$ are treated as missing data. At each iteration, the complete-data log posterior is replaced by its conditional expectation given the observed data and the current parameter estimates; this forms the E-Step. The expected complete-data log posterior is then maximized with respect to (w.r.t.) $\bchi^{(t)} = (\beta_{kj}^{(t,m)}, \blambda^{(t,m)}, {\nu^{-2}_{kj}}^{(t,m)}, \Delta^{(t)})$ for all $m,j,k$, which constitutes the M-Step. Iterating between the E-Step and M-Step generates a sequence of parameter estimates, which converge monotonically to the local maximum of $\pi(\bchi^{(t)}| \y^{(t)})$, under appropriate regularity conditions.
	
	The expression for the complete-data log posterior is given in Section S2.1. For the $u$th iteration of the algorithm, the conditional expectation of the complete-data log posterior is given by
	\begin{eqnarray}
		Q\Big(\beta_{kj}^{(t,m)}, \blambda^{(t,m)}, {\nu^{-2}_{kj}}^{(t,m)}, \Delta^{(t)} | \beta_{kj}^{(t,m),(u)}, \blambda^{(t,m),(u)}, {\nu^{-2}_{kj}}^{(t,m),(u)}, \Delta^{(t),(u)}\Big) = \nonumber \\
		:= E_{\bzeta^{(t)} | \bchi^{(t),(u)}}\Big[ \log\big( \pi(\beta_{kj}^{(t,m)}, \zeta_k^{(t,m)}, \blambda^{(t,m)}, {\nu^{-2}_{kj}}^{(t,m)}, \Delta^{(t)} | \y^{(t)}) \big) \Big], \label{eq: q_fn}
	\end{eqnarray}
	where $E_{\bzeta^{(t)} | \bchi^{(t),(u)}}$ is a conditional expectation with $\bchi^{(t),(u)} = (\beta_{kj}^{(t,m),(u)}, \blambda^{(t,m),(u)}, {\nu^{-2}_{kj}}^{(t,m),(u)}, \Delta^{(t),(u)})$ for all $m,j,k$ and $\bzeta^{(t)} = (\zeta_k^{(t,m)})_{m,k}$. For the $(u+1)$-th iteration in the EM algorithm, we (i) find the expectation in Equation (\ref{eq: q_fn}) to obtain the $Q$-function, and (ii) maximize the $Q$-function w.r.t. $\bchi^{(t)}$ to obtain new estimates $\bchi^{(t),(u+1)}$. 
	
	\subsubsection{The E-Step}\label{subsubsec: e-step}
	As the conditional expectation is computed w.r.t. $\zeta^{(t,m)}_k$, we note that the only two terms in the complete-data log posterior involving $\zeta^{(t,m)}_k$ are the terms arising from the prior for $\beta^{(t,m)}_{j}$ and the prior for $\zeta^{(t,m)}_k$. After some algebra (see Section S2.2), we see that these two terms are given by
	\begin{equation*}
		- \sum\limits_{m\in M} \sum\limits_{k=1}^g \sum\limits_{j = 1}^{p^{(t,m)}}  \frac{1}{2}\log({\nu^2_{kj}}^{(t,m)}) + \frac{(\beta_{kj}^{(t,m)})^2}{2{\nu^2_{kj}}^{(t,m)}}E_{\bzeta^{(t)} \big| \bchi^{(t),(u)}}\Big[\frac{1}{(1-\zeta_k^{(t,m)})v_0 + \zeta_k^{(t,m)}v_1} \Big],\ \text{and}
	\end{equation*}
	\begin{equation*}
		\sum\limits_{m\in M} \sum\limits_{k=1}^g  \log(1-\Phi(\lambda_k^{(t,m)})) + E_{\bzeta^{(t)} \big| \bchi^{(t),(u)}}[\zeta_k^{(t,m)}]\log \frac{\Phi(\lambda_k^{(t,m)})}{1-\Phi(\lambda_k^{(t,m)})}.
	\end{equation*}
	To complete the E-Step, the two expectations we need to compute are given by 
	\[w_k^{(t,m)} := E_{\bzeta^{(t)} \big| \bchi^{(t),(u)}}[\zeta_k^{(t,m)}],\text{ and  }d_k^{(t,m)}:=E_{\bzeta^{(t)} \big| \bchi^{(t),(u)}}\Big[\frac{1}{(1-\zeta_k^{(t,m)})v_0 + \zeta_k^{(t,m)}v_1} \Big] .\]
	Note then that $w_k^{(t,m)} = P(\zeta_k^{(t,m)}=1 \big| \bchi^{(t),(u)} ) = \frac{a_k^{(t,m)}}{a_k^{(t,m)} + b_k^{(t,m)}}$,
	where $a_k^{(t,m)}$ and $b_k^{(t,m)}$ are defined in Equations (S1) and (S2), respectively. To compute $d_k^{(t,m)}$, we note than when 
	$\zeta_k^{(t,m)} \in \{0,1\}$, we have $((1-\zeta_k^{(t,m)})v_0 + \zeta_k^{(t,m)}v_1)^{-1} = \frac{1-\zeta_k^{(t,m)}}{v_0} + \frac{\zeta_k^{(t,m)}}{v_1}$, almost surely. Hence, $d_k^{(t,m)} = \frac{1-w_k^{(t,m)}}{v_0} + \frac{w_k^{(t,m)}}{v_1}$. 
	
	\subsubsection{The M-Step}\label{subsubsec: m-step}
	Once we compute $w_k^{(t,m)}$ and $d_k^{(t,m)}$ in the E-step, we focus on maximizing the expected log posterior, i.e., the $Q$-function, w.r.t. other parameters. The $Q$-function, which is a function of $\bchi^{(t)} = (\beta_{kj}^{(t,m)}, \blambda^{(t,m)}, {\nu^{-2}_{kj}}^{(t,m)}, \Delta^{(t)})$ for all $m,j,k$, can be split into three terms as 
	\begin{eqnarray*}
		Q(\bchi^{(t)} | \bchi^{(t),(u)}) & = & C + Q_1(\beta_{kj}^{(t,m)}, {\nu^{-2}_{kj}}^{(t,m)}, \Delta^{(t)} | \beta_{kj}^{(t,m),(u)}, {\nu^{-2}_{kj}}^{(t,m),(u)}, \Delta^{(t),(u)}) + Q_2(\blambda^{(t,m)} | \blambda^{(t,m),(u)}),
	\end{eqnarray*}
	where $Q_1$ only depends on $\beta_{kj}^{(t,m)}, {\nu^{-2}_{kj}}^{(t,m)}$ and $\Delta^{(t)}$, and $Q_2$ depends on $\blambda^{(t,m)}$. The complete forms of $Q_1$ and $Q_2$ are given in Section S2.3. Completing the EM-based estimation procedure, we maximize $Q$ w.r.t. $\bchi^{(t)}$, i.e., we maximize $Q_1$ w.r.t. $\beta_{kj}^{(t,m)}, {\nu^{-2}_{kj}}^{(t,m)}$ and $\Delta^{(t)}$, and $Q_2$ w.r.t. $\blambda^{(t,m)}$. The objective functions for all of the maximization problems are provided in Section S2.4. 
	
	\textbf{Maximizing $Q_1$ w.r.t. $\beta_{kj}^{(t,m)}$.} Assuming we have the parameters ${\nu^2_{kj}}^{(t,m),(u)}, \Delta^{(t),(u)}$ at iteration $u$ and the value $d_k^{(t,m),(u+1)}$ at iteration $(u+1)$, maximizing Equation (S3) is equivalent to finding the maximum a-posteriori (MAP) estimate of a Bayesian regression $\y^{(t)} \sim N_{np^{(t)}}(Z\bbeta^{(t)}, \Theta^{(t)})$ with the prior $\bbeta^{(t)} \sim N(\0,\bGamma^{(t)})$. Thus, the update (MAP estimate) for $\bbeta^{(t)}$ is given by
	\begin{equation*}
		\bbeta^{(t),(u+1)} = \bSigma^{(t),(u)} Z^\top (\Theta^{(t),(u)})^{-1}\y \text{, where } \bSigma^{(t),(u)} = \big( Z^\top (\Theta^{(t),(u)})^{-1} Z + (\bGamma^{(t),(u)})^{-1}\big)^{-1},
	\end{equation*}
	where $\bGamma^{(t)} = diag(\nu_{kj}^{(t,m)}/d_k^{(t,m)})$ and $\bGamma^{(t),(u)} = diag(\nu_{kj}^{(t,m),(u)}/d_k^{(t,m),(u+1)})$.
	
	\textbf{Maximizing $Q_1$ w.r.t. ${\nu^{-2}_{kj}}^{(t,m)}$.} Each term inside the product in Equation (S4) is the kernel of a Gamma distribution for ${\nu^{-2}_{kj}}^{(t,m)}$. Assuming we have the parameters $\bbeta^{(t),(u+1)}, \Delta^{(t),(u)}$ at iteration $(u+1)$, the update for ${\nu^{-2}_{kj}}^{(t,m)}$ (maximizer of Equation (S4)) for each $m,k,j$ is given by
	\begin{equation*}
		{\nu^{-2}_{kj}}^{(t,m),(u+1)} = \Big( a_1-\frac{1}{2} \Big)\Big/\Big(a_2 + \frac{(\beta_{kj}^{(t,m),(u+1)})^2d_k^{(t,m),(u+1)}}{2}\Big), \text{ where } a_1 > \frac{1}{2}.
	\end{equation*}
	
	\textbf{Maximizing $Q_1$ w.r.t. $\Delta^{(t)}$.} Equation (S5) is the kernel of a Wishart distribution for $(\Delta^{(t)})^{-1}$; thus, its maximizer is the mode of the corresponding Wishart distribution. Assuming we have the parameters $B^{(t),(u+1)}$ at iteration $(u+1)$, the update for $(\Delta^{(t)})^{-1}$ is given by 
	\begin{equation*}
		(\Delta^{(t),(u+1)})^{-1} = (n+\delta+p^{(t)}+1)\big[\Psi + (Y^{(t)} - XB^{(t)})^\top(Y^{(t)} - XB^{(t)})\big]^{-1}.
	\end{equation*}
	
	\textbf{Maximizing $Q_2$ w.r.t. ${\blambda}^{(t,m)}$.} The posterior for ${\blambda}^{(t,m)}$ does not arise from a standard distribution. Thus, we resort to numerical methods, specifically gradient descent, to find the solution for the optimization in Equation (S6), to compute the update for ${\blambda}^{(t,m)}$ at iteration $(u+1)$. The first derivative of the objective function $F({\blambda}^{(t,m)})$ is given by
	\begin{eqnarray*}
		\nabla F({\blambda}^{(t,m)}) &=& ({\blambda^{(t,m)}-\bmu^{(t,m)}})^\top\Lambda^{-1} - (s_1^{(t,m)},\ldots,s_g^{(t,m)}), \text{ where} \\
		s_k^{(t,m)} &=& -\frac{(1-w_k^{(t,m)})\phi(\lambda_k^{(t,m)})}{1-\Phi(\lambda_k^{(t,m)})}+\frac{w_k^{(t,m)}\phi(\lambda_k^{(t,m)})}{\Phi(\lambda_k^{(t,m)})}.
	\end{eqnarray*}
	Then, the update at the $(r+1)$-th iteration of the gradient descent algorithm is given by
	\begin{equation*}
		{\blambda}^{(t,m),(u+1,r+1)} = {\blambda}^{(t,m),(u+1,r)} - \kappa \nabla F({\blambda}^{(t,m),(u+1,r)}),
	\end{equation*}
	where $\kappa$ is the learning rate. This learning rate can be modified within each iteration to speed up convergence. The initial value ${\blambda}^{(t,m),(u+1,0)}$ at iteration $(u+1)$ of the EM algorithm can be set as the solution of Equation (S6) at iteration $u$. This solution for ${\blambda}^{(t,m)}$ is used to assign the prior mean for ${\blambda}^{(t+1,m)}$ as $\bmu^{(t+1,m)} = \alpha\hat{\blambda}^{(t,m)}_+$.
	
	One of the major drawbacks of the EM algorithm is its potential to get trapped in a local maximum. A possible solution for this is to run the algorithm with different starting values. In our estimation approach, we consider the deterministic annealing variant of the EM algorithm \citep{ueda1998deterministic}, referred to as the DAEM algorithm, which improves the chance of finding the global maximum (see Section S2.5 for details). All of the results presented in this paper are based on the DAEM algorithm. We denote the DAEM algorithm-based estimates for the parameters in our model by $\hat{\bchi}^{(t)}$.
	
	\subsection{Model Selection}\label{subsec: modelselection}
	Once we have the estimates of the parameters $\hat{\bchi}^{(t)}$, we can compute the posterior inclusion probabilities for predictor $k$ in layer $t$ and imaging sequence $m$ as $\hat{w}^{(t,m)}_k = P(\zeta^{(t,m)}_k = 1 | \hat{\bchi}^{(t)})$ using Equation (S8). In terms of variable selection, we set $\hat{\zeta}^{(t,m)}_k = 1$ when $P(\zeta^{(t,m)}_k = 1 | \hat{\bchi}^{(t)}) > 0.5$, and set $\hat{\zeta}^{(t,m)}_k = 0$ otherwise. Note that $v_0$ and $v_1$ are fixed hyperparameters and their choice determines the solution set $\hat{\bzeta}^{(t)}$. The speed of the EM algorithm makes it feasible to obtain solution sets $\hat{\bzeta}^{(t)}$ for multiple choices of $v_0$ while fixing $v_1$. As $v_0$ is increased, we see that more coefficients with smaller magnitude get absorbed by the spike distribution, leading to sparser models. We consider a grid of values for $v_0$, fix $v_1$ to be significantly larger than $v_0$, and obtain solutions for each combination of $v_0$ and $v_1$. We denote the solution set corresponding to the choice of $v_0$ as $S(v_0) = \big\{(m,k,1),\ldots,(m,k,p^{(t,m)}) | \zeta^{(t,m)}_k = 1 ~\forall ~m \in M, k \in \{1,\ldots,g\} \text{ and } t \in \{1,\ldots,\tau\} \big\}$.
	
	Once we have identified the set of coefficients to be included in the models, we compute the Bayesian Information Criterion (BIC) across all layers $t$, given by $BIC = \sum_{t=1}^\tau -2\big[ K^{(t)}\log(n) + \log \pi(Y^{(t)}|\hat{\beta}^{(t,m)}_{kj},\widehat{\Delta^{(t)}})\big]$, where $K^{(t)} = \#\{\hat{\beta}^{(t,m)}_{kj} \neq 0 | m,k,j \}$ is the number of nonzero coefficients included in the model for layer $t$. The final model is based on the inclusion of coefficients, which minimize the $BIC$. We outline the overall step-by-step procedure from constructing the PC scores to final model selection in Algorithm \ref{algo: outline}. 
	\begin{algorithm}[!h]
		\caption{Outline of the statistical framework}\label{algo: outline}
		\begin{algorithmic}[1]
			\State Construct layer-wise kernel density estimates $f_i^{(t,m)}$ as described in Algorithm S1 for all subjects $i = 1,\ldots,n$, spherical layers $t = 1,\ldots,\tau$ and MRI sequence $m \in M$.
			\State Consider $X \in \mathbb{R}^{n \times g}$ as the gene expression matrix for LGG-genes. 
			\For {each layer $t=1,\ldots,\tau$}
			\For {each MRI sequence $m \in M$}
			\State Compute the PC scores $Y^{(t,m)} \in \mathbb{R}^{n \times p^{(t,m)}}$ using PCA in Algorithm S3.
			\EndFor
			\State Define $Y^{(t)} = [Y^{(t,FLAIR)}; Y^{(t,T1)}; Y^{(t,T1Gd)}; Y^{(t,T2)}] \in \mathbb{R}^{n \times p^{(t)}}$.
			\EndFor
			\For {each $v_0 \in$ Set of values to consider}
			\For {each layer $t=1,\ldots,\tau$}
			\State Bayesian Modeling:
			\begin{eqnarray*}
				Y^{(t)} \sim N_{n,p^{(t)}}(XB^{(t)}, I_n \otimes \Delta^{(t)}); & 
				\beta_{kj}^{(t,m)} \stackrel{ind}{\sim} N\big(0, \big[(1-\zeta_k^{(t,m)})v_0 + \zeta_k^{(t,m)}v_1\big]{\nu^2_{kj}}^{(t,m)}\big); \\
				\zeta_k^{(t,m)} \stackrel{ind}{\sim} Ber(\Phi(\lambda_k^{(t,m)})); &
				\blambda^{(t,m)} \stackrel{ind}{\sim} N(\bmu^{(t,m)}, \Lambda);\\
				{\nu^{-2}_{kj}}^{(t,m)} \stackrel{iid}{\sim} Gamma(a_1, a_2); &
				\Delta^{(t)} \stackrel{iid}{\sim} IW(\delta, \Psi).
			\end{eqnarray*}
			\State Estimation (Section \ref{sec: estimation}) with deterministic annealing:
			\begin{algsubstates}
				\State ~~~~~~~~~~~~E-Step: Compute the E-Step as described in Section \ref{subsubsec: e-step}.
				\State ~~~~~~~~~~~~~M-Step: Sequentially maximize $Q_1$ w.r.t. $\beta_{kj}^{(t,m)}, {\nu^{-2}_{kj}}^{(t,m)}, \Delta^{(t)}$ and $Q_2$ w.r.t. ${\blambda}^{(t,m)}$.
			\end{algsubstates} 
			\EndFor
			\EndFor
			\State Model Selection: Choose the model corresponding to that $v_0$, which minimizes the BIC as described in Section \ref{subsec: modelselection}.
		\end{algorithmic}
	\end{algorithm}
	
	\section{Simulation Study}\label{sec: simulation}
	
	In this section, we demonstrate the effectiveness of our approach through simulation studies under different scenarios. We do not directly simulate the MRI images; instead we simulate $Y^{(t)}$ corresponding to the PC scores for each layer $t$. To generate the matrix $X = [\x_1,\ldots,\x_n]^\top \in \mathbb{R}^{n \times g}$ of $g$ genomic markers for $n$ subjects, we simulate $\x_i$ independently from $N_g(\0,\Sigma^{(x)})$ for $i=1,\ldots,n$, where $\Sigma^{(x)}$ is the covariance matrix which can be specified to induce different correlation structures between the columns of $X$. We broadly consider two cases for our simulation study: the response $Y^{(t)}$ is (i) generated from the model, and (ii) generated directly based on fixed coefficients $B^{(t)}$.
	
	\textbf{Case 1: $Y^{(t)}$ is generated from the model.} In this case, we want to assess the performance of the proposed approach when the response $Y^{(t)}$ is generated directly from the model in Equation (\ref{eq: model_y}), with the prior structure in Equation (\ref{eq: prior}). The step-by-step procedure to generate the data is described as an algorithm in Section S3. We construct subcases defined by different choices of $\Psi$, which incorporates the correlation structure between the columns of $Y^{(t)}$. For simplicity, we consider $\Psi = \sigma^2I_p$, where $p = p^{(t)}$ for all $t$, indicating that the number of columns of $Y^{(t)}$ is the same for all $t$. Here, $\sigma^2$ acts as the noise parameter that controls the variance in the columns of $Y^{(t)}$. The dependence structure across the layers $t$ and imaging sequences $m$ is incorporated through the parameter $\bmu^{(t,m)}$, as described in Section \ref{sec: model}.
	
	We use $n = 100$ samples, $p = 12$ columns in each $Y^{(t)}$ with $p^{(t,m)} = 3$ for all $m \in M$ and $t \in T$, $g = 20$ genomic markers, and $\tau = 3$ layers. We consider two choices for the covariance in the columns of $X$: (i) $\Sigma^{(x)} = I_g$, indicating no correlation between the genomic markers, and (ii) $\Sigma^{(x)}$ as a $(2 \times 2)$-block matrix, where $\Sigma^{(x)}_{11} = \big((\Sigma^{(x)}_{11})\big)_{1 \leq i,j \leq 10}$ with $(\Sigma^{(x)}_{11})_{ij} = 10$ if $i=j$ and $9$ otherwise, $\Sigma^{(x)}_{12} = \Sigma^{(x)}_{21} = \0$ and $\Sigma^{(x)}_{22} = I_{10}$. The second choice of $\Sigma^{(x)}$ indicates two groups of genomic markers, with one group having high between-group correlation and the other group being uncorrelated. For the noise parameter, we consider $\sigma^2 = 1,10,20,30$. For each combination of $\sigma^2$ and $\Sigma^{(x)}$, we replicate the simulation 30 times.
	
\begin{figure}[!t]
		\centering
		\resizebox{\textwidth}{!}{
			\begin{tabular}{|c|c|c|c|}
				\hline
				\includegraphics[page=1]{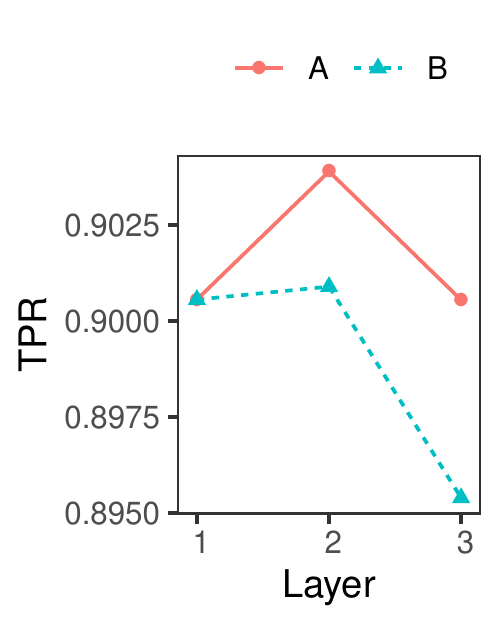} &
				\includegraphics[page=2]{plots/TPR_case1_Uncor.pdf} &
				\includegraphics[page=3]{plots/TPR_case1_Uncor.pdf} &
				\includegraphics[page=4]{plots/TPR_case1_Uncor.pdf} \\
				(a) $\sigma^2 = 1$ & 
				(b) $\sigma^2 = 10$ &
				(c) $\sigma^2 = 20$ &
				(d) $\sigma^2 = 30$ \\ \hline
			\end{tabular}
		}
		\caption{Average TPR in each of the three layers for Case 1 when $\Sigma^{(x)} = I_g$. Model (A): $\bmu^{(t,m)} \neq \0; \Lambda = cor(X) \approx I_g$. Model (B): $\bmu^{(t,m)} = \0; \Lambda = cor(X) \approx I_g$.}
		\label{fig: sim_case_1_Uncor}
	\end{figure}

	Table S1 presents the complete results of EM-based estimation and subsequent model selection (as described in Section \ref{sec: model}) for the simulated data in Case 1. We include the true positive selection rate (TPR), the average absolute error for the inclusion probabilities $w^{(t,m)}_k$, i.e., $E_w = \frac{1}{g |M|\tau}\sum_{k,m,t} |\zeta^{(t,m)}_k - \hat{w}^{(t,m)}_k|$, and the average squared error for the estimates of $\beta$, i.e., $E_\beta = \frac{1}{gp |M|\tau}\sum_{k,j,m,t} (\beta^{(t,m)}_{kj} - \hat{\beta}^{(t,m)}_{kj})^2$. In Figure \ref{fig: sim_case_1_Uncor}, we plot the average TPR across 30 simulations for the four scenarios based on different choices of $\sigma^2$ for the case $\Sigma^{(x)} = I_g$, i.e., assuming no correlation between the genomic markers. As the magnitude of the noise $\sigma^2$ increases, the TPR decreases across all layers. In this case, we also evaluate the performance of estimation while ignoring the dependence between the layers, by setting $\bmu^{(t,m)} = \0$ for all $t$; this result is shown as model (B) in Figure \ref{fig: sim_case_1_Uncor}. From the results for layers 2 and 3, we see that including dependence across layers within the prior of $\blambda^{(t)}$ improves the performance in terms of the TPR for all noise levels. Note that these two scenarios perform similarly in layer 1 as we have $\bmu^{(1,m)} = \0$. The first panel in Table S1 shows these rates in more detail along with $E_w$ and $E_\beta$. Since $\Sigma^{(x)} = I_g$ and we use $\Lambda$ to denote the correlation matrix of $X$, i.e., $\Lambda = cor(X)$, we expect $\Lambda \approx I_g$. 
	
	\begin{figure}[!t]
		\centering
		\resizebox{\textwidth}{!}{
			\begin{tabular}{|c|c|c|c|}
				\hline
				\includegraphics[page=1]{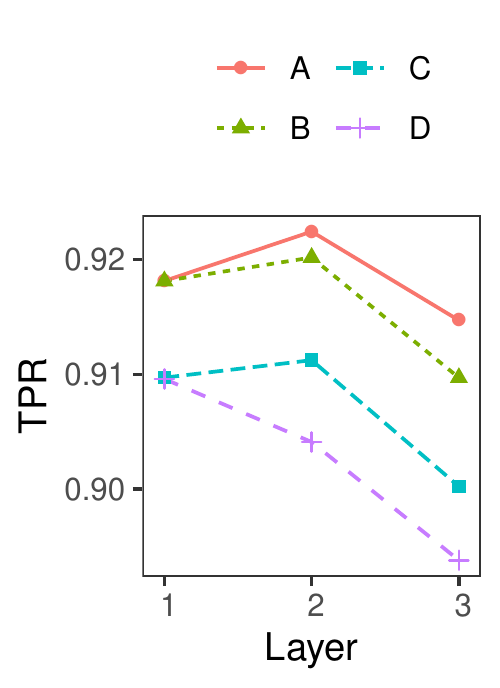} &
				\includegraphics[page=2]{plots/TPR_case1.pdf} &
				\includegraphics[page=3]{plots/TPR_case1.pdf} &
				\includegraphics[page=4]{plots/TPR_case1.pdf} \\
				(a) $\sigma^2 = 1$ & 
				(b) $\sigma^2 = 10$ &
				(c) $\sigma^2 = 20$ &
				(d) $\sigma^2 = 30$ \\ \hline
			\end{tabular}
		}
		\caption{Average TPR in each of the three layers for Case 1 when $\Sigma^{(x)}$ is a $(2 \times 2)$-block matrix. Model (A): $\bmu^{(t,m)} \neq \0; \Lambda = cor(X)$. Model (B): $\bmu^{(t,m)} = \0; \Lambda = cor(X)$. Model (C): $\bmu^{(t,m)} \neq \0; \Lambda = I_g$. Model (D): $\bmu^{(t,m)} = \0; \Lambda = I_g$.}
		\label{fig: sim_case_1_cor}
	\end{figure}
	
	\textbf{Comparative analysis.} In Figure \ref{fig: sim_case_1_cor}, we consider the case where $\Sigma^{(x)}$ has a block structure and evaluate performance of the model under four scenarios: (A) including both $\bmu^{(t,m)}$ and $\Lambda$, (B) ignoring only $\bmu^{(t,m)}$, (C) ignoring only $\Lambda$, and (D) ignoring both $\bmu^{(t,m)}$ and $\Lambda$. Ignoring $\bmu^{(t,m)}$ corresponds to choosing $\bmu^{(t,m)} = \0$ for all $t$; ignoring $\Lambda$ implies choosing $\Lambda = I_g$. Scenario (A) refers to incorporating the structure as described in Section \ref{sec: model}; Scenario (B) is similar in spirit to other variable selection approaches \citep{li2010bayesian,vannucci2010bayesian} that incorporate dependence in the covariates through the indicator variable $\zeta$ (while ignoring the layer structure); Scenarios (C) and (D) correspond to vanilla Bayesian variable selection for each layer with uniform priors on selection probabilities for $\zeta$ \citep{george1997approaches}. To establish comparison, we use the same data replications across all four scenarios. We see that incorporating the structure in $X$ through the hyperparameter $\Lambda = cor(X)$, and borrowing information from previous layers for the mean $\bmu^{(t,m)}$, yields better estimation performance in terms of TPR, compared to having no structure by choosing $\bmu^{(t,m)} = \0 ~\forall ~t$ or $\Lambda = I_g$, or both. This improvement in performance holds true for all choices of the noise parameter $\sigma^2$. Detailed results are shown in the second panel of Table S1. We have not included the false positive rates in the results as they were less than $0.1\%$ on average in all scenarios. We also compared the estimation under a multivariate-multiple regression with LASSO penalty, and observed high TPRs but also substantially higher false positive rates (for both Cases 1 and 2). Results presented in Table S1 report the averages obtained across 30 replications and the corresponding standard deviations. Details of the various hyperparameter choices in the data generation and model estimation processes are given in Section S3.
	
	\textbf{Case 2: $Y^{(t)}$ is generated directly based on fixed $B^{(t)}$.} In this case, we evaluate the performance of our approach in identifying the associations if they are indeed sequentially dependent across layers. Similar to Case 1, we evaluate performance of the model under the four scenarios (A)-(D). Each of these scenarios determines the choices for $\bmu^{(t,m)}$ and $\Lambda$ as described in Case 1.
	
	We assume $p = p^{(t)} ~\forall ~t$ and consider fixed values of $B^{(t)}$ for all layers $t=1,\ldots,\tau=3$. We further assume that $p^{(t,m)} = 3$ for all $m,t$ and $g = 20$. The choice of $B^{(t)}$ is considered so that the genomic markers which are associated with layer $t$ are based on the associations in layers $1,\ldots,(t-1)$. These associations are built so that they do not necessarily need to hold across all imaging sequences. Specifically, we consider $B^{(1)} = \text{block-diag}(B^{(1)}_{15,9}, \0)$, $B^{(1)} = \text{block-diag}(B^{(2)}_{10,6}, \0)$, and $B^{(1)} = \text{block-diag}(B^{(3)}_{5,3}, \0)$, where $B^{(t)}_{r,s}$ is an $(r \times s)$-matrix whose entries are filled by randomly sampling $rs$ values from a double-exponential distribution centered at $0$ with scale parameter $\theta$. The value of $\theta$ acts as the effect size, and we consider four choices: $\theta = 0.7,0.8,0.9,1$. We use $m=1,\ldots,4$ to index the four MRI sequences. Here $B^{(1)}$ indicates that the first 15 genomic markers are associated with all three PC scores corresponding to $m=1,2,3$. As indicated by $B^{(2)}$, the first 10 of these 15 markers are further associated with all three PC scores for $m=1,2$. Similarly, from $B^{(3)}$ we see that the first five markers are associated with all three PC scores for $m=1$. We construct $X$ such that the first 15 columns are positively correlated with each other and uncorrelated with the next five (16-20) columns. The last five columns are also considered to be positively correlated with each other. That is, we consider $\Sigma^{(x)}$ as a $(4 \times 4)$-block matrix as shown in Figure S6 of Section S3. To generate $Y^{(t)}$, we first simulate $\Delta^{(t)} \sim IW(\delta = p^{(t)}, \Psi = \sigma^2I_g)$ with $\sigma^2 = 20$, which is one of the higher choices for the noise parameter in Case 1. We then generate $vec(Y^{(t)}) \sim N_{np^{(t)}}(vec(XB^{(t)}), I_n \otimes \Delta^{(t)})$ and appropriately reshape $vec(Y^{(t)})$ to $Y^{(t)} \in \mathbb{R}^{n \times p^{(t)}}$. This step-by-step procedure of generating the data under Case 2 is provided in Section S3. 
	
	\begin{figure}[!t]
		\centering
		\resizebox{\textwidth}{!}{
			\begin{tabular}{|c|c|c|c|}
				\hline
				\includegraphics[page=1]{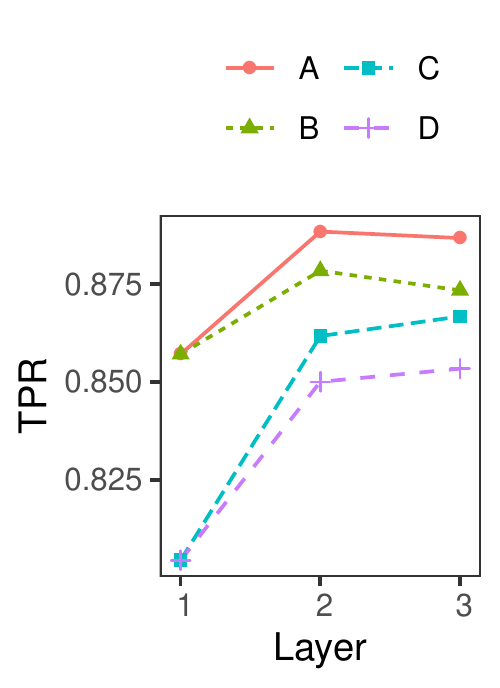} &
				\includegraphics[page=2]{plots/TPR_case2.pdf} &
				\includegraphics[page=3]{plots/TPR_case2.pdf} &
				\includegraphics[page=4]{plots/TPR_case2.pdf} \\
				(a) $\theta = 1$ & 
				(b) $\theta = 0.9$ &
				(c) $\theta = 0.8$ &
				(d) $\theta = 0.7$ \\ \hline
			\end{tabular}
		}
		\caption{Average TPR in each of the three layers for Case 2 when $\Sigma^{(x)}$ is a $(4 \times 4)$-block matrix. Model (A): $\bmu^{(t,m)} \neq \0; \Lambda = cor(X)$. Model (B): $\bmu^{(t,m)} = \0; \Lambda = cor(X)$. Model (C): $\bmu^{(t,m)} \neq \0; \Lambda = I_g$. Model (D): $\bmu^{(t,m)} = \0; \Lambda = I_g$.}
		\label{fig: sim_case_2}
	\end{figure}
	
	We now discuss the results of the simulation study under Case 2. Similar to Case 1, we present the TPR, $E_w$ and $E_\beta$ for each of the $\tau=3$ layers separately in Table S2. For each choice of the effect size $\theta$, we present results corresponding to four scenarios, which are determined based on incorporating (or ignoring) the structure through $\bmu^{(t,m)}$ and $\Lambda$. Figure \ref{fig: sim_case_2} shows the average TPR under all four choices of $\theta$. We note that including the structure through $\bmu^{(t,m)}$ and $\Lambda$ provides higher values for TPR thereby improving performance, even for smaller effect sizes. From Table S2, we see that by including this structure we also obtain lower values for error metrics $E_w$ and $E_\beta$. We note that including the dependence structure in the columns of $X$, through $\Lambda$, has a pronounced influence on the performance compared to ignoring it by considering $\Lambda = I_g$. Including the structure between the layers through $\bmu^{(t,m)}$ also has an incremental effect in the performance for all choices of $\theta$. The results for layer 1 are similar for scenarios with $\bmu^{(t,m)} = 0$ and $\bmu^{(t,m)} \neq 0$ for fixed $\Lambda$ as $\bmu^{(1,m)} = 0$. Table S2 reports the averages obtained across 30 replications, and are accompanied by the standard deviations; false positive rates are not reported as they were negligible across all scenarios. Details of the hyperparameter choices are provided in Section S3.
	
	\section{Radiogenomic Analysis in Lower Grade Gliomas}\label{sec: radiogenomic}
	We leverage the radiology-based imaging data from TCIA and the genomic data from The Cancer Genome Atlas (TCGA; \url{www.cancer.gov/tcga}) in the context of LGG, to understand the associations between imaging phenotypes and genomic markers. In this section, we describe the data acquisition and pre-processing steps, and present results from our Bayesian modeling approach. %A \texttt{R} package is provided with the supplementary material that includes relevant data and code. %A \texttt{R} package \textit{MBVSem} is provided with the online supplementary material that includes relavant data and code. (also available at \url{www.github.com/bayesrx/MBVSem}).
	
	\textbf{Imaging data.} The imaging data acquired from TCIA contain publicly available pre-operative multi-institutional MRI scans for the TCGA LGG cohort. Specifically, we obtained MRI scans and the corresponding tumor segmentation labels for 65 LGG subjects from \cite{bakas2017lggsegmentation,bakas2017advancing}. The segmentation labels were constructed using an automated segmentation method called GLISTRboost \citep{bakas2015glistrboost}. One of the hurdles posed by MRI for statistical analysis is that the voxel intensity values are not comparable either between study visits for the same subject or across different subjects. Hence, it is required to pre-process these images to perform intensity value normalization. We employ a biologically motivated normalization technique using the \emph{R} package \emph{WhiteStripe} for intensity normalization \citep{shinohara2014statistical}. However, we discard imaging data for two subjects due to issues with segmentation masks and intensity value normalization, leaving us with a total of 63 LGG subjects with imaging data.
	
	\textbf{Imaging phenotypes.} Gliomas contain various heterogeneous subregions broadly categorized into three groups: necrosis and non-enhancing core, edema, and enhancing core. These subregions reflect differences in tumor biology, have variable histologic and genomic phenotypes, and exhibit highly variable clinical prognosis \citep{bakas2017advancing}. Typically, the area of dead cells in the center is referred to as necrosis or non-enhancing core, and the tissue swelling caused in the outer region due to accumulation of fluid is referred to as edema (\url{mayfieldclinic.com/pe-braintumor.htm}). Following suit, we divide the tumor region for each subject into $\tau=3$ spherical shells, i.e., one inner sphere and two spherical shells. For each subject, we follow the procedure described in Algorithm S1 to construct the PDFs corresponding to the three spherical shells. The PDFs estimated in each layer are used to construct the PC scores $Y^{(t)}$, which act as the imaging phenotypes. 	
	
	\textbf{Genomic data.} We obtained the genomic data from LinkedOmics \citep{vasaikar2017linkedomics}, a publicly available portal that includes multi-omics data from multiple cancer types in TCGA. The genomic data we acquire are normalized gene-level RNA sequencing data from the Illumina HiSeq system (high-throughput sequencing) with the expression values on the $\log_2$ scale. This data is obtained for all of the 63 matched LGG subjects across 20086 genes. We focus our analysis on {\em LGG-genes}, which are genes specifically reported as cancer drivers in LGG \citep{bailey2018comprehensive}. The cancer driver genes were cataloged as driver genes based on a PanCancer and PanSoftware analysis (comprising of all 33 TCGA projects and several computational tools). The set of cancer driver genes for LGG includes the 24 genes reported in Table S3, whose expression profile was retrieved from the expression of the full set of genes from LinkedOmics. We work with the gene expression data corresponding to these 24 LGG-genes as the genomic markers for the 63 subjects.
	
	\textbf{Prior elicitation.} For our statistical analysis of the LGG data, we choose $\bmu^{(t,m)} = \alpha\hat{\blambda}^{(t-1,m)}_+$, which prioritizes selection of genes in the previous layer, and $\Lambda$ as the correlation matrix of $X$ capturing dependence structure between the genes. The choices of some of the other hyperparameters include $v_1 = 100, a_1 = 4, a_2 = 5, \alpha = 0.5, \Psi = I_g, \delta = p^{(t)}$. As $v_1$ is expected to be very large, we have compensated by placing an informative prior on $\nu^{-2(t,m)}_{kj}$ with more mass around the value $1$ based on our choice of $a_1$ and $a_2$. The choice of $\delta$ and $\Psi$ are made such that we have a non-informative prior around $\Delta^{(t)}$. Within the estimation for each model for layer $t$, we choose $10^{-5}$ as the threshold for convergence. The choice of $\alpha$ is such that we place the prior mean at the midpoint of zero and $\hat{\blambda}^{(t-1,m)}$ for the selected genes. Further details about hyperparameter choices are provided in Section S4.4. For model selection, we estimate the model on a grid of values for $v_0 \in \{ 0.001+ (i-1)0.001 | i=1,\ldots,40 \}$ and compute the BIC based on each value of $v_0$ to choose the best model as the one with lowest BIC. 
	
\begin{figure}[!t]
		\centering
		\resizebox{\textwidth}{!}{
			\begin{tabular}{c}
				\includegraphics[trim = 1cm 0.25cm 0.45cm 0.2cm, clip]{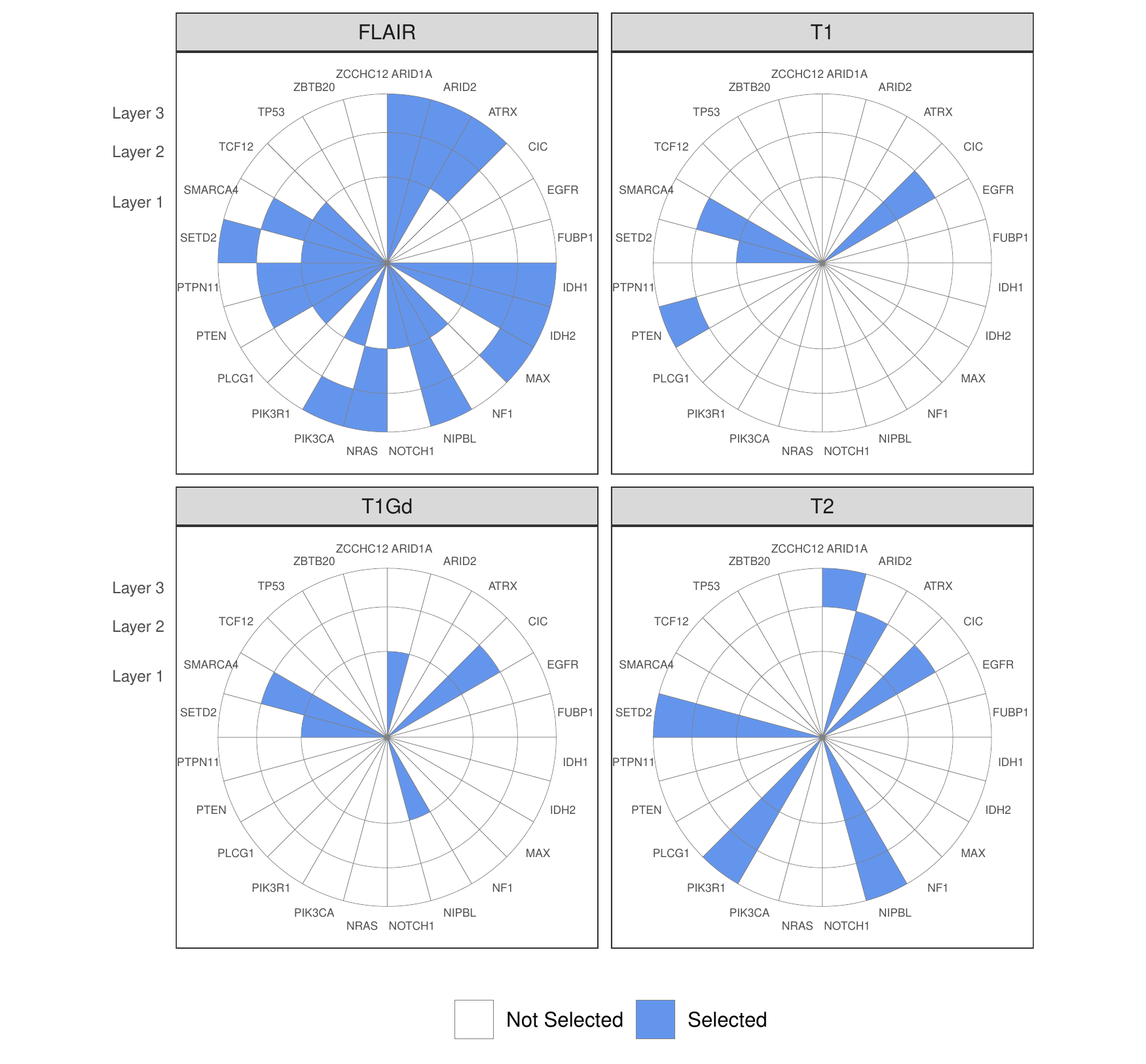}
			\end{tabular}
		}
		\caption{Associations of layer-wise PC scores from the four imaging sequences with the gene expression. Each panel corresponds to an imaging sequence, and a block filled with color indicates association for that gene with one of the three specific layers ($\tau = 3$).}
		\label{fig: sig_genes_3}
	\end{figure}

	We implement the multiple-multivariate Bayesian regression model with the gene expression of the LGG-genes as predictors and the PC scores from the MRI images as responses. Figure \ref{fig: sig_genes_3} presents the results of our analysis, where each panel corresponds to one of the four imaging sequences (FLAIR, T1, T1Gd and T2). The inner sphere from the tumor region is represented as the inner circle (layer 1). Similarly, the two spherical shells are represented as concentric annuli (layers 2 and 3). The circular region in the plot is divided into 24 sectors, with each representing one of the 24 LGG-genes under consideration. Consequently, each sector defined by a gene is divided into three concentric blocks corresponding to the three layers. If $\hat{\zeta}^{(t,m)}_k = 1$ for gene $k$ in layer $t$ of imaging sequence $m$, the corresponding block is colored in blue. Based on our estimation and model selection, if $\hat{\zeta}^{(t,m)}_k = 1$, the expression of gene $k$ has significant association with the PC scores corresponding to layer $t$ from the imaging sequence $m$. For example, in FLAIR, the block in layer 3 corresponding to the sector for gene ATRX is colored, indicating significant association of the expression of ATRX with PC scores from layer 3 in this imaging sequence. 
	
	Figure S13 shows plots corresponding to the $Q$-function maximization in the EM algorithm. In terms of computation time, our algorithm requires about $20$ seconds for estimation and model selection across $\tau = 3$ layers with $g = 24$ genes and $p^{(1)} = 22, p^{(2)} = 18, p^{(3)} = 17$ PC scores. The analysis was executed on a computer with an Intel(R) Core(TM) i7-7700 CPU @ 3.60GHz, 3601 Mhz, 4 cores, 8 logical processors with 32 GB RAM; model selection for $40$ values of $v_0$ was done in parallel across $4$ processors. Under the same settings, generating 200 MCMC samples of the model parameters for each layer sequentially would require about 50 seconds. We demonstrate the utility of our EM-based estimation compared to a MCMC-based approach. Details of this comparison under different choices for the number of parameters are provided in Section S5.
	
	\textbf{Biological findings and implications.}\label{subsec: biological} 
	We now describe the biological significance of the radiogenomic associations identified by our model. These results are presented separately for each imaging sequence in Figure \ref{fig: sig_genes_3}. Most of the radiogenomic associations appear in FLAIR, which presents most of the tumor regions in LGG with relative hypointensity (dark signal on MRI), except for a hyperintense (bright) rim \citep{weerakkody2020anaplastic}. The other three sequences usually present a hypointense signal compared to other brain tissues, with exceptions at certain times. We specifically focus on the following identified associations: 
	\begin{enumerate}
		\item The genes IDH1 and IDH2 demonstrate significant associations with all three layers for the FLAIR sequence. Several recent studies \citep{yan2009idh1,leu2013idh,leu2016idh} have highlighted the prognostic importance of IDH mutations in gliomas as well as their distinctive genetic and clinical characteristics. Mutations that affect IDH1 were identified in more than 70\% of low grade gliomas and in glioblastomas that developed from lower-grade lesions; most of the other tumors often had mutations affecting the gene IDH2 \citep{yan2009idh1}. These mutations were also shown to be strong predictors for survival, specifically in the case of LGG \citep{leu2013idh}. During the developmental stage of most LGG tumors, one of the earliest genetic events is the acquisition of IDH1 or IDH2 mutation \citep{leu2016idh}.
		
		\item The PI3K pathway (group of genes), which includes genes such as PTEN, PIK3CA and PIK3R1, is one of the most deregulated and druggable pathways in human cancer \citep{arafeh2019pik3ca}. We see associations of the genes PTEN, PIK3CA and PIK3R1 with the spherical shells in both T2 and FLAIR. The PI3K pathway controls multiple cellular processes including metabolism, motility, proliferation, growth and survival \citep{janku2018targeting}. Somatic mutations in PIK3R1 act as oncogenic driver events in gliomas and are known to increase signaling through the PI3K pathway; they also promote tumorigenesis of primary normal human astrocytes in an orthotopic xenograft model \citep{quayle2012somatic}.
		
		\item Another significant association across all of the three spherical shells in the FLAIR imaging sequence, and partially in T2, is with the genes ARID1A and ARID2. Studies have shown that somatic mutations in the chromatin remodeling gene ARID1A occur in several tumor types \citep{jones2012somatic}. ARID1A is known to be a tumor suppressor and inhibits glioma cell proliferation via the PI3K pathway \citep{zeng2013arid1a}. ARID2 is a gene associated with chromatin organization and has been predicted to be a glioma driver \citep{ceccarelli2016molecular}. Moreover, recent studies have identified ARID2 as a direct target and functional effector in other cancers such as oral cancer \citep{wu2020mir}.
		
		\item The genes PTPN11 and NIBPL also present significant associations across all three spherical shells in the FLAIR sequence. Additionally, NIPBL presents association across all spherical shells in T2. NIPBL is a somatically altered glioma gene that is known to be a crucial adherin subunit, and is essential for loading cohesins on chromatin \citep{ceccarelli2016molecular}. PTPN11 is known to mediate gliomagenesis in both mice and humans \citep{liu2011shp}.
		
		\item NRAS is a member of the RAS oncogene family, which encodes small enzymes involved in cellular signal transduction \citep{fiore2016mir}. In our analysis, NRAS shows significant association with the exterior layers of the tumor in FLAIR. RAS pathways, when activated, trigger downstream signaling pathways such as MAPKs and PI3K/AKT, which modulate cell growth and survival \citep{schubbert2007hyperactive}. NRAS was also identified as a gene which promoted oncogenesis in glioma stem cell \citep{gong2016knockdown}.
		
		\item We see significant associations of the gene CIC with the inner and middle layers in T1, T1Gd and T2. CIC is a transcriptional repressor that counteracts activation of genes downstream of receptor tyrosine kinase (RTK)/RAS/ERK signaling pathways \citep{bunda2019cic}. Loss of CIC potentiated the formation and reduced the latency in tumor development in an orthotopic mouse model of glioma \citep{yang2017cic}. Tumorigenesis in high-grade gliomas is attributed to hyperactive RTK/RAS/ERK signaling, while CIC has a tumor suppressive function \citep{bunda2019cic}. This indicates an inverse functional relationship between CIC and the genes in RTK/RAS/ERK pathways (e.g., NRAS). Interestingly, we see that associations with CIC are picked only when there are no significant associations with NRAS.
	\end{enumerate}
	These findings indicate several associations between the LGG-genes, and related pathways, with imaging phenotypes from the spherical shells. It is promising that several of these genes have been found to play important roles in the genesis of glioma. It is not surprising that we see most of the associations in the FLAIR sequence, as it exhibits the best contrast between normal brain tissue and presumed infiltrating tumor margins \citep{grier2006low}. It is also the principal imaging sequence for assessment of LGG growth \citep{bynevelt2001flair}. These findings encourage a deeper corroboration, which could potentially indicate intricate characteristics of tumor growth.
	
	In the analysis presented above, the tumor region was divided into three spherical shells to conform with the number of natural subregions: necrosis, edema and enhancing tumor. However, we also deployed our framework with other choices for the number of spherical shells including $\tau = 4, 5,$ and $6$. We plot these results in Figures S10-S12 in Section S4.2 of the supplementary material. We see that the results are reasonably robust, i.e., most of the genes associated with the layers from the imaging sequences are consistent (except for a few additions/deletions) across different choices of $\tau$. This indicates that our approach is broadly able to identify the underlying radiogenomic associations between the radiomic phenotypes and gene expression. 
	
	\section{Discussion and Future Work}\label{sec: discussion}
	
	In this paper, we propose a new framework for integrating radiological imaging and genomics data that is biologically-informed, mimics the tumor evolution process, and accounts for (structural) tumor heterogeneity. In the context of LGG,  we identify  molecular determinants of this tumor evolution process by borrowing information sequentially from concentric spherical shells; this process tries to emulate tumor growth since 3D spheroids are known to better mimic real tumors \citep{breslin2013three}. We build layer-wise sequential Bayesian regression models that consider (i) the PC scores constructed from the PDFs of the MRI-based tumor voxel intensities as the responses, which effectively capture tumor heterogeneity, and (ii) the transcriptomic profiling data in terms of the gene expression of LGG-genes as predictors. Information about the multiple imaging sequences, multiple PC scores from a given sequence, and the correlation between the genes was encoded into the prior structure of the regression model.  The estimation is based on an computationally-efficient EM algorithm, and it identifies genes which have significant association with tumor heterogeneity in each spherical shell. We have proposed novel methodology in terms of multiple-multivariate regression to simultaneously incorporate complex dependence structure between (i) the genomic covariates, and (ii) the components of the multivariate response. Our model incorporates a sequential variable selection strategy to borrow information across layers of the tumor. Furthermore, this sequential strategy can be deployed to model similar structured data, e.g., any imaging data that mimics a growth process or data where the sequencing arises due to time. %We have also extended the EM-based estimation for Bayesian variable selection approaches by \cite{rovckova2014emvs} to a more general setting with multivariate response and multiple covariates. 
	
	An important aspect of our analysis is that we borrow information about the selection of genes starting from the interior sphere and moving in an outward direction. This approach captures the process of tumor growth which starts as a single cancerous cell and grows radially outward to form the entire tumor. The choice of spherical shapes for the different tumor layers is an uninformative one, due to the lack of prior knowledge about the tumor's growth process. If the tumor's growth dynamics are available, this new information can be used to build the layers. This growth process is incorporated through the parameter $\bmu^{(t,m)}$, necessitating sequential estimation. However, we are only interested in the posterior inclusion probabilities for each of the LGG-genes to identify the radiogenomic associations. Hence, instead of using MCMC sampling techniques, we perform estimation using an EM-algorithm, which iteratively searches for the posterior mode of the inclusion probabilities; this, in turn, reduces computational time of estimation. Information about the selection of genes across spherical shells is propagated, and many significant associations in the inner sphere carry their effect forward to subsequent spherical shells. Clearly, this is not always the case, which is reassuring, as these results are not dominated by our choice of prioritizing the previously selected genes through $\bmu^{(t,m)}$. Additionally, dependence structure between the genes is incorporated using $\Lambda$. We see that gene pairs, such as NIBPL and ARID2 or NF1 and ZBTB20, have high positive correlation (Figure S9), and are either jointly selected or jointly not selected in most layers and imaging sequences. Similarly, the genes ATRX and CIC have negative correlation, and their selection alternates, i.e., one of them is not selected while the other one is selected.
	
	Our analysis identifies several associations between image-based tumor characteristics and expression profiles of LGG-genes. The genes associated with the imaging phenotypes in the inner spherical layers can act as potential biomarkers for early events. Monitoring these selected genes from the onset, or the time of diagnosis, could lead to a better understanding of the molecular underpinnings as well as provide effective diagnostic options prior to invasive approaches such as a biopsy. One of the limitations of our data is that the gene expression might not be derived from the spherical layers separately during the biopsy. This can be potentially addressed in the future using more modern technologies that facilitate analyses using spatial transcriptomics \citep{burgess2019spatial} - an area which is still maturing. In the proposed model setup, the parameter space is high-dimensional due to the multivariate regression setting. This results in a high computational burden during estimation while incorporating an extremely large number of genomic markers within the sequential multivariate regression framework; this issue needs to be explored further using highly scalable or approximate estimation algorithms. 
	
	\textbf{Future work.} Our work tries to identify radiogenomic associations between radiomic phenotypes and genomic markers in LGGs by incorporating the biological structure of the tumor into the modeling framework. However, this could be further explored in several directions such as identifying the tumor regions/spherical shells based on known segmentation of the tumor subregions. This could additionally be informed by the shape of the tumor, which the PDFs do not capture, as they only assess the heterogeneity in the tumor voxel values. Also, our model explores linear relationships between the imaging-based PC scores and the gene expression, which could potentially be extended to study nonlinear associations (e.g. using semi-parametric additive models \citep{scheipl2012spike,ni2015bayesian}). Our future work will include integrating these radiogenomic findings into predictive models for clinical outcomes such as overall survival, time to progression.
	
	\section*{Acknowledgments}
	All of the authors acknowledge support by the NIH-NCI grant R37-CA214955. SM was partially supported by Precision Health at The University of Michigan (U-M). SM and AR were partially supported by CCSG P30 CA046592, U-M Institutional Research Funds and ACS RSG-16-005-01. SK was partially supported by NSF grants CCF-1740761, CCF-1839252 and DMS-2015226. SK and KB were partially supported by NSF DMS-1613054. KB was partially supported by NSF DMS-2015374. VB was supported by NIH grants R01-CA160736,  R21-CA220299, and P30 CA46592, NSF grant 1463233, and start-up funds from the U-M Rogel Cancer Center and School of Public Health.
	
	\bibliography{bibliography}
	
	\newpage
	%\appendix
	\setcounter{figure}{0}
	\renewcommand{\thefigure}{S\arabic{figure}}
	\setcounter{section}{0}
	\renewcommand{\thesection}{S\arabic{section}}
	\setcounter{algorithm}{0}
	\renewcommand{\thealgorithm}{S\arabic{algorithm}}
	\setcounter{table}{0}
	\renewcommand{\thetable}{S\arabic{table}}
	\setcounter{equation}{0}
	\renewcommand{\theequation}{S\arabic{equation}}

	\begin{center}
		\LARGE
		\textbf{Supplement for ``Tumor Radiogenomics with Bayesian Layered Variable Selection''}
	\end{center}
	
	We provide several details for the methods and results from the main paper including (i) the construction of layer-wise radiomic phenotypes and the computation of the principal component scores from a sample of probability density functions (Section \ref{app: details_imaging}), (ii) the full posterior of the model, and its logarithm, as well as some computational details related to the Expectation-Maximization (EM) algorithm used for estimation (Section \ref{app: details_est}), (iii) details related to the simulation study, (iv) details related to the gene expression (Section \ref{app: details_sim}), and some additional results for the The Cancer Genome Atlas lower grade glioma data (Section \ref{app: details_case}), and (v) comparison of computational burden for the EM-based estimation and model selection versus a Markov Chain Monte Carlo sampling-based approach.
	
	\section{Computation of Imaging Phenotypes}\label{app: details_imaging}
\begin{figure}[!h]
		\centering
		\resizebox{\textwidth}{!}{%
			\begin{tabular}{|c|c|c|c|}
				\hline
				\includegraphics[trim=1cm 1cm 1cm 2cm, clip, width=1.75in, height=1.75in]{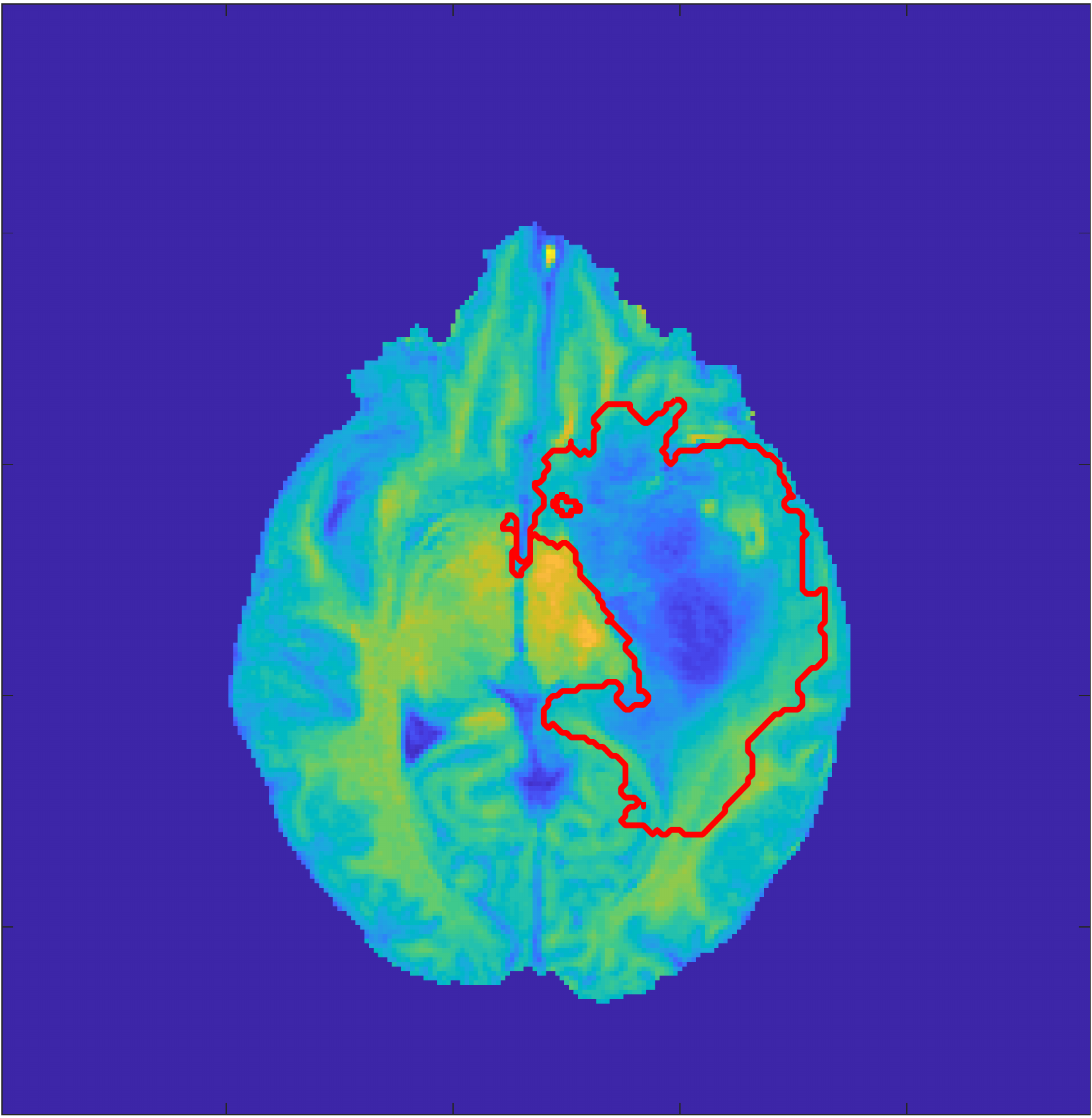} &
				\includegraphics[trim=1cm 1cm 1cm 2cm, clip, width=1.75in, height=1.75in]{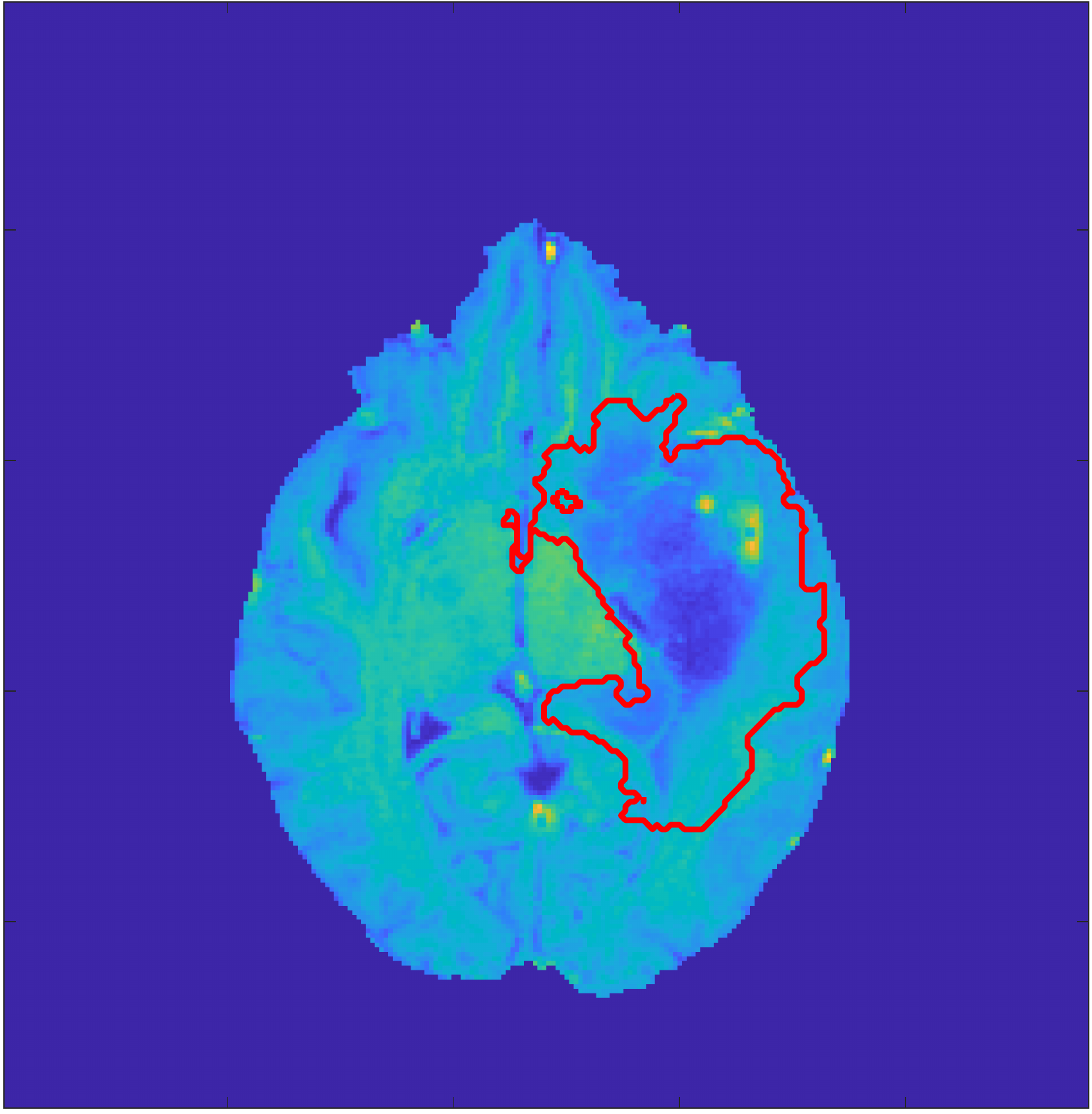} &
				\includegraphics[trim=1cm 1cm 1cm 2cm, clip, width=1.75in, height=1.75in]{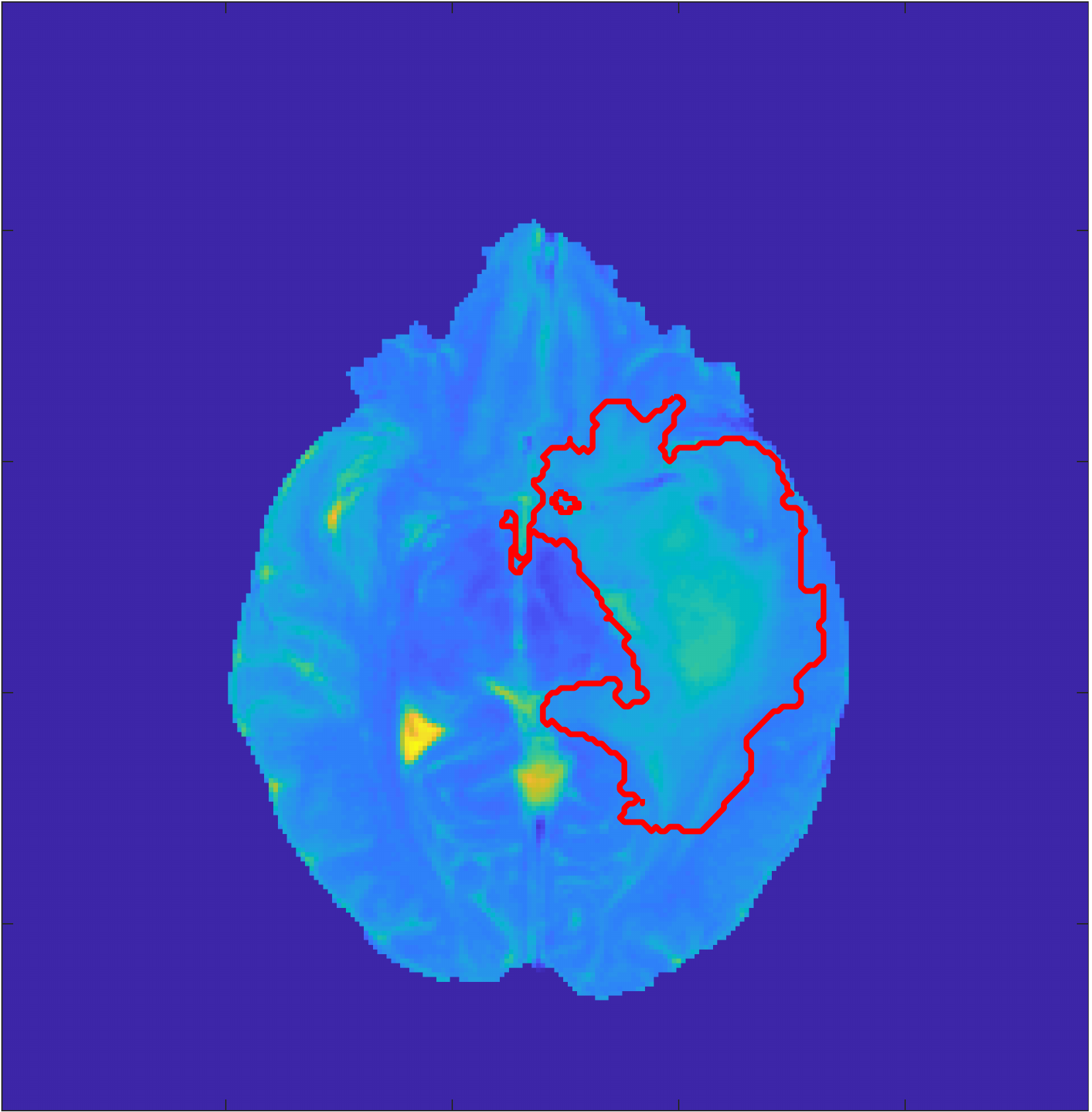} &
				\includegraphics[trim=1cm 1cm 1cm 2cm, clip, width=1.75in, height=1.75in]{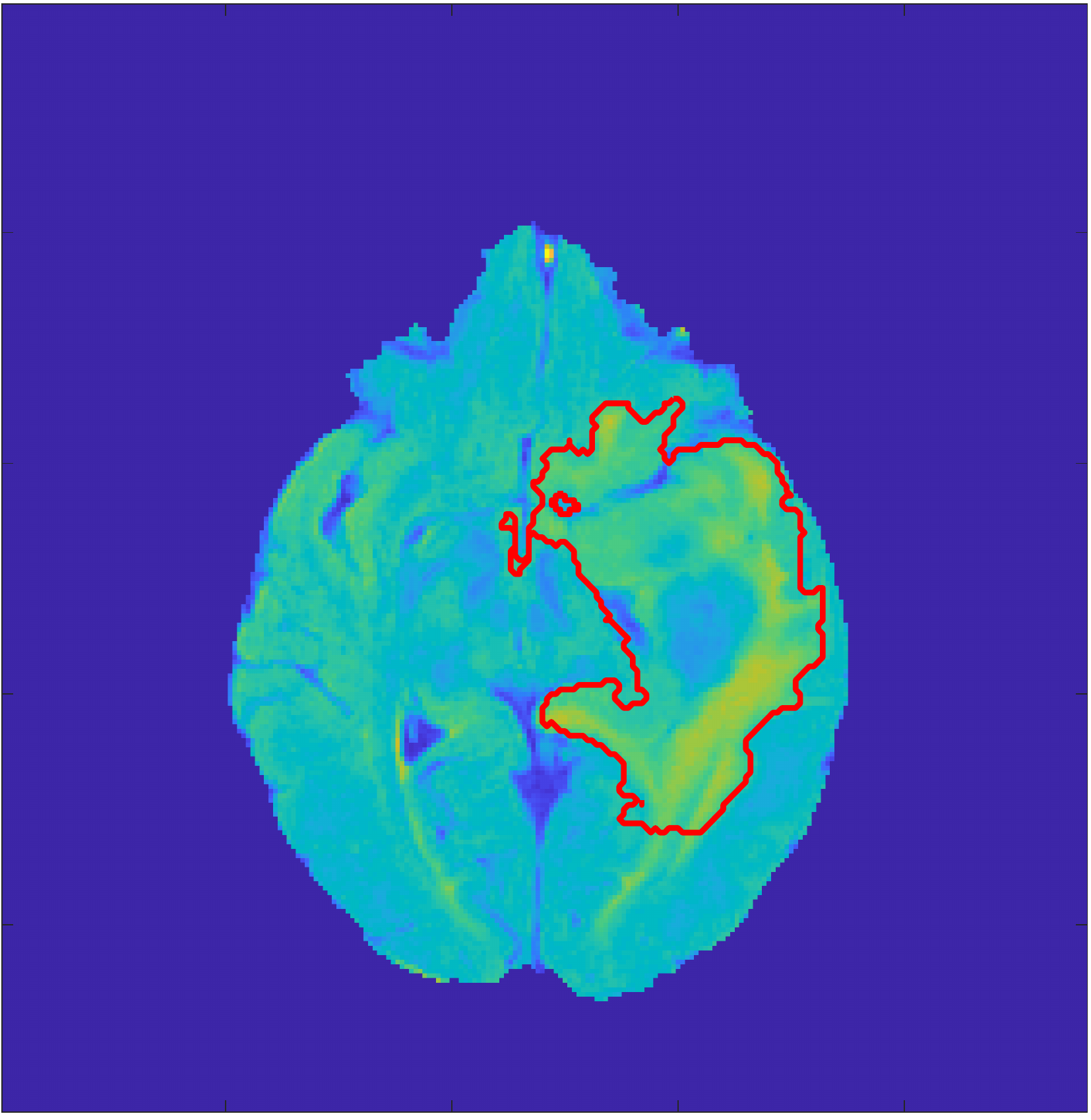} \\
				{\small(a) T1} & {\small(b) T1Gd} &
				{\small (c) T2} & {\small (d) FLAIR}\\
				\hline
			\end{tabular}
		}
		\caption{\small Axial slice of a brain MRI for a subject with LGG for the four imaging sequences: (a) T1, (b) T1Gd, (c) T2, and (d) FLAIR. The region inside the red boundary corresponds to the segmented tumor region.}
		\label{fig: mri}
	\end{figure}

	Figure \ref{fig: mri} shows examples of axial slices of the whole brain magnetic resonance imaging (MRI) for an lower grade glioma (LGG) patient, with the tumor region highlighted using a red boundary. The four images correspond to the T1, T1Gd, T2 and FLAIR imaging sequences, respectively.
	
	\subsection{Construction of Layer-wise Radiomic Phenotypes} %\vb{[Make sure this is labeled appendix]}
	The construction of kernel density estimates $f^{(t,m)}_i$ corresponding to spherical shell $t \in \{1,\ldots,\tau\}$ from MRI sequence $m \in M = \{T1,T1Gd,T2,FLAIR\}$ for subject $i$ are described in Algorithm \ref{algo: densities}. These kernel density estimates are computed using the \emph{density} function in \texttt{R} using the default bandwidth selection method, also commonly referred to as Silverman's rule, which is an optimal bandwidth for estimating Gaussian densities \citep{silverman1986density}.
	%\kb{In Algorithm 1 have Input and Output steps. Define $N_i$ as well}
	\begin{algorithm}[!h]
		\caption{Construction of layer-wise kernel density estimates}\label{algo: densities}
		\begin{algorithmic}[1]
			%\Procedure{MyProcedure}{}
			\State \textit{Input:} MRI scans from $M$ sequences and corresponding tumor segmentation for $n$ subjects.
			\State \textit{Output:} PDFs corresponding to $\tau$ layers from $M$ imaging sequences for $n$ subjects.
			\For {each subject $i=1,\ldots,n$}
			\State Define $N_i$ as the tumor volume (number of tumor voxels).
			\For {each MRI sequence $m \in M$}
			\For {each layer $t=1,\ldots,\tau$}
			\If {$t=1$} 
			\State \multiline{Define $S^{(1)}_i$ as the sphere with $\c_i$ as its center and radius $r^{(1)}_i = \inf\limits_{r \in \mathbb{R}^+} r$ such that $r$ is the radius of all spheres centered at $\c_i$ and containing at least $N_i/\tau$ voxels.}
			\State \multiline{Construct $f^{(1,m)}_i$ as the kernel density estimate based on the voxels in sequence $m$ belonging to the sphere $S^{(1)}_i$.}
			\EndIf
			\If {$t>1$} 
			\State \multiline{Define the spherical shell $D^{(t)}_i = S^{(t)}_i\backslash S^{(t-1)}_i$, where $S^{(t)}_i$ is the sphere with center $\c_i$ and radius $r^{(t)}_i = \inf\limits_{r \in \mathbb{R}^+} r$ such that $r$ is the radius of all spheres centered at $\c_i$ and containing at least $tN_i/\tau$ voxels.}
			\State \multiline{Construct $f^{(t,m)}_i$ as the kernel density estimate based on the voxels in sequence $m$ belonging to the spherical shell $D^{(t)}_i$}.
			\EndIf
			\EndFor
			\EndFor
			\EndFor
		\end{algorithmic}
	\end{algorithm}
	
	\subsection{Mapping Probability Density Functions to PC Scores}\label{subapp: mapping}
	
	The kernel density estimates are probability density functions (PDFs) and belong to the Banach manifold of all PDFs. We focus on PDFs on the domain $[0,1]$, however the following description can be applied to other domains with simple adjustments. Let $\calF = \big\{f:[0,1] \rightarrow \mathbb{R}_{> 0} \big| \int_0^1 f(x) dx = 1 \big\}$. We use the Fisher-Rao (F-R) Riemannian metric \citep{rao1992information,kass2011geometrical,srivastava2007riemannian} on $\calF$ to make it a Riemannian manifold and facilitate computation. This metric has nice statistical properties such as invariance to bijective and smooth transformations of the PDF domain \citep{cencov1982statistical}; it is also closely related to the Fisher information matrix. However, the metric is difficult to employ for computation so we skip the specific formula here for brevity. To build some basic statistical tools, e.g., the distance between a pair of densities, Karcher mean for a sample of densities, and PCA on the space of densities, we work on a transformed space of PDFs using a square-root transformation \citep{bhattacharyya1943measure}. Consequently, the space of PDFs becomes the positive orthant of the unit sphere in $\mathbb{L}^2$, whose geometry is well-known, and the F-R metric flattens to the standard $\mathbb{L}^2$ metric.
	
	\noindent\paragraph{Distances Between PDFs.} The square-root transformation (SRT) of a PDF is defined as a function $h = +\sqrt{f}$ (we omit the $+$ sign hereafter for notational convenience). The inverse mapping is unique and is simply given by $f = h^2$ \citep{kurtek2015bayesian}. The space of SRTs is given by $\mathcal{H} = \{h : [0,1] \rightarrow \mathbb{R}_{>0} | \int_0^1 h(x)^2 dx = 1 \}$, which represents the positive orthant of a unit Hilbert sphere \citep{lang2012fundamentals}. The $\mathbb{L}^2$ Riemannian metric on $\mathcal{H}$ is defined as $\langle\langle \delta h_1 , \delta h_2 \rangle\rangle= \int_0^1 \delta h_1(t)\delta h_2(t) dt$, where $\delta h_1,\delta h_2 \in T_h(\mathcal{H})$ and $T_{h}(\mathcal{H}) = \{ \delta h : [0,1] \rightarrow \mathbb{R} | \int_0^1 h(x)\delta h(x)dx = 0\}$. The geodesic paths and lengths can now be analytically computed due to the Riemannian geometry of $\mathcal{H}$ equipped with the $\mathbb{L}^2$ metric ($\mathcal{H}$ is a Riemannian manifold and the $\mathbb{L}^2$ metric on $\calH$ corresponds to the F-R metric on $\calF$). The geodesic distance between $h_1,h_2\in\mathcal{H}$ is simply given by $d(h_1,h_2)=\theta=\cos^{-1}\big(\int_0^1 h_{1}(x)h_{2}(x) dx\big)$. Under this setup, the geodesic distance between two densities $f_1,f_2 \in \calF$, represented by their SRTs $h_1,h_2 \in \calH$ is defined as $d(f_1,f_2)=d(h_1,h_2)$.
	
	\begin{algorithm}[!t]
		\caption{Sample Karcher mean of densities}\label{algo: karcher}
		\begin{algorithmic}[1]
			%\Procedure{MyProcedure}{}
			\State $\bar{h}_0$ (initial estimate for the Karcher mean) $\gets$ any one of the densities in the sample OR the extrinsic average. Set $j \gets 0$ and $\epsilon_1, \epsilon_2 > 0$ small.
			\State For $i=1,\ldots,n$ compute $u_i = \exp^{-1}_{\bar{h}_j}(h_i)$.
			\State Compute the average direction in the tangent space $\bar{u} = \frac{1}{n} \sum_{i=1}^n u_i$.
			\If {$||\bar{u}||_{L^2} < \epsilon_1$} 
			\State \Return $\bar{h}_j$ as the Karcher mean.
			\Else {} 
			\State$\bar{h}_{j+1} = \exp_{\bar{h}_j}(\epsilon_2 \bar{u})$.
			\State Set $j \gets j+1$.
			\State Return to step $2$.
			\EndIf
		\end{algorithmic}
	\end{algorithm}
	
	\noindent\paragraph{Karcher Mean for a Sample of PDFs.} The geometry of the space of SRTs can be used to define an average PDF of a sample of PDFs, which allows us to efficiently summarize and visualize the sample. The average PDF can be computed using a generalized version of a mean on a metric space called the \textit{Karcher} mean \citep{karcher1977riemannian,dryden1998statistical}. Suppose we have $n$ pdfs $f_1,\ldots,f_n$ and their corresponding SRTs $h_1,\ldots,h_n$. The sample Karcher mean $\bar{h}\in \mathcal{H}$ is defined as the minimizer of the Karcher variance $\rho(\bar{h}) \propto \sum\limits_{i=1}^n d(h_i,\bar{h})^2$, i.e., $\bar{h} = \text{argmin}_{h \in \mathcal{H}} ~\rho(h)$. Algorithm \ref{algo: karcher} presents a gradient-based approach to compute the Karcher mean on $\mathcal{H}$. The Karcher mean of the sample PDFs is an intrinsic average that is computed directly on $\mathcal{H}$ (or equivalently $\mathcal{F}$). For the computation of the Karcher mean, we require tools from differential geometry called the exponential and inverse-exponential maps. The exponential map at a point $h_1 \in \mathcal{H}$, denoted by $\exp: T_{h_1}(\mathcal{H}) \mapsto \mathcal{H}$, is defined as $ \exp_{h_1}(\delta h) = \cos(\| \delta h \|)h_1+ \sin(\| \delta h \|)(\delta h/\|\delta h\|)$,
	where $\|\delta h\|=\big(\int_0^1 \delta h(x)^2 dx\big)^{1/2}$. The inverse-exponential map, denoted by $\exp^{-1}_{h_1}: \mathcal{H} \mapsto T_{h_1}(\mathcal{H})$, is given by $\exp^{-1}_{h_1}(h_2) = (\theta/\sin(\theta))(h_2 - h_1\cos(\theta))$.
	
	\noindent\paragraph{Principal Component Analysis (PCA) for a Sample of PDFs.} Under standard settings, visualizing variability on the space of PDFs is not straightforward. PCA is an effective method to explore the variability in the PDFs through the primary modes of variation in the data. Note that the tangent space is a (Euclidean) vector space, hence PCA can be implemented there, as in standard multivariate problems. Algorithm \ref{algo: pca} describes the computation of PCA on the space generated by SRTs $h_1,\ldots,h_n$ corresponding to the sample of PDFs $f_1,\ldots,f_n$, assuming that they have been sampled using $m$ points on the domain $[0,1]$.
	
	\begin{algorithm}[!t]
		\caption{PCA on $T_{\bar{h}}(\mathcal{H})$}\label{algo: pca}
		\begin{algorithmic}[1]
			%\Procedure{MyProcedure}{}
			\State Compute the Karcher mean of $h_1,\ldots,h_n$ as $\bar{h}$ using Algorithm \ref{algo: karcher}.
			\For {$i=1,\ldots,n$} 
			\State Compute projections ($v_i = \exp^{-1}_{\bar{h}} (h_i)$) of $h_i$ onto $T_{\bar{h}}(\mathcal{H})$.
			\EndFor
			\State Evaluate sample covariance matrix $K = \frac{1}{n-1} \sum\limits_{i=1}^n v_i v_i^\top \in \mathbb{R}^{m \times m}$.
			\State Compute the SVD of $K = U\Sigma U^\top$.
		\end{algorithmic}
	\end{algorithm}
	
	The first $r$ columns of $U$ (denoted by $\tilde{U} \in \mathbb{R}^{m \times r}$) span the $r$-dimensional principal subspace. We can compute the principal component (PC) scores as $Y = V\tilde{U}$, where $V^\top = [v_1~ v_2~ \ldots~ v_n] \in \mathbb{R}^{m \times n}$. The PC scores in $Y_i$ act as Euclidean coordinates corresponding to the density $f_i$, and can be used for downstream modelling. In our case, for each imaging sequence $m$ and each tumor layer $t$, we perform PCA with the sample of PDFs $f^{(t,m)}_1,\ldots,f^{(t,m)}_n$ to obtain the PC scores $Y^{(t,m)}_1,\ldots,Y^{(t,m)}_n$.
	
	\section{Estimation Details}\label{app: details_est}
	\subsection{Full Posterior and Log Posterior}\label{app: full}
	The posterior distribution, $\pi(\beta_{kj}^{(t,m)}, \zeta_k^{(t,m)}, \blambda^{(t,m)}, {\nu^{-2}_{kj}}^{(t,m)}, {\theta^{-2}}^{(t,m)}, \Delta^{(t)} | \y^{(t)}, X)$, corresponding to the model in Equation (1), with the prior structure in Equation (2), is proportional to
	\allowdisplaybreaks{
		\begin{small}
			\begin{eqnarray*}
				&|\Delta^{(t)}|^{-\frac{n}{2}}\exp\Big( -\frac{1}{2}(\y^{(t)}-Z\bbeta^{(t)})^\top (\Theta^{(t)})^{-1} (\y^{(t)}-Z\bbeta^{(t)}) \Big) \times |\Delta^{(t)}|^{-\frac{\delta+p^{(t)}+1}{2}} \exp\Big(-\frac{1}{2}tr(\Psi(\Delta^{(t)})^{-1}\Big) \times \\
				& \times \prod\limits_{m\in M} \prod\limits_{j = 1}^{p^{(t,m)}} \prod\limits_{k=1}^g \Big[ ({\eta^2_{kj}}^{(t,m)})^{-1/2} \exp\Big( -\frac{(\beta_{kj}^{(t,m)})^2}{2{\eta^2_{kj}}^{(t,m)}} \Big) \Big] \times \prod\limits_{m\in M} \prod\limits_{k=1}^g  (1-\Phi(\lambda_k^{(t,m)}))^{(1-\zeta_k^{(t,m)})}\Phi(\lambda_k^{(t,m)})^{\zeta_k^{(t,m)}} \times \\
				& \times \prod\limits_{m\in M}  \exp\Big( -\frac{({\blambda^{(t,m)}-\bmu^{(t,m)}})^\top\Lambda^{-1}(\blambda^{(t,m)}-\bmu^{(t,m)})}{2}  \Big) \times \prod\limits_{m\in M} \prod\limits_{j = 1}^{p^{(t,m)}} \prod\limits_{k=1}^g ({\nu^{-2}_{kj}}^{(t,m)})^{a_1 - 1}\exp( -a_2{\nu^{-2}_{kj}}^{(t,m)} ). 
			\end{eqnarray*}
		\end{small}
	}
	
	\noindent The complete-data log posterior as described for the EM-based estimation is given by
	\allowdisplaybreaks{
		\begin{small}
			\begin{eqnarray*}
				&&\log\big( \pi(\beta_{kj}^{(t,m)}, \zeta_k^{(t,m)}, \blambda^{(t,m)}, {\nu^{-2}_{kj}}^{(t,m)}, \Delta^{(t)} | \y^{(t)}, X) \big) = c -\frac{1}{2}(\y^{(t)}-Z\bbeta^{(t)})^\top (\Theta^{(t)})^{-1} (\y^{(t)}-Z\bbeta^{(t)}) + \\
				&+& \sum\limits_{m\in M} \sum\limits_{j = 1}^{p^{(t,m)}} \sum\limits_{k=1}^g \Big[ -\frac{1}{2} \log({\eta^2_{kj}}^{(t,m)}) -\frac{(\beta_{kj}^{(t,m)})^2}{2{\eta^2_{kj}}^{(t,m)}} \Big] + \\
				&+& \sum\limits_{m\in M} \sum\limits_{k=1}^g  {(1-\zeta_k^{(t,m)})} \log(1-\Phi(\lambda_k^{(t,m)})) + {\zeta_k^{(t,m)}}\log \Phi(\lambda_k^{(t,m)}) - \\
				&-&\sum\limits_{m\in M} 
				\frac{({\blambda^{(t,m)}-\bmu^{(t,m)}})^\top\Lambda^{-1}(\blambda^{(t,m)}-\bmu^{(t,m)})}{2} + \sum\limits_{m\in M} \sum\limits_{j = 1}^{p^{(t,m)}} \sum\limits_{k=1}^g -(a_1 - 1) \log({\nu^{2}_{kj}}^{(t,m)}) -a_2{\nu^{-2}_{kj}}^{(t,m)}- \\
				&-&\frac{n+\delta+p^{(t)}+1}{2} \log|\Delta^{(t)}| -\frac{1}{2}tr(\Psi(\Delta^{(t)})^{-1}).
			\end{eqnarray*}
		\end{small}
	}
	
	\subsection{E-Step Computation}\label{subapp: e-step}
	We have two terms which involve $\zeta_k^{(t,m)}$ and hence the expectation  $E_{\bzeta^{(t)} | \bchi^{(t),(u)}}$. The first one is the term for the prior on $\beta_{kj}^{(t,m)}$, which is given by
	\begin{small}
		\begin{eqnarray*}
			&&-\sum\limits_{m\in M} \sum\limits_{k=1}^g \sum\limits_{j = 1}^{p^{(t,m)}} E_{\bzeta^{(t)} \big| \bchi^{(t),(u)}}\Big[ \frac{1}{2} \log({\eta^2_{kj}}^{(t,m)}) + \frac{(\beta_{kj}^{(t,m)})^2}{2{\eta^2_{kj}}^{(t,m)}} \Big] = \\
			&=& -\sum\limits_{m\in M} \sum\limits_{k=1}^g \sum\limits_{j = 1}^{p^{(t,m)}} E_{\bzeta^{(t)} \big| \bchi^{(t),(u)}}\Big[ \frac{1}{2} \log(\big[(1-\zeta_k^{(t,m)})v_0 + \zeta_k^{(t,m)}v_1\big])+\frac{1}{2}\log({\nu^2_{kj}}^{(t,m)}) + \frac{(\beta_{kj}^{(t,m)})^2}{2{\eta^2_{kj}}^{(t,m)}} \Big] = \\
			&=& c - \sum\limits_{m\in M} \sum\limits_{k=1}^g \sum\limits_{j = 1}^{p^{(t,m)}}  \frac{1}{2}\log({\nu^2_{kj}}^{(t,m)}) + \frac{(\beta_{kj}^{(t,m)})^2}{2{\nu^2_{kj}}^{(t,m)}}E_{\bzeta^{(t)} \big| \bchi^{(t),(u)}}\Big[\frac{1}{(1-\zeta_k^{(t,m)})v_0 + \zeta_k^{(t,m)}v_1} \Big].
		\end{eqnarray*}
	\end{small}
	\noindent Note that we do not need to compute $E_{\bzeta^{(t)} \big| \bchi^{(t),(u)}}\Big[\log(\big[(1-\zeta_k^{(t,m)})v_0 + \zeta_k^{(t,m)}v_1\big])\Big]$ as it is a scalar and does not involve any other parameters. The second term arises from the prior for $\zeta_k^{(t,m)}$ which is given by
	\begin{small}
		\begin{eqnarray*}
			&&\sum\limits_{m\in M} \sum\limits_{k=1}^g  E_{\bzeta^{(t)} \big| \bchi^{(t),(u)}}[(1-\zeta_k^{(t,m)})] \log(1-\Phi(\lambda_k^{(t,m)})) + E_{\bzeta^{(t)} \big| \bchi^{(t),(u)}}[\zeta_k^{(t,m)}]\log \Phi(\lambda_k^{(t,m)})=\\
			&=& \sum\limits_{m\in M} \sum\limits_{k=1}^g  \log(1-\Phi(\lambda_k^{(t,m)})) + E_{\bzeta^{(t)} \big| \bchi^{(t),(u)}}[\zeta_k^{(t,m)}]\log \frac{\Phi(\lambda_k^{(t,m)})}{1-\Phi(\lambda_k^{(t,m)})}.
		\end{eqnarray*}
	\end{small}
	
	\noindent Additionally, we have
	\begin{small}
		\begin{eqnarray}
			a_k^{(t,m)} & = & P(\zeta_k^{(t,m)}=1 \big| \lambda_k^{(t,m),(u)}) \prod\limits_{j=1}^{p^{(t,m)}} \pi(\beta_{kj}^{(t,m),(u)} \big| {\nu^{2}_{kj}}^{(t,m),(u)}, \zeta_k^{(t,m)}=1 ) \nonumber\\
			& = & \Phi(\lambda_k^{(t,m),(u)})\prod\limits_{j=1}^{p^{(t,m)}}\phi(\beta_{kj}^{(t,m),(u)} \big| 0,v_1{\nu^{2}_{kj}}^{(t,m),(u)}), \text{ and } \label{eq: a}\\
			b_k^{(t,m)} & = & P(\zeta_k^{(t,m)}=0 \big| \lambda_k^{(t,m),(u)}) \prod\limits_{j=1}^{p^{(t,m)}}\pi(\beta_{kj}^{(t,m),(u)} \big| {\nu^{2}_{kj}}^{(t,m),(u)}, \zeta_k^{(t,m)}=0 ) \nonumber\\
			& = & \big( 1-\Phi(\lambda_k^{(t,m),(u)}) \big) \prod\limits_{j=1}^{p^{(t,m)}}\phi(\beta_{kj}^{(t,m),(u)} \big| 0,v_0{\nu^{2}_{kj}}^{(t,m),(u)}). \label{eq: b}
		\end{eqnarray}
	\end{small}
	\noindent Here, $\phi(.|\theta,\sigma^2)$ denotes the PDF of a normal distribution with mean $\theta$ and variance $\sigma^2$, and $\Phi(.)$ denotes the cumulative distribution function of a standard normal distribution.
	
	\subsection{$Q$-Function}\label{subapp: q_fn}
	The $Q$-function is separable in the parameters as $Q_1$ and $Q_2$ are given by
	\begin{small}
		\begin{eqnarray*}
			&&Q_1(\beta_{kj}^{(t,m)}, {\nu^{-2}_{kj}}^{(t,m)}, \Delta^{(t)} | \beta_{kj}^{(t,m),(u)}, {\nu^{-2}_{kj}}^{(t,m),(u)}, \Delta^{(t),(u)}) = \\	&=&-\frac{1}{2}(\y^{(t)}-Z\bbeta^{(t)})^\top (\Theta^{(t)})^{-1} (\y^{(t)}-Z\bbeta^{(t)})- \\
			&-& \sum\limits_{m\in M} \sum\limits_{k=1}^g \sum\limits_{j = 1}^{p^{(t,m)}}  \frac{1}{2}\log({\nu^2_{kj}}^{(t,m)}) + \frac{(\beta_{kj}^{(t,m)})^2}{2{\nu^2_{kj}}^{(t,m)}}d_k^{(t,m)} +\\
			&+&\sum\limits_{m\in M} \sum\limits_{j = 1}^{p^{(t,m)}} \sum\limits_{k=1}^g -(a_1 - 1) \log({\nu^{2}_{kj}}^{(t,m)}) -a_2{\nu^{-2}_{kj}}^{(t,m)}- \\
			&-&\frac{n+\delta+p^{(t)}+1}{2} \log|\Delta^{(t)}| -\frac{1}{2}tr(\Psi(\Delta^{(t)})^{-1}),\text{ and}
		\end{eqnarray*}
	\end{small}
	\begin{small}
		\begin{eqnarray*}
			&&Q_2(\blambda^{(t,m)} | \blambda^{(t,m),(u)}) = 	\sum\limits_{m\in M} \sum\limits_{k=1}^g  \log(1-\Phi(\lambda_k^{(t,m)})) + E_{\bzeta^{(t)} \big| \bchi^{(t),(u)}}[\zeta_k^{(t,m)}]\log \frac{\Phi(\lambda_k^{(t,m)})}{1-\Phi(\lambda_k^{(t,m)})}- \\
			& -&\sum\limits_{m\in M} %-\frac{g}{2}\log({\theta^{2}}^{(t,m)})  
			\frac{({\blambda^{(t,m)}-\bmu^{(t,m)}})^\top\Lambda^{-1}(\blambda^{(t,m)}-\bmu^{(t,m)})}{2}.
		\end{eqnarray*}
	\end{small}
	
	\subsection{M-Step Computation}\label{subapp: m-step}
	\paragraph{Maximizing $Q_1$ w.r.t. $\beta_{kj}^{(t,m)}$.} This involves maximizing
	\begin{small}
		\begin{eqnarray}
			&&\max\limits_{\bbeta^{(t)}} -\frac{1}{2}(\y^{(t)}-Z\bbeta^{(t)})^\top (\Theta^{(t)})^{-1} (\y^{(t)}-Z\bbeta^{(t)}) - \sum\limits_{m\in M} \sum\limits_{k=1}^g \sum\limits_{j = 1}^{p^{(t,m)}} \frac{(\beta_{kj}^{(t,m)})^2}{2{\nu^2_{kj}}^{(t,m)}}d_k^{(t,m)}\nonumber =\\
			&=& \max\limits_{\bbeta^{(t)}} -\frac{1}{2}(\y^{(t)}-Z\bbeta^{(t)})^\top (\Theta^{(t)})^{-1} (\y^{(t)}-Z\bbeta^{(t)}) - \frac{1}{2}{\bbeta^{(t)}}^\top (\bGamma^{(t)})^{-1}\bbeta^{(t)}, \label{eq: max_beta}
		\end{eqnarray}
	\end{small}
	where $\bGamma^{(t)} = diag(\nu_{kj}^{(t,m)}/d_k^{(t,m)})$.
	
	\paragraph{Maximizing $Q_1$ w.r.t. ${\nu^{-2}_{kj}}^{(t,m)}$.} This involves maximizing
	\begin{small}
		\begin{equation*}
			\max\limits_{{\nu^{-2}_{kj}}^{(t,m)}} - \sum\limits_{m\in M} \sum\limits_{k=1}^g \sum\limits_{j = 1}^{p^{(t,m)}} \bigg[ (a_1-\frac{1}{2})\log({\nu^2_{kj}}^{(t,m)}) + (a_2 + \frac{(\beta_{kj}^{(t,m)})^2d_k^{(t,m)}}{2}){\nu^{-2}_{kj}}^{(t,m)} \bigg],
		\end{equation*}
	\end{small}
	which is the same as maximizing
	\begin{small}
		\begin{equation}
			\max\limits_{{\nu^{-2}_{kj}}^{(t,m)}} \prod\limits_{m\in M} \prod\limits_{k=1}^g \prod\limits_{j = 1}^{p^{(t,m)}}  ({\nu^{-2}_{kj}}^{(t,m)})^{(a_1-\frac{1}{2})}\exp\bigg( -\Big(a_2 + \frac{(\beta_{kj}^{(t,m)})^2d_k^{(t,m)}}{2}\Big){\nu^{-2}_{kj}}^{(t,m)}\bigg). \label{eq: max_nu}
		\end{equation}
	\end{small}
	
	\paragraph{Maximizing $Q_1$ w.r.t. $\Delta^{(t)}$.} This involves maximizing
	\begin{small}
		\begin{eqnarray*}
			&&\max\limits_{(\Delta^{(t)})^{-1}} -\frac{1}{2}(\y^{(t)}-Z\bbeta^{(t)})^\top (\Theta^{(t)})^{-1} (\y^{(t)}-Z\bbeta^{(t)}) -\frac{n+\delta+p^{(t)}+1}{2} \log|\Delta^{(t)}| -\frac{1}{2}tr(\Psi(\Delta^{(t)})^{-1}) \\
			&=& \max\limits_{(\Delta^{(t)})^{-1}} -\frac{n+\delta+p^{(t)}+1}{2} \log|\Delta^{(t)}| -\frac{1}{2}tr(\big[\Psi + (Y^{(t)} - XB^{(t)})^\top(Y^{(t)} - XB^{(t)})\big](\Delta^{(t)})^{-1}),
		\end{eqnarray*}
	\end{small}
	which is the same as maximizing
	\begin{small}
		\begin{equation}
			\max\limits_{(\Delta^{(t)})^{-1}} |(\Delta^{(t)})^{-1}|^{\frac{n+\delta+p^{(t)}+1}{2}} \exp\bigg( -\frac{1}{2}tr\big(\big[\Psi + (Y^{(t)} - XB^{(t)})^\top(Y^{(t)} - XB^{(t)})\big](\Delta^{(t)})^{-1}\big) \bigg).
			\label{eq: max_Delta}
		\end{equation}
	\end{small}
	Equation (\ref{eq: max_Delta}) is the kernel of a Wishart distribution for $(\Delta^{(t)})^{-1}$; hence, its maximizer is the mode of the corresponding Wishart distribution.
	
	\paragraph{Maximizing $Q_2$ w.r.t. ${\blambda}^{(t,m)}$.} The terms in $Q_2$ involving ${\blambda}^{(t,m)}$ are given by
	\begin{small}
		\begin{eqnarray*}
			&&\max\limits_{{\blambda}^{(t,m)}} \sum\limits_{m\in M} \sum\limits_{k=1}^g (1-w_k^{(t,m)}) \log(1-\Phi(\lambda_k^{(t,m)})) + w_k^{(t,m)}\log \Phi(\lambda_k^{(t,m)}) \\ %\nonumber \\
			&+&\sum\limits_{m\in M} -\frac{({\blambda^{(t,m)}-\bmu^{(t,m)}})^\top\Lambda^{-1}(\blambda^{(t,m)}-\bmu^{(t,m)})}{2}. %\label{eq: max_lambda}
		\end{eqnarray*}
	\end{small}
	\noindent These terms are separable in $m$ and thus let us consider the maximization problem for each $m$, which can be termed as a minimization problem given by
	\begin{small}
		\begin{eqnarray}
			&&\min\limits_{{\blambda}^{(t,m)}} F({\blambda}^{(t,m)}) = \min\limits_{{\blambda}^{(t,m)}} -\sum\limits_{k=1}^g (1-w_k^{(t,m)}) \log(1-\Phi(\lambda_k^{(t,m)})) + w_k^{(t,m)}\log \Phi(\lambda_k^{(t,m)}) \nonumber \\
			&+& \frac{({\blambda^{(t,m)}-\bmu^{(t,m)}})^\top\Lambda^{-1}(\blambda^{(t,m)}-\bmu^{(t,m)})}{2}. \label{eq: max_lambda_2}
		\end{eqnarray}
	\end{small}
	
	\subsection{Deterministic Annealing to Mitigate Multi-modality}\label{subapp: deterministic}
	In our estimation approach, we consider the deterministic annealing variant of the EM algorithm \citep{ueda1998deterministic}, referred to as the DAEM algorithm, which improves the chance of finding the global maximum. Using the DAEM algorithm, we aim to maximize a tempered version of the complete-data log posterior, which is the same as computing the E-step as
	\begin{equation}
		\tilde{Q}(\bchi^{(t)} | \bchi^{(t),(u)}) = \int \log\big( \pi(\bchi^{(t)}, \zeta_k^{(t,m)} | \y^{(t)}) \big) \frac{\pi(\bchi^{(t),(u)}, \zeta_k^{(t,m)} | \y^{(t)})^q}{\int \pi(\bchi^{(t),(u)}, \zeta_k^{(t,m)} | \y^{(t)})^q d\bzeta^{(t)}} d\bzeta^{(t)},
		\label{eq: tempered}
	\end{equation}
	where $0 < q \leq 1$ and $1/q$ corresponds to a temperature parameter which determines the degree of separation between the multiple modes of $\tilde{Q}(\bchi^{(t)} | \bchi^{(t),(u)})$. Large values of temperature smooth the local modes of $\tilde{Q}(\bchi^{(t)} | \bchi^{(t),(u)})$; the modes begin to reappear as the temperature decreases, i.e., as $q$ gets close to $1$. This can be incorporated in the iterative process by increasing the value of $q$ in each iteration of the DAEM algorithm (bounded above by $1$), which decreases the sensitivity to initialization.  For our analyses, we start with $q=0.01$ and increase it by 10\% with each iteration until $q=1$. This change in the E-step does not affect the M-step. The computation of the E-step in Equation (\ref{eq: tempered}) is straightforward: the two terms requiring evaluation of expectation remain the same as described in Section 4.1.1. However, their computations change as follows:
	\begin{eqnarray}
		\tilde{w}_k^{(t,m)} & = & E_{\bzeta^{(t)} \big| \bchi^{(t),(u)}}[\zeta_k^{(t,m)}] = \frac{{a_k^{(t,m)}}^q}{{a_k^{(t,m)}}^q + {b_k^{(t,m)}}^q} \label{eq: expectation_1a}\text{, and} \\ \tilde{d}_k^{(t,m)} & = & E_{\bzeta^{(t)} \big| \bchi^{(t),(u)}}\Big[\frac{1}{(1-\zeta_k^{(t,m)})v_0 + \zeta_k^{(t,m)}v_1} \Big] = \frac{1-\tilde{w}_k^{(t,m)}}{v_0} + \frac{\tilde{w}_k^{(t,m)}}{v_1}, \nonumber
		%\label{eq: expectation_1b}
	\end{eqnarray} 
	where $a_k^{(t,m)}$ and $b_k^{(t,m)}$ are given in Equations (\ref{eq: a}) and (\ref{eq: b}), respectively. Other details about the limiting behavior w.r.t. the temperature, and the superiority of DAEM over the regular EM approach in the context of variable selection, are discussed in \cite{rovckova2014emvs}.
	
	\section{Details of Simulation Study}\label{app: details_sim}
	\subsection{Case 1}
	Algorithm \ref{algo: simcase1} describes the procedure to generate the simulated data under Case 1. To generate data from the model, we make the following choices for the hyperparameters: $\delta = p^{(t)} = 12, \Psi = \sigma^2 I_g, a_1 = 5, a_2 = 5, v_0 = 0.01, v_1 = 1$ and $\alpha = 0.8$. We use the same choices for all cases of $\sigma^2 = 1,10,20,30$.
	
\begin{figure}[!t]
		\centering
		\resizebox{\textwidth}{!}{
			\begin{tabular}{|c|c|c|c|}
				\hline
				\includegraphics[page=1]{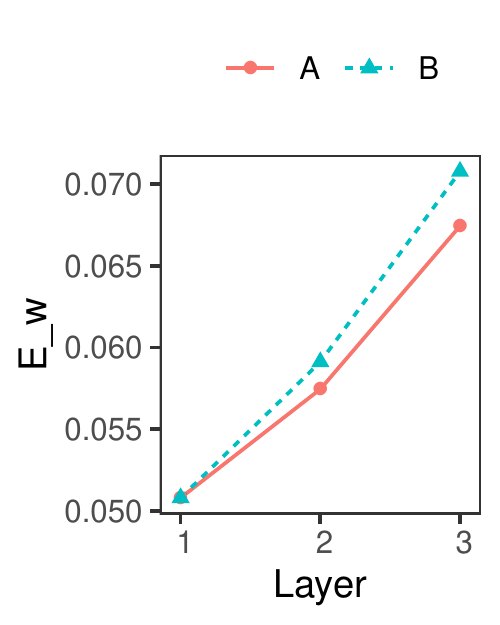} &
				\includegraphics[page=2]{plots/E_w_case1_Uncor.pdf} &
				\includegraphics[page=3]{plots/E_w_case1_Uncor.pdf} &
				\includegraphics[page=4]{plots/E_w_case1_Uncor.pdf} \\
				(a) $\sigma^2 = 1$ & 
				(b) $\sigma^2 = 10$ &
				(c) $\sigma^2 = 20$ &
				(d) $\sigma^2 = 30$ \\ \hline
			\end{tabular}
		}
		\caption{Average values of $E_w$ in each of the three layers for Case 1: $\Sigma^{(x)} = I_g$. Model (A): $\bmu^{(t,m)} \neq \0; \Lambda = cor(X) \approx I_g$. Model (B): $\bmu^{(t,m)} = \0; \Lambda = cor(X) \approx I_g$.}
		\label{fig: sim_case_1_Uncor_E_w}
	\end{figure}

\begin{figure}[!t]
		\centering
		\resizebox{\textwidth}{!}{
			\begin{tabular}{|c|c|c|c|}
				\hline
				\includegraphics[page=1]{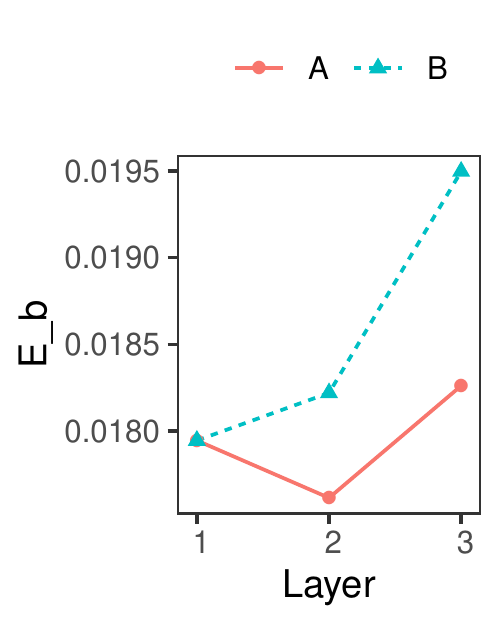} &
				\includegraphics[page=2]{plots/E_b_case1_Uncor.pdf} &
				\includegraphics[page=3]{plots/E_b_case1_Uncor.pdf} &
				\includegraphics[page=4]{plots/E_b_case1_Uncor.pdf} \\
				(a) $\sigma^2 = 1$ & 
				(b) $\sigma^2 = 10$ &
				(c) $\sigma^2 = 20$ &
				(d) $\sigma^2 = 30$ \\ \hline
			\end{tabular}
		}
		\caption{Average values of $E_\beta$ in each of the three layers for Case 1: $\Sigma^{(x)} = I_g$. Model (A): $\bmu^{(t,m)} \neq \0; \Lambda = cor(X) \approx I_g$. Model (B): $\bmu^{(t,m)} = \0; \Lambda = cor(X) \approx I_g$.}
		\label{fig: sim_case_1_Uncor_E_b}
	\end{figure}

\begin{figure}[!t]
		\centering
		\resizebox{\textwidth}{!}{
			\begin{tabular}{|c|c|c|c|}
				\hline
				\includegraphics[page=1]{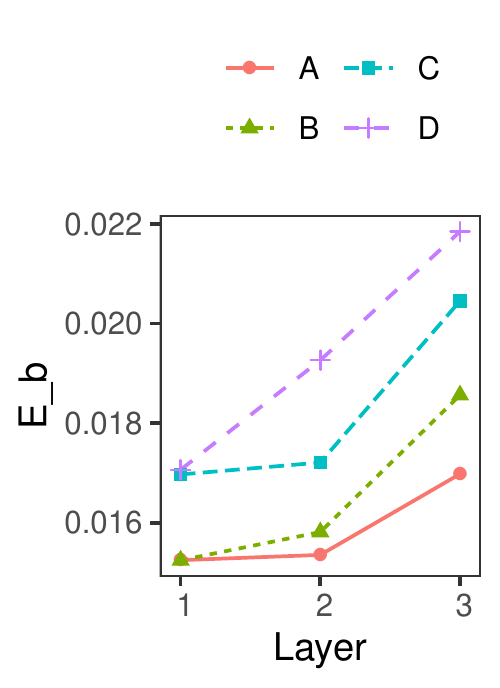} &
				\includegraphics[page=2]{plots/E_b_case1.pdf} &
				\includegraphics[page=3]{plots/E_b_case1.pdf} &
				\includegraphics[page=4]{plots/E_b_case1.pdf} \\
				(a) $\sigma^2 = 1$ & 
				(b) $\sigma^2 = 10$ &
				(c) $\sigma^2 = 20$ &
				(d) $\sigma^2 = 30$ \\ \hline
			\end{tabular}
		}
		\caption{Average values of $E_\beta$ in each of the three layers for Case 1: $\Sigma^{(x)}$ is a $(2 \times 2)$-block matrix. Model (A): $\bmu^{(t,m)} \neq \0; \Lambda = cor(X)$. Model (B): $\bmu^{(t,m)} = \0; \Lambda = cor(X)$. Model (C): $\bmu^{(t,m)} \neq \0; \Lambda = I_g$. Model (D): $\bmu^{(t,m)} = \0; \Lambda = I_g$.}
		\label{fig: sim_case_1_cor_E_b}
	\end{figure}

\begin{figure}[!t]
		\centering
		\resizebox{\textwidth}{!}{
			\begin{tabular}{|c|c|c|c|}
				\hline
				\includegraphics[page=1]{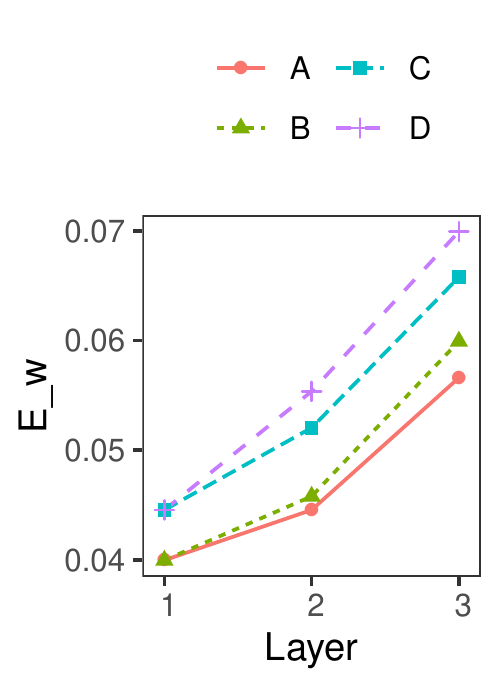} &
				\includegraphics[page=2]{plots/E_w_case1.pdf} &
				\includegraphics[page=3]{plots/E_w_case1.pdf} &
				\includegraphics[page=4]{plots/E_w_case1.pdf} \\
				(a) $\sigma^2 = 1$ & 
				(b) $\sigma^2 = 10$ &
				(c) $\sigma^2 = 20$ &
				(d) $\sigma^2 = 30$ \\ \hline
			\end{tabular}
		}
		\caption{Average values of $E_w$ in each of the three layers for Case 1: $\Sigma^{(x)}$ is a $(2 \times 2)$-block matrix. Model (A): $\bmu^{(t,m)} \neq \0; \Lambda = cor(X)$. Model (B): $\bmu^{(t,m)} = \0; \Lambda = cor(X)$. Model (C): $\bmu^{(t,m)} \neq \0; \Lambda = I_g$. Model (D): $\bmu^{(t,m)} = \0; \Lambda = I_g$.}
		\label{fig: sim_case_1_cor_E_w}
	\end{figure}

	For estimation, we choose $a_1 = 4, a_2 = 5, \alpha = 0.5, \Psi = I_g, \delta = p^{(t)}$ and set $v_1$ as the smallest power of 10 greater than the maximum magnitude of the entries in the ordinary least square estimate of $B^{(t)}$. For model selection, we first perform estimation on a grid of values for $v_0$. From the estimation based on each value of $v_0$, we compute the Bayesian information criterion (BIC) and choose the best model based on the lowest BIC value. In Table \ref{tab: sim_case_1}, we show the average true positive rates (TPR), $E_w$ and $E_\beta$ after model selection based on 30 replications. We include results for both of the choices for $\Sigma^{(x)}$ as described in the main manuscript. Figures \ref{fig: sim_case_1_Uncor_E_w} and \ref{fig: sim_case_1_cor_E_w} show the plots of $E_w$ across layers for various choices of $\sigma^2$ when $\Sigma^{(x)} = I_g$ and the $(2\times 2)$-block matrix structure, respectively. Moreover, we also include plots for $E_\beta$ in Figures \ref{fig: sim_case_1_Uncor_E_b} and \ref{fig: sim_case_1_cor_E_b} for the cases when $\Sigma^{(x)} = I_g$ and the $(2\times 2)$-block matrix structure, respectively. We considered the grid of values for $v_0$ as $\{ 0.001 + (v-1)0.001 | v=1,\ldots,10 \}$.
	
	\begin{algorithm}[!t]
		\caption{Simulation from the model}\label{algo: simcase1}
		\begin{algorithmic}[1]
			%\Procedure{MyProcedure}{}
			\State Specify the choices for $n=100, p^{(t)}=12,p^{(t,m)}=3, m=4, \tau=3, g=20, \hat{\lambda}_k^{(t,m)} = 0 ~\forall k,m,t$.
			\State Specify the hyperparameters $\delta, \Psi, a_1, a_2, v_0, v_1, \alpha$ and $\sigma^2 \in \{1,10,20,30\}$.
			\State Generate $X = [\x_1,\ldots,\x_n]^\top \in \mathbb{R}^{n \times g}$, where $\x_i \stackrel{iid}{\sim} N_g(\0,\Sigma^{(x)})$ for $i=1,\ldots,n$.
			\State $\Lambda \leftarrow \Sigma^{(x)}$.
			\For {$t=1,\ldots,\tau$} 
			\State Generate $\Delta^{(t)} \sim IW(\delta, \Psi)$.
			\State Generate ${\nu^{-2}_{kj}}^{(t,m)} \stackrel{iid}{\sim} Gamma(a_1, a_2)$ for all $j,k,t,m$.
			\If {$t=1$} 
			\State \Return $\bmu^{(t,m)}=\0$.
			\Else {}
			\State $\bmu^{(t,m)} = \alpha\hat{\blambda}^{(t-1,m)}_+ $, where $\hat{\lambda}^{(t-1,m)}_{k,+} = \max(\hat{\lambda}^{(t-1,m)}_k, 0)$ for $k=1,\ldots,g$ and $\alpha \in [0,1]$.
			\EndIf
			\State Generate $\blambda^{(t,m)} \stackrel{ind}{\sim} N(\bmu^{(t,m)}, \Lambda)$ for all $t,m$.
			\State Generate $\zeta_k^{(t,m)} \stackrel{ind}{\sim} Ber(\Phi(\lambda_k^{(t,m)}))$ for all $k,t,m$.
			\State Generate $\beta_{kj}^{(t,m)} \stackrel{ind}{\sim} N\big(0, \big[(1-\zeta_k^{(t,m)})v_0 + \zeta_k^{(t,m)}v_1\big]{\nu^2_{kj}}^{(t,m)}\big)$ for all $k,j,t,m$.
			\State Generate $vec(Y^{(t)}) \sim N_{np^{(t)}}(vec(XB^{(t)}), I_n \otimes \Delta^{(t)})$ and appropriately reshape $Y^{(t)} \in \mathbb{R}^{n \times p^{(t)}}$.
			\EndFor
		\end{algorithmic}
	\end{algorithm}
	
	\begin{landscape}
		\begin{small}
			\begin{table}[!t]
				\centering
				\resizebox{1.3\textwidth}{0.45\textheight}{%
					\begin{tabular}{ccccccccccccccc}
						\hline
						&  &  &  & \multicolumn{3}{c}{Layer 1} &  & \multicolumn{3}{c}{Layer 2} &  & \multicolumn{3}{c}{Layer 3} \\ \hline
						\multicolumn{1}{|c|}{$\Sigma^{(x)}$} & \multicolumn{1}{c|}{$\bmu^{(t,m)}, \Lambda$} & \multicolumn{1}{c|}{$\sigma^2$} & \multicolumn{1}{c|}{} & \multicolumn{1}{c|}{TPR} & \multicolumn{1}{c|}{$E_w$} & \multicolumn{1}{c|}{$E_\beta$} & \multicolumn{1}{c|}{} & \multicolumn{1}{c|}{TPR} & \multicolumn{1}{c|}{$E_w$} & \multicolumn{1}{c|}{$E_\beta$} & \multicolumn{1}{c|}{} & \multicolumn{1}{c|}{TPR} & \multicolumn{1}{c|}{$E_w$} & \multicolumn{1}{c|}{$E_\beta$} \\ \cline{1-3} \cline{5-7} \cline{9-11} \cline{13-15} 
						&  &  &  &  &  &  &  &  &  &  &  &  &  &  \\ \cline{1-3} \cline{5-7} \cline{9-11} \cline{13-15} 
						\multicolumn{1}{|c|}{\multirow{9}{*}{\begin{tabular}[c]{@{}c@{}}\rotatebox{90}{Identity Matrix}\end{tabular}}} & \multicolumn{1}{c|}{\multirow{4}{*}{\begin{tabular}[c]{@{}c@{}}$\bmu^{(t,m)} \neq \0$\\ $\Lambda = cor(X)$\end{tabular}}} & \multicolumn{1}{c|}{1} & \multicolumn{1}{c|}{} & \multicolumn{1}{c|}{0.901 (0.048)} & \multicolumn{1}{c|}{0.051 (0.024)} & \multicolumn{1}{c|}{0.018 (0.006)} & \multicolumn{1}{c|}{} & \multicolumn{1}{c|}{0.904 (0.039)} & \multicolumn{1}{c|}{0.057 (0.024)} & \multicolumn{1}{c|}{0.018 (0.006)} & \multicolumn{1}{c|}{} & \multicolumn{1}{c|}{0.901 (0.038)} & \multicolumn{1}{c|}{0.067 (0.027)} & \multicolumn{1}{c|}{0.018 (0.008)} \\ \cline{3-3} \cline{5-7} \cline{9-11} \cline{13-15} 
						\multicolumn{1}{|c|}{} & \multicolumn{1}{c|}{} & \multicolumn{1}{c|}{10} & \multicolumn{1}{c|}{} & \multicolumn{1}{c|}{0.804 (0.074)} & \multicolumn{1}{c|}{0.101 (0.04)} & \multicolumn{1}{c|}{0.036 (0.012)} & \multicolumn{1}{c|}{} & \multicolumn{1}{c|}{0.787 (0.075)} & \multicolumn{1}{c|}{0.125 (0.047)} & \multicolumn{1}{c|}{0.046 (0.016)} & \multicolumn{1}{c|}{} & \multicolumn{1}{c|}{0.818 (0.067)} & \multicolumn{1}{c|}{0.122 (0.047)} & \multicolumn{1}{c|}{0.046 (0.019)} \\ \cline{3-3} \cline{5-7} \cline{9-11} \cline{13-15} 
						\multicolumn{1}{|c|}{} & \multicolumn{1}{c|}{} & \multicolumn{1}{c|}{20} & \multicolumn{1}{c|}{} & \multicolumn{1}{c|}{0.746 (0.084)} & \multicolumn{1}{c|}{0.126 (0.045)} & \multicolumn{1}{c|}{0.055 (0.015)} & \multicolumn{1}{c|}{} & \multicolumn{1}{c|}{0.732 (0.065)} & \multicolumn{1}{c|}{0.157 (0.044)} & \multicolumn{1}{c|}{0.065 (0.02)} & \multicolumn{1}{c|}{} & \multicolumn{1}{c|}{0.711 (0.075)} & \multicolumn{1}{c|}{0.192 (0.056)} & \multicolumn{1}{c|}{0.079 (0.025)} \\ \cline{3-3} \cline{5-7} \cline{9-11} \cline{13-15} 
						\multicolumn{1}{|c|}{} & \multicolumn{1}{c|}{} & \multicolumn{1}{c|}{30} & \multicolumn{1}{c|}{} & \multicolumn{1}{c|}{0.669 (0.089)} & \multicolumn{1}{c|}{0.168 (0.056)} & \multicolumn{1}{c|}{0.077 (0.024)} & \multicolumn{1}{c|}{} & \multicolumn{1}{c|}{0.735 (0.057)} & \multicolumn{1}{c|}{0.154 (0.036)} & \multicolumn{1}{c|}{0.075 (0.017)} & \multicolumn{1}{c|}{} & \multicolumn{1}{c|}{0.711 (0.07)} & \multicolumn{1}{c|}{0.186 (0.045)} & \multicolumn{1}{c|}{0.089 (0.021)} \\ \cline{2-3} \cline{5-7} \cline{9-11} \cline{13-15} 
						\multicolumn{1}{|c|}{} &  &  &  &  &  &  &  &  &  &  &  &  &  &  \\ \cline{2-3} \cline{5-7} \cline{9-11} \cline{13-15} 
						\multicolumn{1}{|c|}{} & \multicolumn{1}{c|}{\multirow{4}{*}{\begin{tabular}[c]{@{}c@{}}$\bmu^{(t,m)} = \0$\\ $\Lambda = cor(X)$\\ $\approx I_g$\end{tabular}}} & \multicolumn{1}{c|}{1} & \multicolumn{1}{c|}{} & \multicolumn{1}{c|}{0.901 (0.048)} & \multicolumn{1}{c|}{0.051 (0.024)} & \multicolumn{1}{c|}{0.018 (0.006)} & \multicolumn{1}{c|}{} & \multicolumn{1}{c|}{0.901 (0.04)} & \multicolumn{1}{c|}{0.059 (0.025)} & \multicolumn{1}{c|}{0.018 (0.006)} & \multicolumn{1}{c|}{} & \multicolumn{1}{c|}{0.895 (0.038)} & \multicolumn{1}{c|}{0.071 (0.027)} & \multicolumn{1}{c|}{0.019 (0.008)} \\ \cline{3-3} \cline{5-7} \cline{9-11} \cline{13-15} 
						\multicolumn{1}{|c|}{} & \multicolumn{1}{c|}{} & \multicolumn{1}{c|}{10} & \multicolumn{1}{c|}{} & \multicolumn{1}{c|}{0.804 (0.074)} & \multicolumn{1}{c|}{0.101 (0.04)} & \multicolumn{1}{c|}{0.036 (0.012)} & \multicolumn{1}{c|}{} & \multicolumn{1}{c|}{0.782 (0.076)} & \multicolumn{1}{c|}{0.128 (0.047)} & \multicolumn{1}{c|}{0.047 (0.017)} & \multicolumn{1}{c|}{} & \multicolumn{1}{c|}{0.808 (0.067)} & \multicolumn{1}{c|}{0.13 (0.048)} & \multicolumn{1}{c|}{0.049 (0.019)} \\ \cline{3-3} \cline{5-7} \cline{9-11} \cline{13-15} 
						\multicolumn{1}{|c|}{} & \multicolumn{1}{c|}{} & \multicolumn{1}{c|}{20} & \multicolumn{1}{c|}{} & \multicolumn{1}{c|}{0.746 (0.084)} & \multicolumn{1}{c|}{0.126 (0.045)} & \multicolumn{1}{c|}{0.055 (0.015)} & \multicolumn{1}{c|}{} & \multicolumn{1}{c|}{0.728 (0.064)} & \multicolumn{1}{c|}{0.16 (0.045)} & \multicolumn{1}{c|}{0.067 (0.02)} & \multicolumn{1}{c|}{} & \multicolumn{1}{c|}{0.707 (0.076)} & \multicolumn{1}{c|}{0.195 (0.056)} & \multicolumn{1}{c|}{0.081 (0.026)} \\ \cline{3-3} \cline{5-7} \cline{9-11} \cline{13-15} 
						\multicolumn{1}{|c|}{} & \multicolumn{1}{c|}{} & \multicolumn{1}{c|}{30} & \multicolumn{1}{c|}{} & \multicolumn{1}{c|}{0.669 (0.089)} & \multicolumn{1}{c|}{0.168 (0.056)} & \multicolumn{1}{c|}{0.077 (0.024)} & \multicolumn{1}{c|}{} & \multicolumn{1}{c|}{0.733 (0.056)} & \multicolumn{1}{c|}{0.155 (0.035)} & \multicolumn{1}{c|}{0.076 (0.016)} & \multicolumn{1}{c|}{} & \multicolumn{1}{c|}{0.71 (0.071)} & \multicolumn{1}{c|}{0.187 (0.045)} & \multicolumn{1}{c|}{0.09 (0.021)} \\ \cline{1-3} \cline{5-7} \cline{9-11} \cline{13-15} 
						&  &  &  &  &  &  &  &  &  &  &  &  &  &  \\ \cline{1-3} \cline{5-7} \cline{9-11} \cline{13-15} 
						\multicolumn{1}{|c|}{\multirow{19}{*}{\begin{tabular}[c]{@{}c@{}}\rotatebox{90}{Block Matrix}\end{tabular}}} & \multicolumn{1}{c|}{\multirow{4}{*}{\begin{tabular}[c]{@{}c@{}}$\bmu^{(t,m)} \neq \0$\\ $\Lambda = cor(X)$\end{tabular}}} & \multicolumn{1}{c|}{1} & \multicolumn{1}{c|}{} & \multicolumn{1}{c|}{0.918 (0.048)} & \multicolumn{1}{c|}{0.04 (0.024)} & \multicolumn{1}{c|}{0.015 (0.007)} & \multicolumn{1}{c|}{} & \multicolumn{1}{c|}{0.922 (0.036)} & \multicolumn{1}{c|}{0.045 (0.021)} & \multicolumn{1}{c|}{0.015 (0.006)} & \multicolumn{1}{c|}{} & \multicolumn{1}{c|}{0.915 (0.043)} & \multicolumn{1}{c|}{0.057 (0.028)} & \multicolumn{1}{c|}{0.017 (0.008)} \\ \cline{3-3} \cline{5-7} \cline{9-11} \cline{13-15} 
						\multicolumn{1}{|c|}{} & \multicolumn{1}{c|}{} & \multicolumn{1}{c|}{10} & \multicolumn{1}{c|}{} & \multicolumn{1}{c|}{0.819 (0.062)} & \multicolumn{1}{c|}{0.091 (0.036)} & \multicolumn{1}{c|}{0.035 (0.013)} & \multicolumn{1}{c|}{} & \multicolumn{1}{c|}{0.81 (0.066)} & \multicolumn{1}{c|}{0.112 (0.041)} & \multicolumn{1}{c|}{0.041 (0.015)} & \multicolumn{1}{c|}{} & \multicolumn{1}{c|}{0.801 (0.071)} & \multicolumn{1}{c|}{0.13 (0.049)} & \multicolumn{1}{c|}{0.046 (0.015)} \\ \cline{3-3} \cline{5-7} \cline{9-11} \cline{13-15} 
						\multicolumn{1}{|c|}{} & \multicolumn{1}{c|}{} & \multicolumn{1}{c|}{20} & \multicolumn{1}{c|}{} & \multicolumn{1}{c|}{0.737 (0.08)} & \multicolumn{1}{c|}{0.132 (0.039)} & \multicolumn{1}{c|}{0.053 (0.016)} & \multicolumn{1}{c|}{} & \multicolumn{1}{c|}{0.734 (0.073)} & \multicolumn{1}{c|}{0.155 (0.042)} & \multicolumn{1}{c|}{0.064 (0.018)} & \multicolumn{1}{c|}{} & \multicolumn{1}{c|}{0.704 (0.066)} & \multicolumn{1}{c|}{0.191 (0.041)} & \multicolumn{1}{c|}{0.076 (0.018)} \\ \cline{3-3} \cline{5-7} \cline{9-11} \cline{13-15} 
						\multicolumn{1}{|c|}{} & \multicolumn{1}{c|}{} & \multicolumn{1}{c|}{30} & \multicolumn{1}{c|}{} & \multicolumn{1}{c|}{0.73 (0.065)} & \multicolumn{1}{c|}{0.134 (0.042)} & \multicolumn{1}{c|}{0.061 (0.017)} & \multicolumn{1}{c|}{} & \multicolumn{1}{c|}{0.699 (0.066)} & \multicolumn{1}{c|}{0.181 (0.044)} & \multicolumn{1}{c|}{0.083 (0.025)} & \multicolumn{1}{c|}{} & \multicolumn{1}{c|}{0.724 (0.06)} & \multicolumn{1}{c|}{0.18 (0.041)} & \multicolumn{1}{c|}{0.085 (0.023)} \\ \cline{2-3} \cline{5-7} \cline{9-11} \cline{13-15} 
						\multicolumn{1}{|c|}{} &  &  &  &  &  &  &  &  &  &  &  &  &  &  \\ \cline{2-3} \cline{5-7} \cline{9-11} \cline{13-15} 
						\multicolumn{1}{|c|}{} & \multicolumn{1}{c|}{\multirow{4}{*}{\begin{tabular}[c]{@{}c@{}}$\bmu^{(t,m)} = \0$\\ $\Lambda = cor(X)$\end{tabular}}} & \multicolumn{1}{c|}{1} & \multicolumn{1}{c|}{} & \multicolumn{1}{c|}{0.918 (0.048)} & \multicolumn{1}{c|}{0.04 (0.024)} & \multicolumn{1}{c|}{0.015 (0.007)} & \multicolumn{1}{c|}{} & \multicolumn{1}{c|}{0.92 (0.037)} & \multicolumn{1}{c|}{0.046 (0.021)} & \multicolumn{1}{c|}{0.016 (0.007)} & \multicolumn{1}{c|}{} & \multicolumn{1}{c|}{0.91 (0.044)} & \multicolumn{1}{c|}{0.06 (0.028)} & \multicolumn{1}{c|}{0.019 (0.009)} \\ \cline{3-3} \cline{5-7} \cline{9-11} \cline{13-15} 
						\multicolumn{1}{|c|}{} & \multicolumn{1}{c|}{} & \multicolumn{1}{c|}{10} & \multicolumn{1}{c|}{} & \multicolumn{1}{c|}{0.82 (0.06)} & \multicolumn{1}{c|}{0.09 (0.035)} & \multicolumn{1}{c|}{0.035 (0.012)} & \multicolumn{1}{c|}{} & \multicolumn{1}{c|}{0.806 (0.066)} & \multicolumn{1}{c|}{0.115 (0.043)} & \multicolumn{1}{c|}{0.042 (0.015)} & \multicolumn{1}{c|}{} & \multicolumn{1}{c|}{0.798 (0.072)} & \multicolumn{1}{c|}{0.132 (0.05)} & \multicolumn{1}{c|}{0.046 (0.015)} \\ \cline{3-3} \cline{5-7} \cline{9-11} \cline{13-15} 
						\multicolumn{1}{|c|}{} & \multicolumn{1}{c|}{} & \multicolumn{1}{c|}{20} & \multicolumn{1}{c|}{} & \multicolumn{1}{c|}{0.737 (0.08)} & \multicolumn{1}{c|}{0.132 (0.039)} & \multicolumn{1}{c|}{0.053 (0.016)} & \multicolumn{1}{c|}{} & \multicolumn{1}{c|}{0.732 (0.073)} & \multicolumn{1}{c|}{0.155 (0.042)} & \multicolumn{1}{c|}{0.064 (0.018)} & \multicolumn{1}{c|}{} & \multicolumn{1}{c|}{0.7 (0.066)} & \multicolumn{1}{c|}{0.193 (0.042)} & \multicolumn{1}{c|}{0.077 (0.018)} \\ \cline{3-3} \cline{5-7} \cline{9-11} \cline{13-15} 
						\multicolumn{1}{|c|}{} & \multicolumn{1}{c|}{} & \multicolumn{1}{c|}{30} & \multicolumn{1}{c|}{} & \multicolumn{1}{c|}{0.73 (0.065)} & \multicolumn{1}{c|}{0.134 (0.042)} & \multicolumn{1}{c|}{0.061 (0.017)} & \multicolumn{1}{c|}{} & \multicolumn{1}{c|}{0.698 (0.066)} & \multicolumn{1}{c|}{0.182 (0.044)} & \multicolumn{1}{c|}{0.083 (0.026)} & \multicolumn{1}{c|}{} & \multicolumn{1}{c|}{0.722 (0.061)} & \multicolumn{1}{c|}{0.181 (0.041)} & \multicolumn{1}{c|}{0.086 (0.023)} \\ \cline{2-3} \cline{5-7} \cline{9-11} \cline{13-15} 
						\multicolumn{1}{|c|}{} &  &  &  &  &  &  &  &  &  &  &  &  &  &  \\ \cline{2-3} \cline{5-7} \cline{9-11} \cline{13-15} 
						\multicolumn{1}{|c|}{} & \multicolumn{1}{c|}{\multirow{4}{*}{\begin{tabular}[c]{@{}c@{}}$\bmu^{(t,m)} \neq \0$\\ $\Lambda = I_g$\end{tabular}}} & \multicolumn{1}{c|}{1} & \multicolumn{1}{c|}{} & \multicolumn{1}{c|}{0.91 (0.047)} & \multicolumn{1}{c|}{0.045 (0.024)} & \multicolumn{1}{c|}{0.017 (0.007)} & \multicolumn{1}{c|}{} & \multicolumn{1}{c|}{0.911 (0.037)} & \multicolumn{1}{c|}{0.052 (0.023)} & \multicolumn{1}{c|}{0.017 (0.007)} & \multicolumn{1}{c|}{} & \multicolumn{1}{c|}{0.9 (0.041)} & \multicolumn{1}{c|}{0.066 (0.027)} & \multicolumn{1}{c|}{0.02 (0.008)} \\ \cline{3-3} \cline{5-7} \cline{9-11} \cline{13-15} 
						\multicolumn{1}{|c|}{} & \multicolumn{1}{c|}{} & \multicolumn{1}{c|}{10} & \multicolumn{1}{c|}{} & \multicolumn{1}{c|}{0.805 (0.052)} & \multicolumn{1}{c|}{0.098 (0.033)} & \multicolumn{1}{c|}{0.037 (0.011)} & \multicolumn{1}{c|}{} & \multicolumn{1}{c|}{0.801 (0.063)} & \multicolumn{1}{c|}{0.118 (0.043)} & \multicolumn{1}{c|}{0.042 (0.015)} & \multicolumn{1}{c|}{} & \multicolumn{1}{c|}{0.796 (0.072)} & \multicolumn{1}{c|}{0.134 (0.051)} & \multicolumn{1}{c|}{0.046 (0.016)} \\ \cline{3-3} \cline{5-7} \cline{9-11} \cline{13-15} 
						\multicolumn{1}{|c|}{} & \multicolumn{1}{c|}{} & \multicolumn{1}{c|}{20} & \multicolumn{1}{c|}{} & \multicolumn{1}{c|}{0.722 (0.077)} & \multicolumn{1}{c|}{0.139 (0.039)} & \multicolumn{1}{c|}{0.058 (0.018)} & \multicolumn{1}{c|}{} & \multicolumn{1}{c|}{0.73 (0.071)} & \multicolumn{1}{c|}{0.157 (0.042)} & \multicolumn{1}{c|}{0.064 (0.018)} & \multicolumn{1}{c|}{} & \multicolumn{1}{c|}{0.697 (0.068)} & \multicolumn{1}{c|}{0.195 (0.042)} & \multicolumn{1}{c|}{0.077 (0.019)} \\ \cline{3-3} \cline{5-7} \cline{9-11} \cline{13-15} 
						\multicolumn{1}{|c|}{} & \multicolumn{1}{c|}{} & \multicolumn{1}{c|}{30} & \multicolumn{1}{c|}{} & \multicolumn{1}{c|}{0.722 (0.067)} & \multicolumn{1}{c|}{0.137 (0.044)} & \multicolumn{1}{c|}{0.062 (0.018)} & \multicolumn{1}{c|}{} & \multicolumn{1}{c|}{0.692 (0.061)} & \multicolumn{1}{c|}{0.185 (0.042)} & \multicolumn{1}{c|}{0.085 (0.025)} & \multicolumn{1}{c|}{} & \multicolumn{1}{c|}{0.716 (0.058)} & \multicolumn{1}{c|}{0.184 (0.039)} & \multicolumn{1}{c|}{0.087 (0.022)} \\ \cline{2-3} \cline{5-7} \cline{9-11} \cline{13-15} 
						\multicolumn{1}{|c|}{} &  &  &  &  &  &  &  &  &  &  &  &  &  &  \\ \cline{2-3} \cline{5-7} \cline{9-11} \cline{13-15} 
						\multicolumn{1}{|c|}{} & \multicolumn{1}{c|}{\multirow{4}{*}{\begin{tabular}[c]{@{}c@{}}$\bmu^{(t,m)} = \0$\\ $\Lambda = I_g$\end{tabular}}} & \multicolumn{1}{c|}{1} & \multicolumn{1}{c|}{} & \multicolumn{1}{c|}{0.91 (0.045)} & \multicolumn{1}{c|}{0.045 (0.024)} & \multicolumn{1}{c|}{0.017 (0.007)} & \multicolumn{1}{c|}{} & \multicolumn{1}{c|}{0.904 (0.042)} & \multicolumn{1}{c|}{0.055 (0.024)} & \multicolumn{1}{c|}{0.019 (0.01)} & \multicolumn{1}{c|}{} & \multicolumn{1}{c|}{0.894 (0.04)} & \multicolumn{1}{c|}{0.07 (0.027)} & \multicolumn{1}{c|}{0.022 (0.008)} \\ \cline{3-3} \cline{5-7} \cline{9-11} \cline{13-15} 
						\multicolumn{1}{|c|}{} & \multicolumn{1}{c|}{} & \multicolumn{1}{c|}{10} & \multicolumn{1}{c|}{} & \multicolumn{1}{c|}{0.805 (0.051)} & \multicolumn{1}{c|}{0.098 (0.033)} & \multicolumn{1}{c|}{0.038 (0.011)} & \multicolumn{1}{c|}{} & \multicolumn{1}{c|}{0.798 (0.065)} & \multicolumn{1}{c|}{0.12 (0.044)} & \multicolumn{1}{c|}{0.044 (0.018)} & \multicolumn{1}{c|}{} & \multicolumn{1}{c|}{0.789 (0.074)} & \multicolumn{1}{c|}{0.138 (0.052)} & \multicolumn{1}{c|}{0.048 (0.017)} \\ \cline{3-3} \cline{5-7} \cline{9-11} \cline{13-15} 
						\multicolumn{1}{|c|}{} & \multicolumn{1}{c|}{} & \multicolumn{1}{c|}{20} & \multicolumn{1}{c|}{} & \multicolumn{1}{c|}{0.725 (0.077)} & \multicolumn{1}{c|}{0.137 (0.037)} & \multicolumn{1}{c|}{0.056 (0.016)} & \multicolumn{1}{c|}{} & \multicolumn{1}{c|}{0.725 (0.07)} & \multicolumn{1}{c|}{0.16 (0.042)} & \multicolumn{1}{c|}{0.065 (0.018)} & \multicolumn{1}{c|}{} & \multicolumn{1}{c|}{0.695 (0.067)} & \multicolumn{1}{c|}{0.197 (0.043)} & \multicolumn{1}{c|}{0.078 (0.019)} \\ \cline{3-3} \cline{5-7} \cline{9-11} \cline{13-15} 
						\multicolumn{1}{|c|}{} & \multicolumn{1}{c|}{} & \multicolumn{1}{c|}{30} & \multicolumn{1}{c|}{} & \multicolumn{1}{c|}{0.722 (0.067)} & \multicolumn{1}{c|}{0.137 (0.044)} & \multicolumn{1}{c|}{0.062 (0.018)} & \multicolumn{1}{c|}{} & \multicolumn{1}{c|}{0.689 (0.064)} & \multicolumn{1}{c|}{0.187 (0.044)} & \multicolumn{1}{c|}{0.086 (0.025)} & \multicolumn{1}{c|}{} & \multicolumn{1}{c|}{0.714 (0.057)} & \multicolumn{1}{c|}{0.186 (0.039)} & \multicolumn{1}{c|}{0.088 (0.022)} \\ \cline{1-3} \cline{5-7} \cline{9-11} \cline{13-15} 
					\end{tabular}%
				}
				\caption{\small Simulation results for Case 1 for each of the three layers. We consider scenarios which include/ignore the structure in $X$. Top: $\Sigma^{(x)} = I_g$. Bottom: $\Sigma^{(x)}$ is a block matrix where components of one of the blocks are highly correlated.}
				\label{tab: sim_case_1}
			\end{table}
		\end{small}
	\end{landscape}
	
	\subsection{Case 2}
	
\begin{figure}[!t]
		\centering
		\includegraphics[scale=0.7]{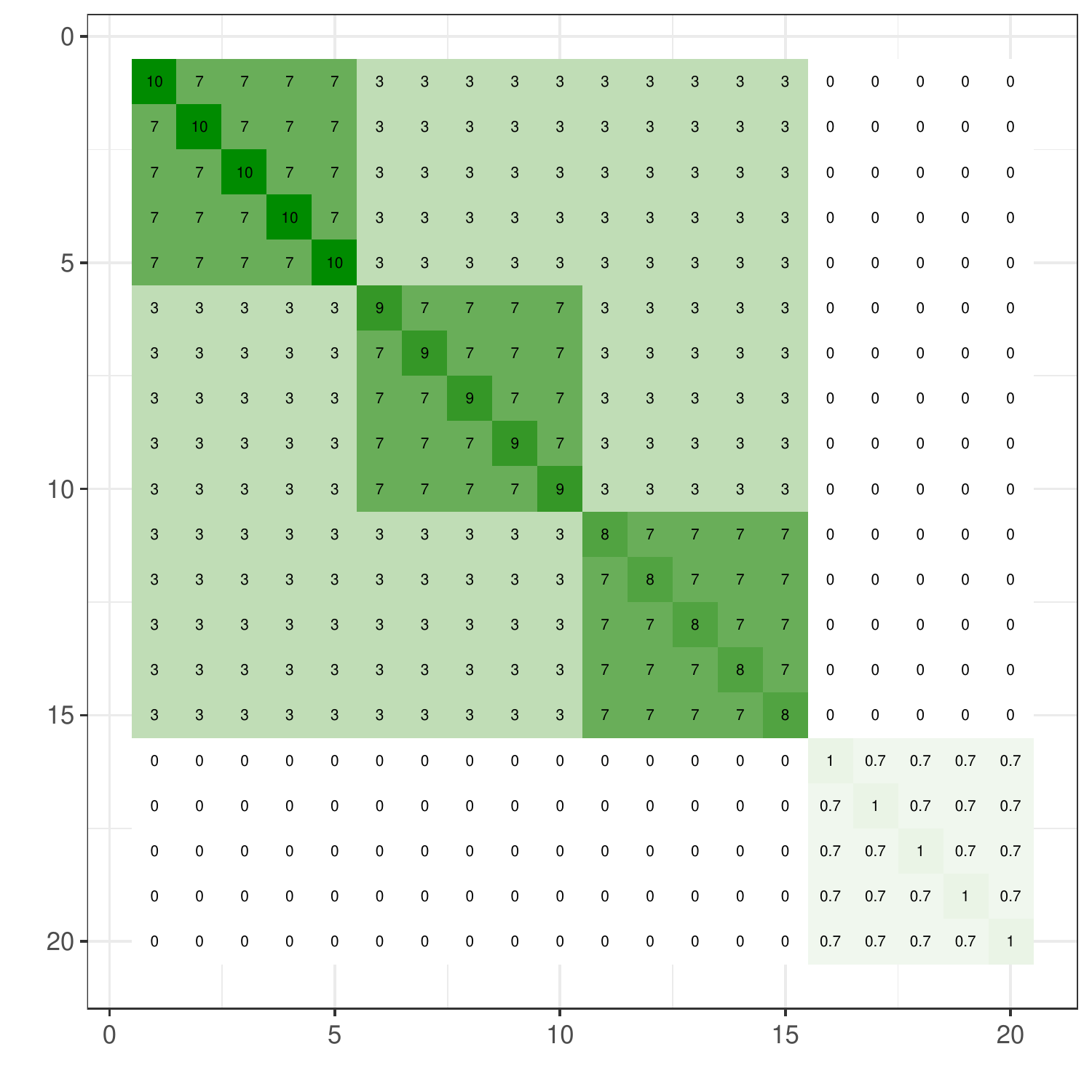}
		\caption{The choice of $\Sigma^{(x)}$, the covariance matrix of $X$, for Case 2. The rows and columns correspond to the $g=20$ genomic markers.}
		\label{fig: sim_case2_sigma}
	\end{figure}

	Figure \ref{fig: sim_case2_sigma} displays the covariance matrix considered under the different scenarios for Case 2. Algorithm \ref{algo: simcase2} describes the procedure to generate the simulated data. We make the following choices for the hyperparameters: $\delta = p^{(t)} = 12, \Psi = \sigma^2 I_g$, where $\sigma^2 = 20$. We use the same choices for all cases of $\theta = 1,0.9,0.8,0.7$.
	
	\begin{algorithm}[!t]
		\caption{Simulation from the model}\label{algo: simcase2}
		\begin{algorithmic}[1]
			%\Procedure{MyProcedure}{}
			\State Specify the choices for $n=100, p^{(t)}=12, p^{(t,m)}=3, m=4, \tau=3, g=20$. 
			\State Specify the hyperparameters $\delta = p^{(t)}, \Psi = 20I_g$ and $\theta \in \{0.7,0.8,0.9,1\}$.
			\State Generate $X = [\x_1,\ldots,\x_n]^\top \in \mathbb{R}^{n \times g}$, where $\x_i \stackrel{iid}{\sim} N_g(\0,\Sigma^{(x)})$ for $i=1,\ldots,n$.
			\For {$t=1,\ldots,\tau$} 
			\State Generate $\Delta^{(t)} \sim IW(\delta, \Psi)$.
			\State Generate $B^{(t)} = \begin{bmatrix}
				(r(t)\times s(t))\text{-matrix Double-Exp(location}=0, \text{scale}=\theta) & \0 \\
				\0 & \0
			\end{bmatrix}$.
			\State Generate $vec(Y^{(t)}) \sim N_{np^{(t)}}(vec(XB^{(t)}), I_n \otimes \Delta^{(t)})$ and appropriately reshape $Y^{(t)} \in \mathbb{R}^{n \times p^{(t)}}$.
			\EndFor
		\end{algorithmic}
	\end{algorithm}
	
	\begin{figure}[!t]
		\centering
		\resizebox{\textwidth}{!}{
			\begin{tabular}{|c|c|c|c|}
				\hline
				\includegraphics[page=1]{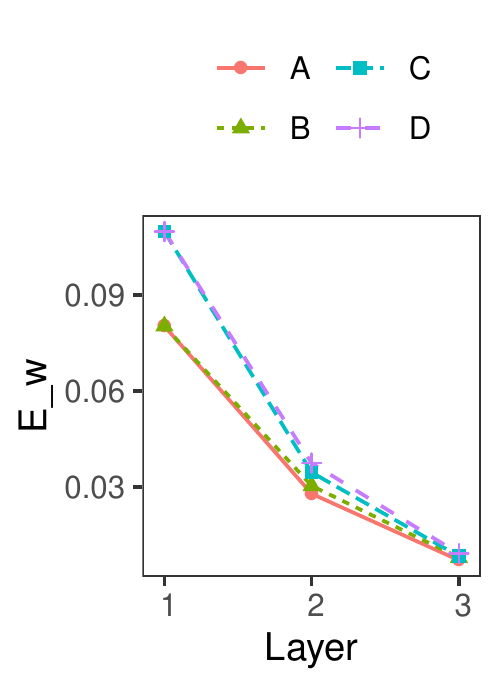} &
				\includegraphics[page=2]{plots/E_w_case2.pdf} &
				\includegraphics[page=3]{plots/E_w_case2.pdf} &
				\includegraphics[page=4]{plots/E_w_case2.pdf} \\
				(a) $\theta = 1$ & 
				(b) $\theta = 0.9$ &
				(c) $\theta = 0.8$ &
				(d) $\theta = 0.7$ \\ \hline
			\end{tabular}
		}
		\caption{Average values of $E_w$ in each of the three layers for Case 2 when $\Sigma^{(x)}$ is a $(2 \times 2)$-block matrix. Model (A): $\bmu^{(t,m)} \neq \0; \Lambda = cor(X)$. Model (B): $\bmu^{(t,m)} = \0; \Lambda = cor(X)$. Model (C): $\bmu^{(t,m)} \neq \0; \Lambda = I_g$. Model (D): $\bmu^{(t,m)} = \0; \Lambda = I_g$.}
		\label{fig: sim_case_2_E_w}
	\end{figure}
	
\begin{figure}[!t]
		\centering
		\resizebox{\textwidth}{!}{
			\begin{tabular}{|c|c|c|c|}
				\hline
				\includegraphics[page=1]{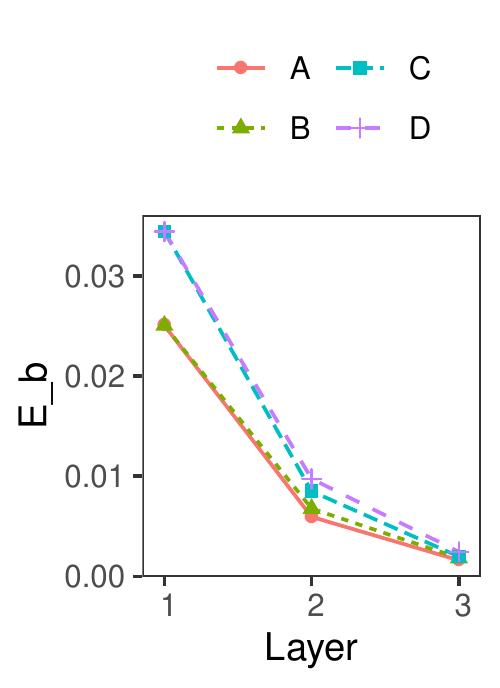} &
				\includegraphics[page=2]{plots/E_b_case2.pdf} &
				\includegraphics[page=3]{plots/E_b_case2.pdf} &
				\includegraphics[page=4]{plots/E_b_case2.pdf} \\
				(a) $\theta = 1$ & 
				(b) $\theta = 0.9$ &
				(c) $\theta = 0.8$ &
				(d) $\theta = 0.7$ \\ \hline
			\end{tabular}
		}
		\caption{Average values of $E_\beta$ in each of the three layers for Case 2 when $\Sigma^{(x)}$ is a $(2 \times 2)$-block matrix. Model (A): $\bmu^{(t,m)} \neq \0; \Lambda = cor(X)$. Model (B): $\bmu^{(t,m)} = \0; \Lambda = cor(X)$. Model (C): $\bmu^{(t,m)} \neq \0; \Lambda = I_g$. Model (D): $\bmu^{(t,m)} = \0; \Lambda = I_g$.}
		\label{fig: sim_case_2_E_b}
	\end{figure}

	For estimation, we choose $a_1 = 4, a_2 = 5, \alpha = 0.5, \Psi = I_g, \delta = p^{(t)}$ and set $v_1$ as the smallest power of 10 greater than the maximum magnitude of the entries in the ordinary least square estimate of $B^{(t)}$. For model selection, we estimate the model on a grid of values for $v_0$. From the estimation based on each value of $v_0$, we compute BIC and choose the best model in terms of the lowest BIC value. Table \ref{tab: sim_case_2} shows the average TPR, $E_w$ and $E_\beta$ after model selection based on 30 replications. Figure \ref{fig: sim_case_2_E_w} shows the plot of average $E_w$ by layer for various choices of the effect size $\theta$. Moreover, we also include plots for $E_\beta$ in Figure \ref{fig: sim_case_2_E_b}. We considered the grid of values for $v_0$ as $\{ 0.001 + (v-1)0.001 | v=1,\ldots,10 \}$.
	
	\begin{landscape}
		\begin{small}
			\begin{table}[!t]
				\centering
				\resizebox{1.3\textwidth}{0.35\textheight}{%
					\begin{tabular}{cccccccccccccc}
						\hline
						&  &  & \multicolumn{3}{c}{Layer 1} &  & \multicolumn{3}{c}{Layer 2} &  & \multicolumn{3}{c}{Layer 3} \\ \hline
						\multicolumn{1}{|c|}{\begin{tabular}[c]{@{}c@{}}$\bmu^{(t,m)}, \bLambda$\end{tabular}} & \multicolumn{1}{c|}{$\theta$} & \multicolumn{1}{c|}{} & \multicolumn{1}{c|}{TPR} & \multicolumn{1}{c|}{$E_w$} & \multicolumn{1}{c|}{$E_\beta$} & \multicolumn{1}{c|}{} & \multicolumn{1}{c|}{TPR} & \multicolumn{1}{c|}{$E_w$} & \multicolumn{1}{c|}{$E_\beta$} & \multicolumn{1}{c|}{} & \multicolumn{1}{c|}{TPR} & \multicolumn{1}{c|}{$E_w$} & \multicolumn{1}{c|}{$E_\beta$} \\ \cline{1-2} \cline{4-6} \cline{8-10} \cline{12-14} 
						&  &  &  &  &  &  &  &  &  &  &  &  &  \\ \cline{1-2} \cline{4-6} \cline{8-10} \cline{12-14} 
						\multicolumn{1}{|c|}{\multirow{4}{*}{\begin{tabular}[c]{@{}c@{}}$\bmu^{(t,m)} \neq  \0$\\ $\bLambda = cor(X)$\end{tabular}}} & \multicolumn{1}{c|}{1} & \multicolumn{1}{c|}{} & \multicolumn{1}{c|}{0.857 (0.065)} & \multicolumn{1}{c|}{0.08 (0.036)} & \multicolumn{1}{c|}{0.025 (0.016)} & \multicolumn{1}{c|}{} & \multicolumn{1}{c|}{0.888 (0.057)} & \multicolumn{1}{c|}{0.028 (0.014)} & \multicolumn{1}{c|}{0.006 (0.004)} & \multicolumn{1}{c|}{} & \multicolumn{1}{c|}{0.887 (0.155)} & \multicolumn{1}{c|}{0.007 (0.01)} & \multicolumn{1}{c|}{0.002 (0.002)} \\ \cline{2-2} \cline{4-6} \cline{8-10} \cline{12-14} 
						\multicolumn{1}{|c|}{} & \multicolumn{1}{c|}{0.9} & \multicolumn{1}{c|}{} & \multicolumn{1}{c|}{0.816 (0.06)} & \multicolumn{1}{c|}{0.103 (0.034)} & \multicolumn{1}{c|}{0.029 (0.01)} & \multicolumn{1}{c|}{} & \multicolumn{1}{c|}{0.832 (0.098)} & \multicolumn{1}{c|}{0.042 (0.024)} & \multicolumn{1}{c|}{0.009 (0.007)} & \multicolumn{1}{c|}{} & \multicolumn{1}{c|}{0.847 (0.163)} & \multicolumn{1}{c|}{0.01 (0.01)} & \multicolumn{1}{c|}{0.002 (0.002)} \\ \cline{2-2} \cline{4-6} \cline{8-10} \cline{12-14} 
						\multicolumn{1}{|c|}{} & \multicolumn{1}{c|}{0.8} & \multicolumn{1}{c|}{} & \multicolumn{1}{c|}{0.743 (0.061)} & \multicolumn{1}{c|}{0.145 (0.035)} & \multicolumn{1}{c|}{0.038 (0.013)} & \multicolumn{1}{c|}{} & \multicolumn{1}{c|}{0.8 (0.086)} & \multicolumn{1}{c|}{0.05 (0.022)} & \multicolumn{1}{c|}{0.011 (0.007)} & \multicolumn{1}{c|}{} & \multicolumn{1}{c|}{0.867 (0.152)} & \multicolumn{1}{c|}{0.008 (0.009)} & \multicolumn{1}{c|}{0.002 (0.002)} \\ \cline{2-2} \cline{4-6} \cline{8-10} \cline{12-14} 
						\multicolumn{1}{|c|}{} & \multicolumn{1}{c|}{0.7} & \multicolumn{1}{c|}{} & \multicolumn{1}{c|}{0.667 (0.1)} & \multicolumn{1}{c|}{0.187 (0.057)} & \multicolumn{1}{c|}{0.044 (0.018)} & \multicolumn{1}{c|}{} & \multicolumn{1}{c|}{0.753 (0.095)} & \multicolumn{1}{c|}{0.062 (0.024)} & \multicolumn{1}{c|}{0.013 (0.006)} & \multicolumn{1}{c|}{} & \multicolumn{1}{c|}{0.753 (0.163)} & \multicolumn{1}{c|}{0.015 (0.01)} & \multicolumn{1}{c|}{0.003 (0.003)} \\ \cline{1-2} \cline{4-6} \cline{8-10} \cline{12-14} 
						&  &  &  &  &  &  &  &  &  &  &  &  &  \\ \cline{1-2} \cline{4-6} \cline{8-10} \cline{12-14} 
						\multicolumn{1}{|c|}{\multirow{4}{*}{\begin{tabular}[c]{@{}c@{}}$\bmu^{(t,m)} = \0$\\ $\bLambda = cor(X)$\end{tabular}}} & \multicolumn{1}{c|}{1} & \multicolumn{1}{c|}{} & \multicolumn{1}{c|}{0.857 (0.065)} & \multicolumn{1}{c|}{0.08 (0.036)} & \multicolumn{1}{c|}{0.025 (0.016)} & \multicolumn{1}{c|}{} & \multicolumn{1}{c|}{0.878 (0.057)} & \multicolumn{1}{c|}{0.03 (0.014)} & \multicolumn{1}{c|}{0.007 (0.004)} & \multicolumn{1}{c|}{} & \multicolumn{1}{c|}{0.873 (0.153)} & \multicolumn{1}{c|}{0.008 (0.01)} & \multicolumn{1}{c|}{0.002 (0.002)} \\ \cline{2-2} \cline{4-6} \cline{8-10} \cline{12-14} 
						\multicolumn{1}{|c|}{} & \multicolumn{1}{c|}{0.9} & \multicolumn{1}{c|}{} & \multicolumn{1}{c|}{0.816 (0.06)} & \multicolumn{1}{c|}{0.103 (0.034)} & \multicolumn{1}{c|}{0.029 (0.01)} & \multicolumn{1}{c|}{} & \multicolumn{1}{c|}{0.818 (0.1)} & \multicolumn{1}{c|}{0.045 (0.025)} & \multicolumn{1}{c|}{0.01 (0.007)} & \multicolumn{1}{c|}{} & \multicolumn{1}{c|}{0.84 (0.161)} & \multicolumn{1}{c|}{0.01 (0.01)} & \multicolumn{1}{c|}{0.002 (0.003)} \\ \cline{2-2} \cline{4-6} \cline{8-10} \cline{12-14} 
						\multicolumn{1}{|c|}{} & \multicolumn{1}{c|}{0.8} & \multicolumn{1}{c|}{} & \multicolumn{1}{c|}{0.744 (0.062)} & \multicolumn{1}{c|}{0.144 (0.035)} & \multicolumn{1}{c|}{0.038 (0.013)} & \multicolumn{1}{c|}{} & \multicolumn{1}{c|}{0.787 (0.101)} & \multicolumn{1}{c|}{0.053 (0.025)} & \multicolumn{1}{c|}{0.012 (0.008)} & \multicolumn{1}{c|}{} & \multicolumn{1}{c|}{0.86 (0.15)} & \multicolumn{1}{c|}{0.009 (0.009)} & \multicolumn{1}{c|}{0.002 (0.002)} \\ \cline{2-2} \cline{4-6} \cline{8-10} \cline{12-14} 
						\multicolumn{1}{|c|}{} & \multicolumn{1}{c|}{0.7} & \multicolumn{1}{c|}{} & \multicolumn{1}{c|}{0.669 (0.099)} & \multicolumn{1}{c|}{0.186 (0.056)} & \multicolumn{1}{c|}{0.044 (0.017)} & \multicolumn{1}{c|}{} & \multicolumn{1}{c|}{0.75 (0.09)} & \multicolumn{1}{c|}{0.062 (0.022)} & \multicolumn{1}{c|}{0.013 (0.006)} & \multicolumn{1}{c|}{} & \multicolumn{1}{c|}{0.747 (0.157)} & \multicolumn{1}{c|}{0.016 (0.01)} & \multicolumn{1}{c|}{0.003 (0.002)} \\ \cline{1-2} \cline{4-6} \cline{8-10} \cline{12-14} 
						&  &  &  &  &  &  &  &  &  &  &  &  &  \\ \cline{1-2} \cline{4-6} \cline{8-10} \cline{12-14} 
						\multicolumn{1}{|c|}{\multirow{4}{*}{\begin{tabular}[c]{@{}c@{}}$\bmu^{(t,m)} \neq \0$\\ $\bLambda = I_g$\end{tabular}}} & \multicolumn{1}{c|}{1} & \multicolumn{1}{c|}{} & \multicolumn{1}{c|}{0.804 (0.064)} & \multicolumn{1}{c|}{0.11 (0.036)} & \multicolumn{1}{c|}{0.034 (0.017)} & \multicolumn{1}{c|}{} & \multicolumn{1}{c|}{0.862 (0.063)} & \multicolumn{1}{c|}{0.035 (0.016)} & \multicolumn{1}{c|}{0.009 (0.004)} & \multicolumn{1}{c|}{} & \multicolumn{1}{c|}{0.867 (0.16)} & \multicolumn{1}{c|}{0.008 (0.01)} & \multicolumn{1}{c|}{0.002 (0.003)} \\ \cline{2-2} \cline{4-6} \cline{8-10} \cline{12-14} 
						\multicolumn{1}{|c|}{} & \multicolumn{1}{c|}{0.9} & \multicolumn{1}{c|}{} & \multicolumn{1}{c|}{0.775 (0.058)} & \multicolumn{1}{c|}{0.127 (0.033)} & \multicolumn{1}{c|}{0.037 (0.01)} & \multicolumn{1}{c|}{} & \multicolumn{1}{c|}{0.8 (0.101)} & \multicolumn{1}{c|}{0.05 (0.025)} & \multicolumn{1}{c|}{0.011 (0.008)} & \multicolumn{1}{c|}{} & \multicolumn{1}{c|}{0.833 (0.158)} & \multicolumn{1}{c|}{0.01 (0.01)} & \multicolumn{1}{c|}{0.002 (0.003)} \\ \cline{2-2} \cline{4-6} \cline{8-10} \cline{12-14} 
						\multicolumn{1}{|c|}{} & \multicolumn{1}{c|}{0.8} & \multicolumn{1}{c|}{} & \multicolumn{1}{c|}{0.713 (0.065)} & \multicolumn{1}{c|}{0.161 (0.037)} & \multicolumn{1}{c|}{0.044 (0.014)} & \multicolumn{1}{c|}{} & \multicolumn{1}{c|}{0.758 (0.102)} & \multicolumn{1}{c|}{0.06 (0.025)} & \multicolumn{1}{c|}{0.014 (0.009)} & \multicolumn{1}{c|}{} & \multicolumn{1}{c|}{0.847 (0.146)} & \multicolumn{1}{c|}{0.01 (0.009)} & \multicolumn{1}{c|}{0.002 (0.003)} \\ \cline{2-2} \cline{4-6} \cline{8-10} \cline{12-14} 
						\multicolumn{1}{|c|}{} & \multicolumn{1}{c|}{0.7} & \multicolumn{1}{c|}{} & \multicolumn{1}{c|}{0.636 (0.078)} & \multicolumn{1}{c|}{0.205 (0.044)} & \multicolumn{1}{c|}{0.05 (0.013)} & \multicolumn{1}{c|}{} & \multicolumn{1}{c|}{0.717 (0.092)} & \multicolumn{1}{c|}{0.071 (0.023)} & \multicolumn{1}{c|}{0.016 (0.007)} & \multicolumn{1}{c|}{} & \multicolumn{1}{c|}{0.713 (0.18)} & \multicolumn{1}{c|}{0.018 (0.011)} & \multicolumn{1}{c|}{0.004 (0.005)} \\ \cline{1-2} \cline{4-6} \cline{8-10} \cline{12-14} 
						&  &  &  &  &  &  &  &  &  &  &  &  &  \\ \cline{1-2} \cline{4-6} \cline{8-10} \cline{12-14} 
						\multicolumn{1}{|c|}{\multirow{4}{*}{\begin{tabular}[c]{@{}c@{}}$\bmu^{(t,m)} = \0$\\   $\bLambda = I_g$\end{tabular}}} & \multicolumn{1}{c|}{1} & \multicolumn{1}{c|}{} & \multicolumn{1}{c|}{0.804 (0.064)} & \multicolumn{1}{c|}{0.11 (0.036)} & \multicolumn{1}{c|}{0.034 (0.017)} & \multicolumn{1}{c|}{} & \multicolumn{1}{c|}{0.85 (0.068)} & \multicolumn{1}{c|}{0.037 (0.017)} & \multicolumn{1}{c|}{0.01 (0.005)} & \multicolumn{1}{c|}{} & \multicolumn{1}{c|}{0.853 (0.166)} & \multicolumn{1}{c|}{0.009 (0.01)} & \multicolumn{1}{c|}{0.002 (0.003)} \\ \cline{2-2} \cline{4-6} \cline{8-10} \cline{12-14} 
						\multicolumn{1}{|c|}{} & \multicolumn{1}{c|}{0.9} & \multicolumn{1}{c|}{} & \multicolumn{1}{c|}{0.774 (0.057)} & \multicolumn{1}{c|}{0.127 (0.032)} & \multicolumn{1}{c|}{0.038 (0.011)} & \multicolumn{1}{c|}{} & \multicolumn{1}{c|}{0.792 (0.102)} & \multicolumn{1}{c|}{0.052 (0.025)} & \multicolumn{1}{c|}{0.012 (0.007)} & \multicolumn{1}{c|}{} & \multicolumn{1}{c|}{0.813 (0.157)} & \multicolumn{1}{c|}{0.012 (0.01)} & \multicolumn{1}{c|}{0.003 (0.003)} \\ \cline{2-2} \cline{4-6} \cline{8-10} \cline{12-14} 
						\multicolumn{1}{|c|}{} & \multicolumn{1}{c|}{0.8} & \multicolumn{1}{c|}{} & \multicolumn{1}{c|}{0.715 (0.066)} & \multicolumn{1}{c|}{0.16 (0.037)} & \multicolumn{1}{c|}{0.044 (0.014)} & \multicolumn{1}{c|}{} & \multicolumn{1}{c|}{0.752 (0.1)} & \multicolumn{1}{c|}{0.062 (0.025)} & \multicolumn{1}{c|}{0.015 (0.009)} & \multicolumn{1}{c|}{} & \multicolumn{1}{c|}{0.847 (0.146)} & \multicolumn{1}{c|}{0.01 (0.009)} & \multicolumn{1}{c|}{0.002 (0.003)} \\ \cline{2-2} \cline{4-6} \cline{8-10} \cline{12-14} 
						\multicolumn{1}{|c|}{} & \multicolumn{1}{c|}{0.7} & \multicolumn{1}{c|}{} & \multicolumn{1}{c|}{0.636 (0.078)} & \multicolumn{1}{c|}{0.205 (0.044)} & \multicolumn{1}{c|}{0.05 (0.013)} & \multicolumn{1}{c|}{} & \multicolumn{1}{c|}{0.695 (0.101)} & \multicolumn{1}{c|}{0.076 (0.025)} & \multicolumn{1}{c|}{0.018 (0.008)} & \multicolumn{1}{c|}{} & \multicolumn{1}{c|}{0.693 (0.172)} & \multicolumn{1}{c|}{0.019 (0.011)} & \multicolumn{1}{c|}{0.004 (0.005)} \\ \cline{1-2} \cline{4-6} \cline{8-10} \cline{12-14} 
					\end{tabular}%
				}
				\caption{\small Simulation results for Case 2 for each of the three layers. Each panel corresponds to a combination of choices for $\bmu^{(t,m)}$ and $\Lambda$, which include/ignore the structure in $X$.}
				\label{tab: sim_case_2}
			\end{table}
		\end{small}
	\end{landscape}
	
	\section{LGG Case Study Details}\label{app: details_case}
	
	The set of cancer driver genes for LGG includes the 24 genes reported in Table \ref{tab: LGG-genes}, whose expression profile was retrieved from the expression of the full set of genes from LinkedOmics\footnote{\url{www.linkedomics.org}}.
	
	\begin{table}[!h]
		\centering
		\begin{tabular}{|llllll|}
			\hline
			ARID1A & EGFR & MAX & NRAS & PTEN & TCF12 \\ 
			ARID2 & FUBP1 & NF1 & PIK3CA & PTPN11 & TP53 \\ 
			ATRX & IDH1 & NIPBL & PIK3R1 & SETD2 & ZBTB20 \\ 
			CIC & IDH2 & NOTCH1 & PLCG1 & SMARCA4 & ZCCHC12 \\ 
			\hline
		\end{tabular}
		\caption{LGG-genes: The LGG-specific cancer driver genes \citep{bailey2018comprehensive}.}
		\label{tab: LGG-genes}
	\end{table}
	
	\subsection{Correlation Between the LGG-Genes}\label{subapp: correlation}
\begin{figure}[!t]
		\centering
		\resizebox{\textwidth}{!}{
			\includegraphics{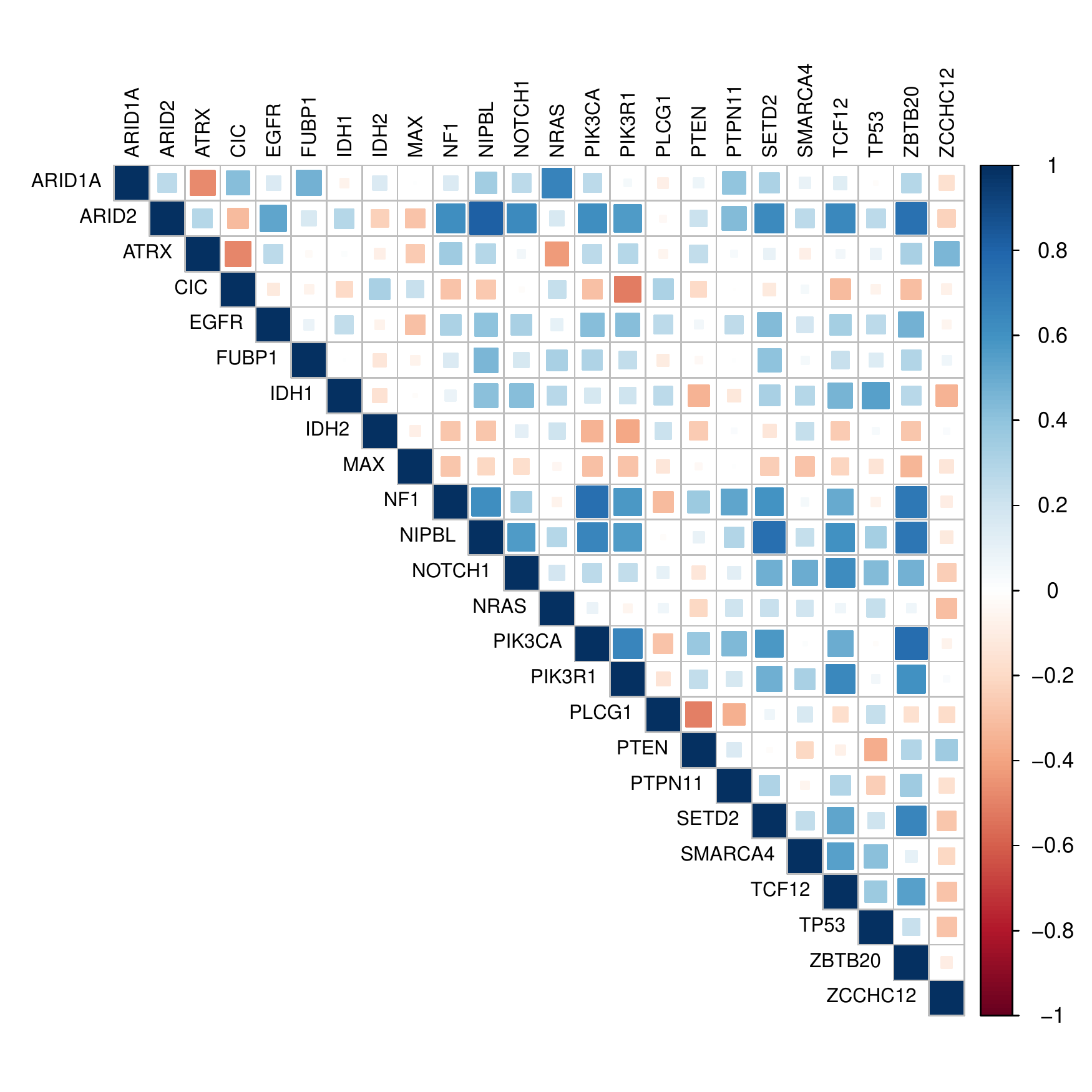}
		}
		\caption{Correlation between the 24 LGG-genes.}
		\label{fig: cor_gene_exp}
	\end{figure}

	Figure \ref{fig: cor_gene_exp} displays the correlation matrix for the 24 LGG-genes considered in the case study.
	
	\subsection{Results for Different Layer Splits}\label{subapp: results}
\begin{figure}[!t]
		\centering
		\resizebox{\textwidth}{!}{
			\begin{tabular}{c}
				\includegraphics[trim = 1cm 0.25cm 0.45cm 0.2cm, clip]{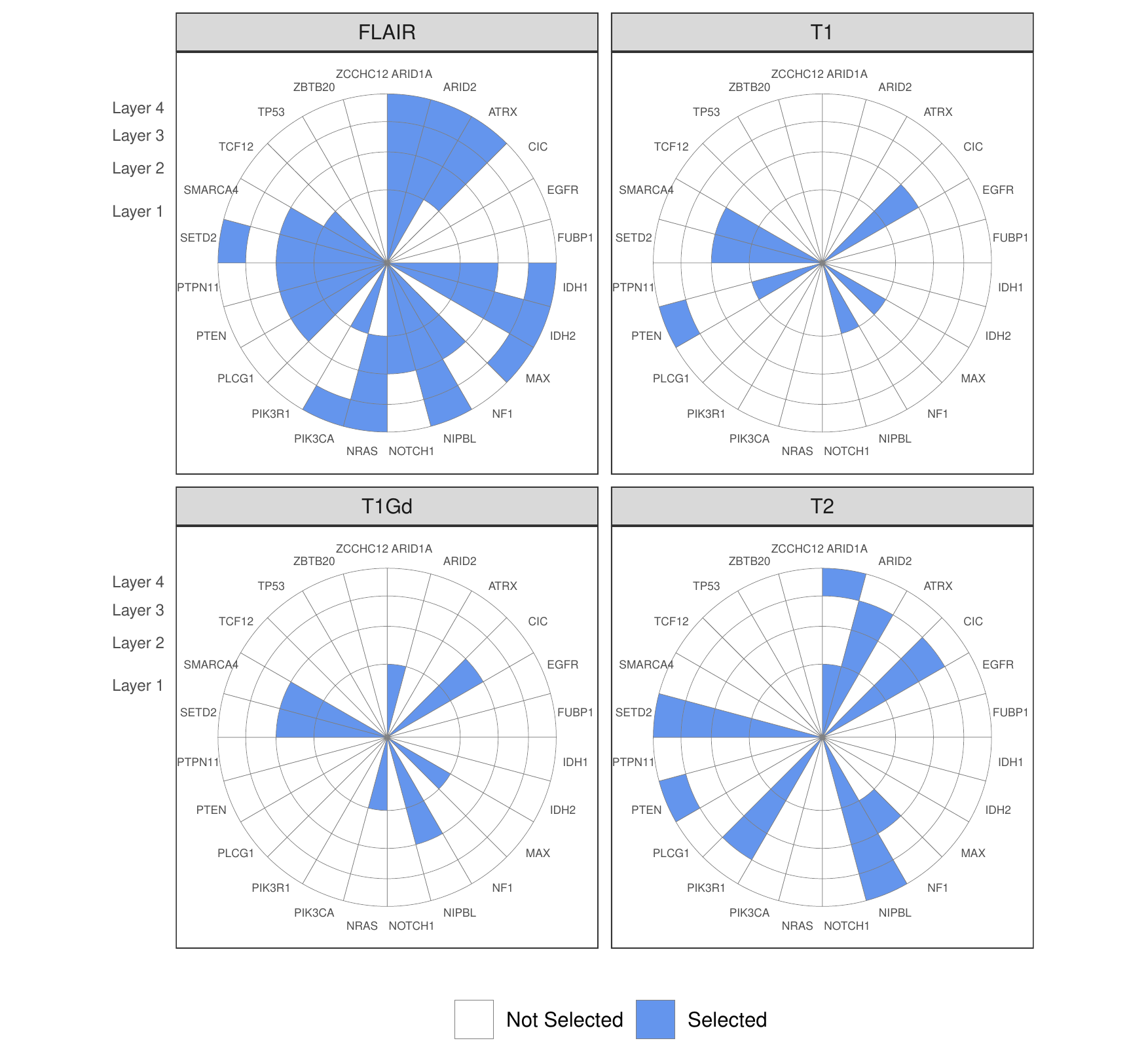}
			\end{tabular}
		}
		\caption{Associations of the layer-wise PC scores from the imaging sequences with the gene expression. Each panel corresponds to an imaging sequence, and a block filled with color indicates association for that gene with one of the three specific layers ($\tau = 4$).}
		\label{fig: sig_genes_4}
	\end{figure}

\begin{figure}[!t]
		\centering
		\resizebox{\textwidth}{!}{
			\begin{tabular}{c}
				\includegraphics[trim = 1cm 0.25cm 0.45cm 0.2cm, clip]{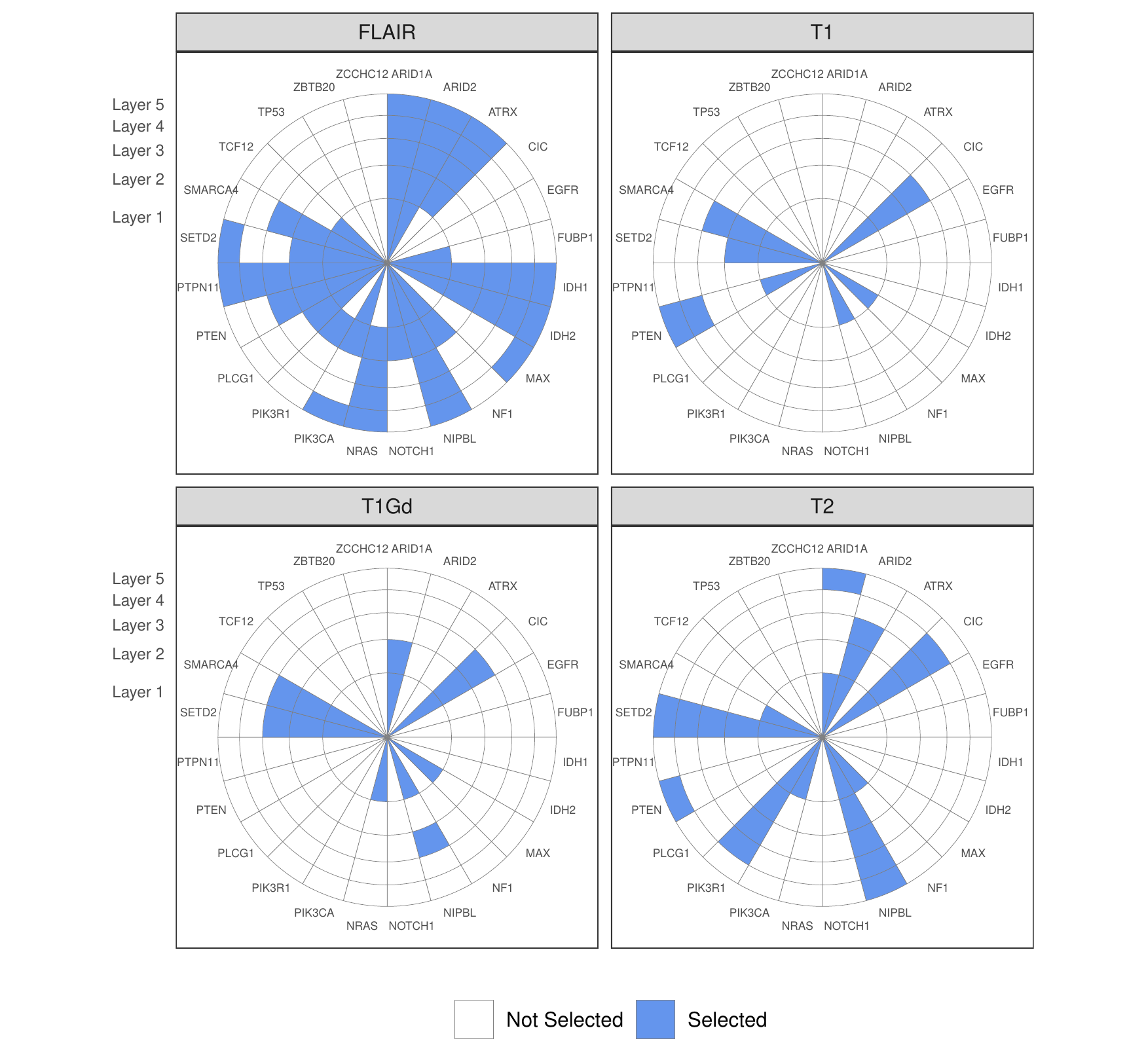}
			\end{tabular}
		}
		\caption{Associations of the layer-wise PC scores from the imaging sequences with the gene expression. Each panel corresponds to an imaging sequence, and a block filled with color indicates association for that gene with one of the three specific layers ($\tau = 5$).}
		\label{fig: sig_genes_5}
	\end{figure}

\begin{figure}[!t]
		\centering
		\resizebox{\textwidth}{!}{
			\begin{tabular}{c}
				\includegraphics[trim = 1cm 0.25cm 0.45cm 0.2cm, clip]{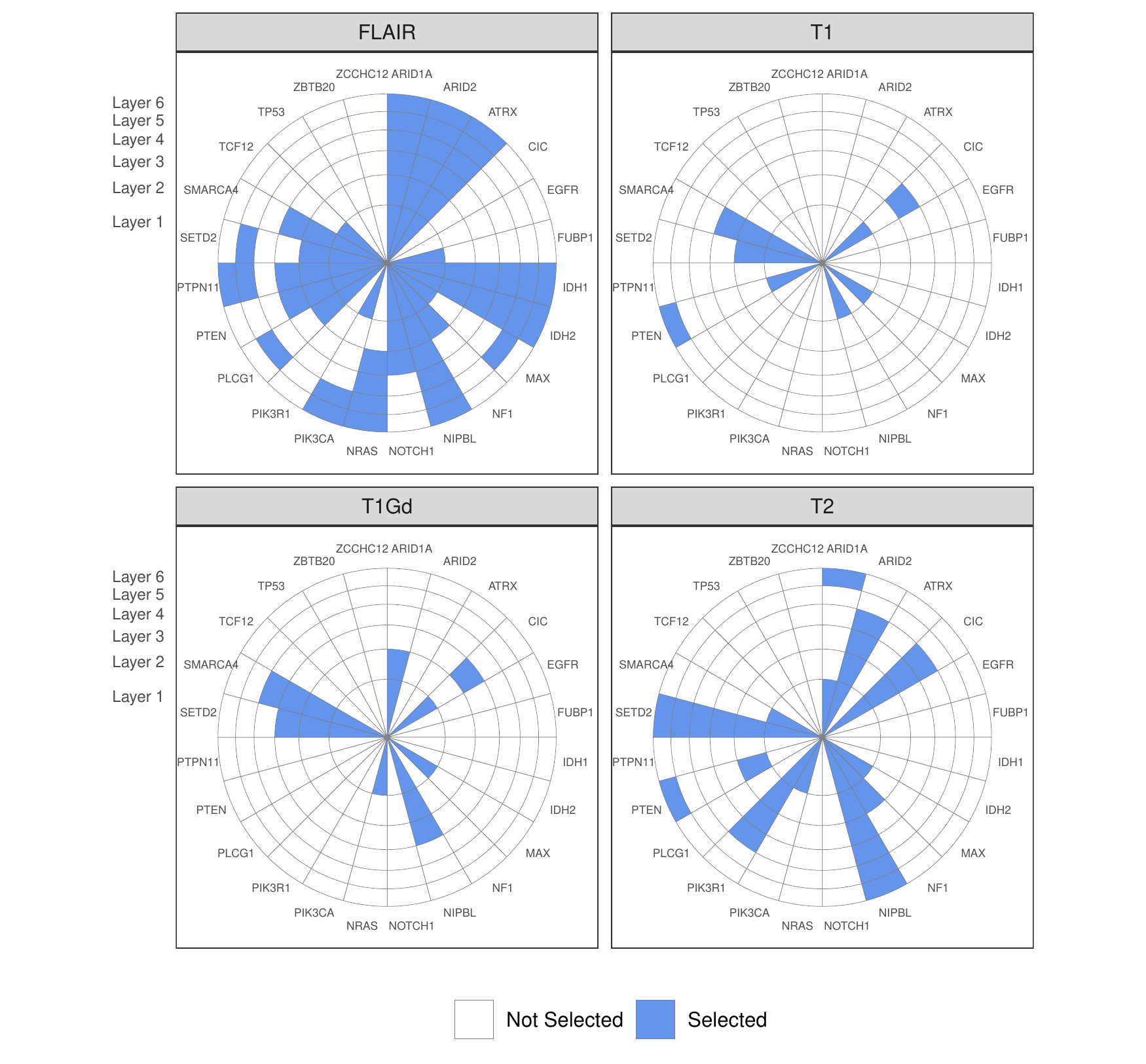}
			\end{tabular}
		}
		\caption{Associations of the layer-wise PC scores from the imaging sequences with the gene expression. Each panel corresponds to an imaging sequence, and a block filled with color indicates association for that gene with one of the three specific layers ($\tau = 6$).}
		\label{fig: sig_genes_6}
	\end{figure}

	We present additional results based on the proposed modeling approach, when the tumor region is split into $\tau = 4, 5$ and $6$ spherical layers. The results for each case are given in Figures \ref{fig: sig_genes_4}, \ref{fig: sig_genes_5} and \ref{fig: sig_genes_6}, respectively. The results suggest that our approach is robust to the choice of the number of layers to compartmentalize the tumor.
	
	\subsection{Maximizing the $Q$-function}\label{subapp: maxq-fn}
\begin{figure}[!t]
		\centering
		\resizebox{\textwidth}{!}{
			\begin{tabular}{cc}
				\includegraphics[trim = 0cm 1.6cm 1cm 1.5cm, clip, page=1]{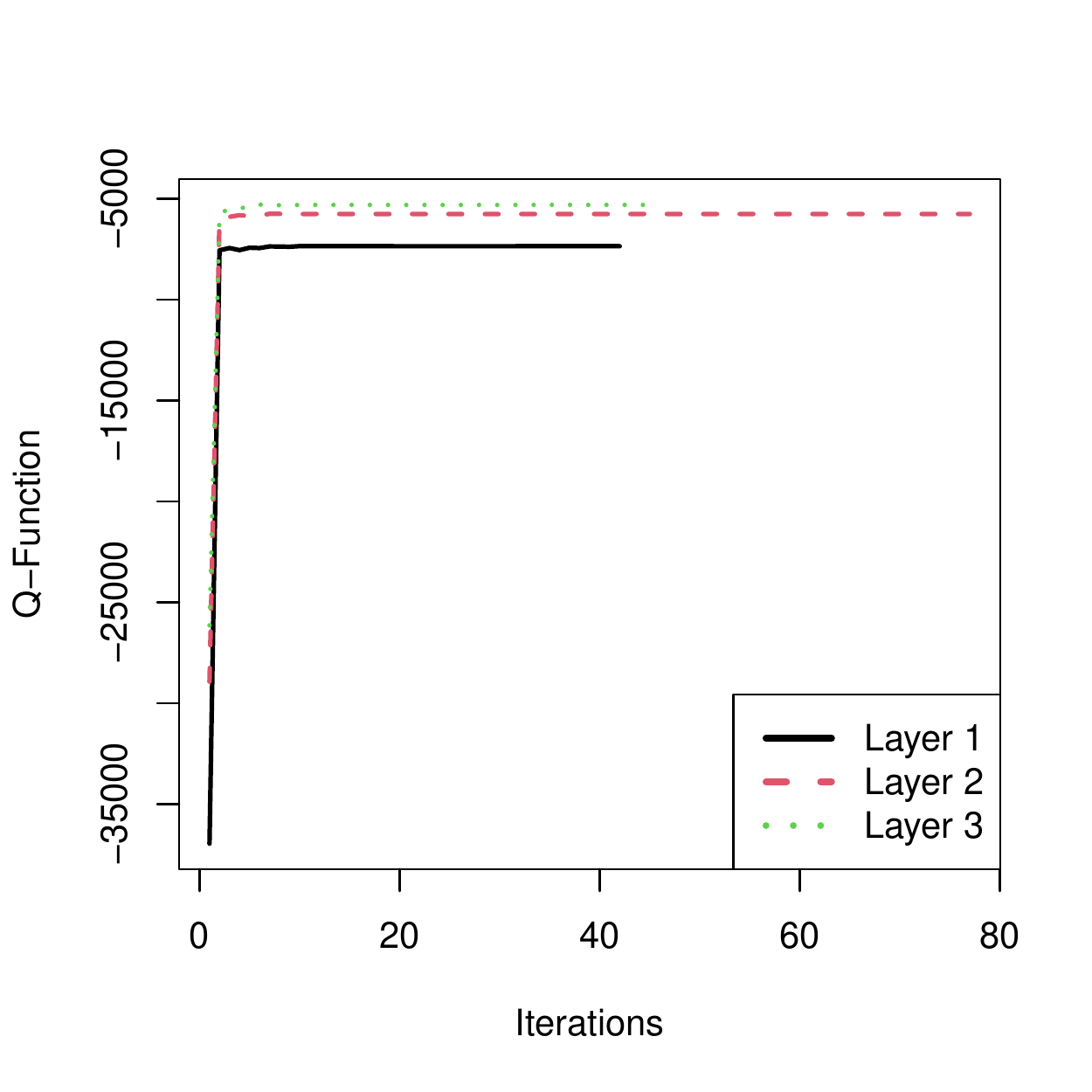} &
				\includegraphics[trim = 1cm 1.6cm 1cm 1.5cm, clip, page=2]{plots/figureS13.pdf} \\
				(a) $\tau = 3$ spherical shells & (b) $\tau = 4$ spherical shells \\
				\includegraphics[trim = 0cm 0.25cm 1cm 1.5cm, clip, page=3]{plots/figureS13.pdf} &
				\includegraphics[trim = 1cm 0.25cm 1cm 1.5cm, clip, page=4]{plots/figureS13.pdf} \\
				(c) $\tau = 5$ spherical shells & (d) $\tau = 6$ spherical shells
			\end{tabular}
		}
		\caption{Convergence of the $Q$-function across iterations of the EM algorithm. Each panel corresponds to different values of the number of tumor layers $\tau$.}
		\label{fig: q-fn}
	\end{figure}

	Figure \ref{fig: q-fn} shows the value of the $Q$-function based on the iterative (EM-based) estimation approach for the models with $\tau =3, 4, 5$ and $6$ spherical layers.
	
	\subsection{Hyperparameter Settings}\label{subapp: hyperparameter}
	In our analysis, we choose $\bmu^{(t,m)} = \alpha\hat{\blambda}^{(t-1,m)}_+$ with $\alpha = 0.5$. This choice was made to lead to a natural mid-point between incorporating no prior information ($\alpha = 0$) about the selection and having complete dependence ($\alpha = 1$) on the knowledge gained from the previous layer. The value of $v_1$ is expected to be large enough to capture effect sizes of higher magnitude. This is complemented by the choice of $a_1$ ($>0.5$ -- to maximize $Q_1$ w.r.t. ${\nu^{-2}_{kj}}^{(t,m)}$) and $a_2$ that leads to an informative prior. Our choice of $a_1 = 4$ and $a_2 = 5$ was made such that the prior on ${\nu^{-2}_{kj}}^{(t,m)}$ (along with $v_1$) is roughly centered around the value $1$ and would still place reasonable prior weight for different magnitudes of effect sizes. The choice of hyperparameters for the degrees of freedom ($\delta$) and the scale matrix ($\Psi$) of the inverse-Wishart prior lead to a non-informative/vague prior around $\Delta^{(t)}$. We consider a grid of values for $v_0$ and clearly as $v_0$ is increased the more effect sizes with higher magnitudes are pushed towards zero.
	
	\FloatBarrier
	\section{EM \textit{versus} MCMC}
	
	We compare the computational burden of our Expectation-Maximization (EM)-based estimation approach with a Markov Chain Monte Carlo (MCMC) sampling-based approach. To do this, we consider the simulation settings similar to Case 1 with $\sigma = 1$ as described in Sections 5 and \ref{app: details_sim}. However, we vary the number of genes as $g = \{10,20,\ldots,60\}$ and the number of principal components as $p^{(t)} = \{12, 24, 36\}$ for each $t = \{1,2,3\}$ (i.e., $p^{(t,m)} = \{3,6,9\}$ for all $m = \{1,2,3,4\}$). Note that these cases lead to the number of model coefficients ($\beta_{kj}^{(t,m)}$) ranging between $360$ and $6480 ~\forall ~k,j,t,m$, while the number of subjects is fixed at $n = 100$.
	
	Under these settings, we first use our proposed EM-based strategy for estimation across all three layers $t = \{1,2,3\}$ and model selection for 10 different values of $v_0$. The computations were performed on a computer with an Intel(R) Core(TM) i7-7700 CPU @ 3.60GHz, 3601 Mhz, 4 cores, 8 logical processors with 32 GB RAM; model selection for $10$ values of $v_0$ was done in parallel across $5$ processors. We obtain the time taken for the complete evaluation (estimation and model selection) and denote it as $T_{EM}$. We then run a MCMC sampler (Gibbs sampling with Metropolis-Hastings sampling for $\lambda$s) for the same data for the layer $t = 1$, and count the number of MCMC samples generated in time $T_{EM}$ for each parameter. Note that, since the estimation has to happen sequentially for each layer $t$, we report the number of samples generated only for the first layer. Additionally, the sampling approach is carried out for a fixed value of $v_0$ and does not perform model selection.
	
\begin{figure}[!t]
		\centering
		\begin{tabular}{|cc|}
			\hline
			&  \\
			\includegraphics[trim = 0cm 0cm 3.7in 0cm, clip, scale=1]{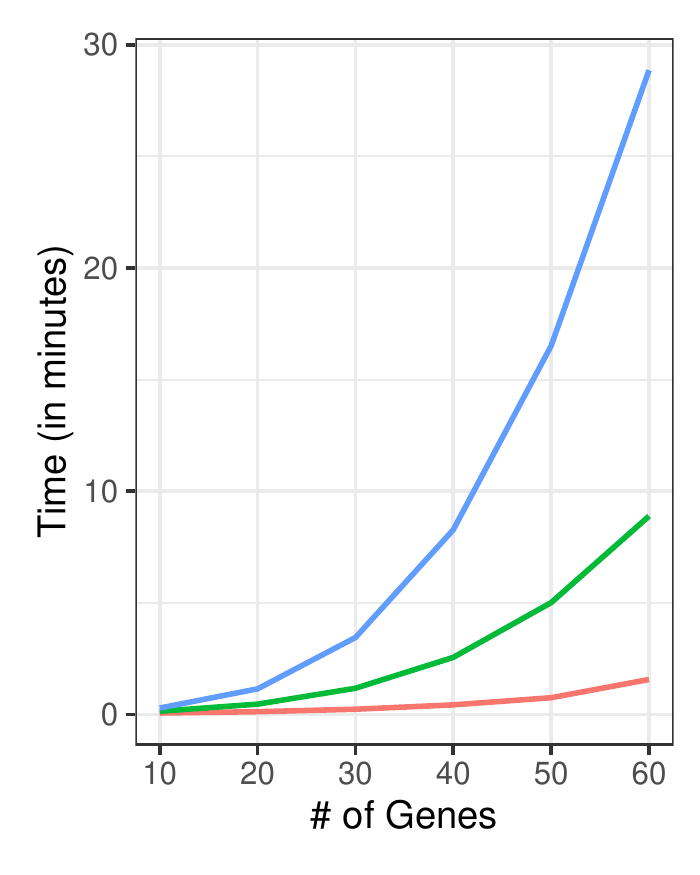} & \includegraphics[trim = 2.75in 0cm 0cm 0cm, clip, scale=1]{plots/em_mcmc_comparison.pdf} \\
			\begin{tabular}[c]{@{}c@{}}(a) Average time taken for EM-based \\ estimation and model selection.\end{tabular} & \begin{tabular}[c]{@{}c@{}}(b) Average number of MCMC samples\\ generated within the time taken in (a).\end{tabular} \\
			\hline
		\end{tabular}
		\caption{Comparison of EM-based estimation (and model selection) and MCMC sampling-based estimation approach. The left panel shows the average time taken to perform EM-based estimation and model selection across 30 replications for different choices of number of covariates ($g$) and number of components in the multivariate response ($p^{(t)}$).}
		\label{fig: em_mcmc}
	\end{figure}

	We repeat the process described above for 30 replications and present the results of this comparison in Figure \ref{fig: em_mcmc}. The left panel in Figure \ref{fig: em_mcmc} shows the average time taken to perform the EM-based estimation (across three layers) and model selection (for 10 values of $v_0$), that is, the average value of $T_{EM}$ across 30 replications. The right panel shows the average number of MCMC samples generated for all parameters of the first layer in time $T_{EM}$ for a fixed value of $v_0$. For example, when the number of genes is 50 and the number of principal components in each layer is 36 (blue curves in Figure \ref{fig: em_mcmc}), then the average time taken for EM-based approach ($T_{EM}$) is about 16 minutes. However, under the same settings, we are only able to generate about 400 MCMC samples for all parameters of layer $t=1$ in the same 16 minutes. Note that we do not perform any sampling for parameters of layers $t = 2$ and $3$. That is, if the tumor region is divided in $\tau$ layers, the sampling time for all the parameters for layers $2$ to $\tau$ will be additive with each additional layer. This clearly shows that the sampling-based approach is not able to generate more than a few hundred samples in the time needed for the EM-based approach to perform estimation and model selection. This demonstrates the underwhelming performance and infeasibility of the sampling-based approach and hence the utility of the EM-based estimation and model selection.
	
	In our analysis of LGGs, we divided the tumor region into $\tau = 3$ layers which leads to $p^{(1)} = 22, p^{(2)} = 18, p^{(3)} = 17$ PC scores and we considered $g = 24$ genes. From Figure \ref{fig: em_mcmc}(a), we note that when the number of principal components $p^{(t)}$ (number of columns in the multivariate response) is small then the average time taken for the EM-based approach does not increase rapidly as the number of genes $g$ (number of covariates) increases. Moreover, the value of $p^{(t)}$ can be controlled based on the choice of the amount of variance to be explained by the PCs, and it is usually not very large as seen in the context of the analysis of LGGs.
	
	\FloatBarrier
	
\end{document}